\documentclass[twocolumn,appendixfloats,twocolappendix]{aastex63}

\usepackage{natbib}
\usepackage{amsmath,amssymb}
\usepackage{xspace}
\usepackage{rotating}
\usepackage{multirow}
\usepackage{dcolumn}
\usepackage{bm}

\definecolor{mpl_blue}{HTML}{1F77B4}
\definecolor{mpl_orange}{HTML}{FF7F0E}
\definecolor{mpl_green}{HTML}{2CA02C}
\definecolor{mpl_red}{HTML}{D62728}
\definecolor{rbp}{HTML}{663399}

\newcommand{\AgwFixedGammaNinetyPercentConfidence}[0]{2.4_{-0.6}^{+0.7}\times10^{-15}}
\newcommand{\AgwVariableGammaNinetyPercentConfidence}[0]{6.4_{-2.7}^{+4.2}\times10^{-15}}
\newcommand{\gammagwVariableGammaNinetyPercentConfidence}[0]{3.2_{-0.6}^{+0.6}}

\newcommand{\RhoGWFixedGammaNinetyPercentConfidence}[0]{7.7_{-3.3}^{+4.8}\times 10^{-17}\,\textrm{ergs\,cm}^{-3}}
\newcommand{\OmegaGWFixedGammaNinetyPercentConfidence}[0]{9.3^{+5.8}_{-4.0} \times 10^{-9}}

\newcommand{\logAgwVariableGammaAreciboSixtyEightPercent}[0]{-14.02^{+0.18}_{-0.22}}
\newcommand{\logAgwVariableGammaGBTSixtyEightPercent}[0]{-14.2^{+0.15}_{-0.17}}

\newcommand{\gammagwAreciboSixtyEightPercent}[0]{2.78^{+0.70}_{-0.64}}
\newcommand{\gammagwGBTSixtyEightPercent}[0]{3.37^{+0.40}_{-0.38}}

\hypersetup{backref, breaklinks, plainpages=false, colorlinks=true, anchorcolor=cyan, linkcolor=mpl_red, citecolor=cyan, urlcolor=cyan, bookmarks=false}
\usepackage[all]{hypcap}
\usepackage[caption=false]{subfig}
\newcommand{\m}[1]{\mathbf{#1}}

\def\be{\begin{equation}}
\def\ee{\end{equation}}
\newcommand{\bb}{\begin{bmatrix}}
\newcommand{\eb}{\end{bmatrix}}
\def\bea{\begin{eqnarray}}
\def\eea{\end{eqnarray}}

\def\R{\color{red}}

\newcommand{\Agw}{\ensuremath{A_\mathrm{GWB}}}

\newcommand{\modelirn}{\textsc{irn}}
\newcommand{\modelcurn}{\textsc{curn}}
\newcommand{\modelcurngamma}{\textsc{curn}${}^\gamma$}
\newcommand{\modelcurngw}{\textsc{curn}${}^{13/3}$}
\newcommand{\modelcurnfree}{\textsc{curn}${}^\mathrm{free}$}
\newcommand{\modelcurnturnover}{\textsc{curn}${}^\mathrm{turnover}$}
\newcommand{\modelcurntps}{\textsc{curn}${}^\mathrm{TPS}$}
\newcommand{\modelhd}{\textsc{hd}}
\newcommand{\modelhdgamma}{\textsc{hd}${}^\gamma$}
\newcommand{\modelhdgw}{\textsc{hd}${}^{13/3}$}
\newcommand{\modelhdfree}{\textsc{hd}${}^\mathrm{free}$}

\newcommand{\modelmonofree}{\textsc{monopole}${}^\mathrm{free}$}
\newcommand{\modeldipolefree}{\textsc{dipole}${}^\mathrm{free}$}
\newcommand{\modelsinusoid}{\textsc{sinusoid}}

\defcitealias{abb+15}{NG9}
\defcitealias{abb+16}{NG9gwb}
\defcitealias{abb+18a}{NG11}
\defcitealias{abb+18b}{NG11gwb}
\defcitealias{aab+20}{NG12}
\defcitealias{abb+20}{NG12gwb}
\defcitealias{aaa+23}{NG15}
\defcitealias{aaa+23_noise}{NG15detchar}

\begin{document}

\title{The NANOGrav 15-year Data Set: Evidence for a Gravitational-Wave Background}

\shorttitle{NANOGrav 15-year Gravitational-Wave Background}
\shortauthors{The NANOGrav Collaboration}

\author[0000-0001-5134-3925]{Gabriella Agazie}
\affiliation{Center for Gravitation, Cosmology and Astrophysics, Department of Physics, University of Wisconsin-Milwaukee,\\ P.O. Box 413, Milwaukee, WI 53201, USA}
\author[0000-0002-8935-9882]{Akash Anumarlapudi}
\affiliation{Center for Gravitation, Cosmology and Astrophysics, Department of Physics, University of Wisconsin-Milwaukee,\\ P.O. Box 413, Milwaukee, WI 53201, USA}
\author[0000-0003-0638-3340]{Anne M. Archibald}
\affiliation{Newcastle University, NE1 7RU, UK}
\author{Zaven Arzoumanian}
\affiliation{X-Ray Astrophysics Laboratory, NASA Goddard Space Flight Center, Code 662, Greenbelt, MD 20771, USA}
\author[0000-0003-2745-753X]{Paul T. Baker}
\affiliation{Department of Physics and Astronomy, Widener University, One University Place, Chester, PA 19013, USA}
\author[0000-0003-0909-5563]{Bence B\'{e}csy}
\affiliation{Department of Physics, Oregon State University, Corvallis, OR 97331, USA}
\author[0000-0002-2183-1087]{Laura Blecha}
\affiliation{Physics Department, University of Florida, Gainesville, FL 32611, USA}
\author[0000-0001-6341-7178]{Adam Brazier}
\affiliation{Cornell Center for Astrophysics and Planetary Science and Department of Astronomy, Cornell University, Ithaca, NY 14853, USA}
\affiliation{Cornell Center for Advanced Computing, Cornell University, Ithaca, NY 14853, USA}
\author[0000-0003-3053-6538]{Paul R. Brook}
\affiliation{Institute for Gravitational Wave Astronomy and School of Physics and Astronomy, University of Birmingham, Edgbaston, Birmingham B15 2TT, UK}
\author[0000-0003-4052-7838]{Sarah Burke-Spolaor}
\affiliation{Department of Physics and Astronomy, West Virginia University, P.O. Box 6315, Morgantown, WV 26506, USA}
\affiliation{Center for Gravitational Waves and Cosmology, West Virginia University, Chestnut Ridge Research Building, Morgantown, WV 26505, USA}
% ...

\collaboration{1000}{The NANOGrav Collaboration}
\noaffiliation

\correspondingauthor{The NANOGrav Collaboration}
\email{comments@nanograv.org}

\begin{abstract}
We report multiple lines of evidence for a stochastic signal that is correlated among 67 pulsars from the 15-year pulsar-timing data set collected by the North American Nanohertz Observatory for Gravitational Waves. The correlations follow the Hellings--Downs pattern expected for a stochastic gravitational-wave background. The presence of such a gravitational-wave background with a power-law--spectrum is favored over a model with only independent pulsar noises with a Bayes factor in excess of $10^{14}$, and this same model is favored over an uncorrelated common power-law--spectrum model with Bayes factors of 200--1000, depending on spectral modeling choices. We have built a statistical background distribution for these latter Bayes factors using a method that removes inter-pulsar correlations from our data set, finding $p = 10^{-3}$ (approx.\ $3 \sigma$) for the observed Bayes factors in the null no-correlation scenario.
A frequentist test statistic built directly as a weighted sum of inter-pulsar correlations yields $p=5 \times 10^{-5}$--$1.9\times10^{-4}$ (approx.\ $3.5$--$4\sigma$). Assuming a fiducial $f^{-2/3}$  characteristic-strain spectrum, as appropriate for an ensemble of binary supermassive black-hole inspirals, the strain amplitude is $2.4_{-0.6}^{+0.7}\times10^{-15}$ (median + 90\% credible interval) at a reference frequency of 1 $\mathrm{yr}^{-1}$.  
The inferred gravitational-wave background amplitude and spectrum are consistent with astrophysical expectations for a signal from a population of supermassive black-hole binaries, although more exotic cosmological and astrophysical sources cannot be excluded. The observation of Hellings--Downs correlations points to the gravitational-wave origin of this signal. 
\end{abstract}


\section{Introduction}
\label{sec:intro}
 
\begin{figure*}[t]
	\begin{tabular}{cc}
		\hspace{32pt} {\normalsize \bf (a)} & \hspace{32pt} {\normalsize \bf (c)} \\
		\includegraphics[width=\columnwidth]{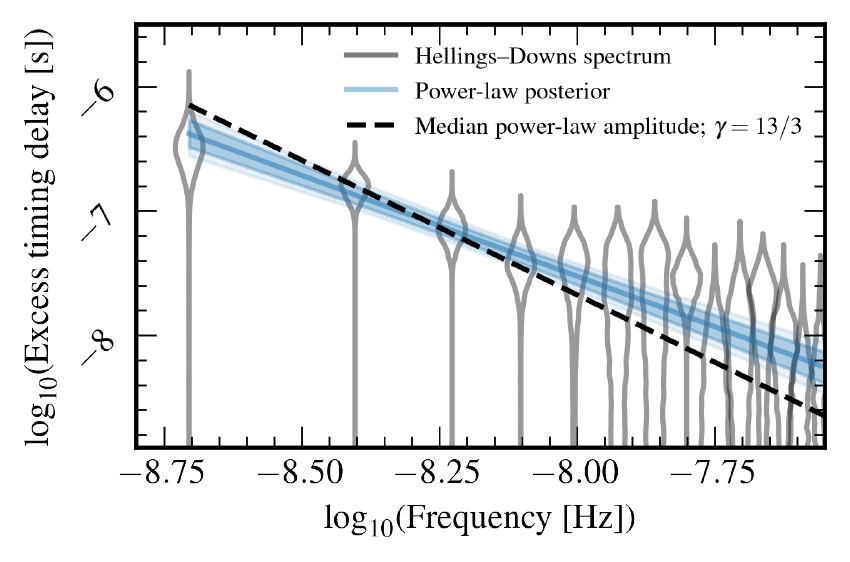} &
		\includegraphics[width=\columnwidth]{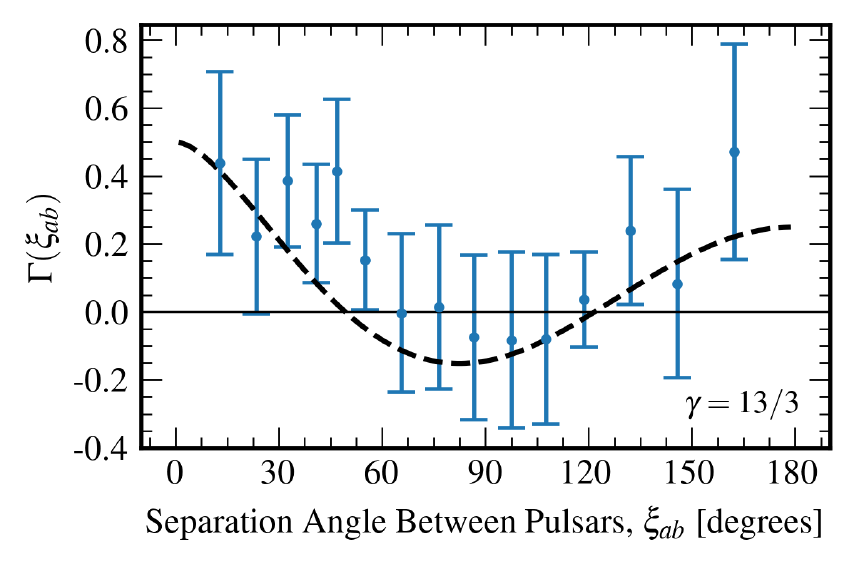} \\
\hspace{32pt} {\normalsize \bf (b)} & \hspace{32pt} {\normalsize \bf (d)} \\[3pt]
		\includegraphics[width=\columnwidth]{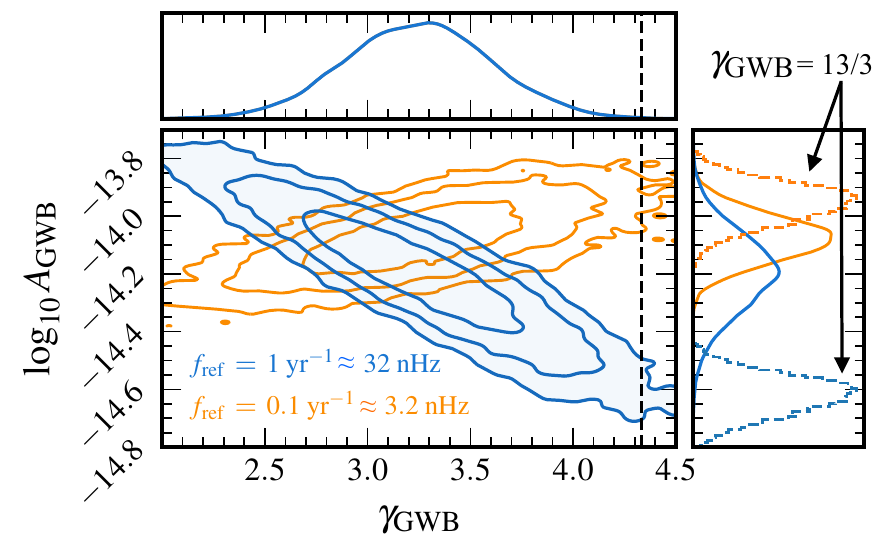} &
		\includegraphics[width=\columnwidth]{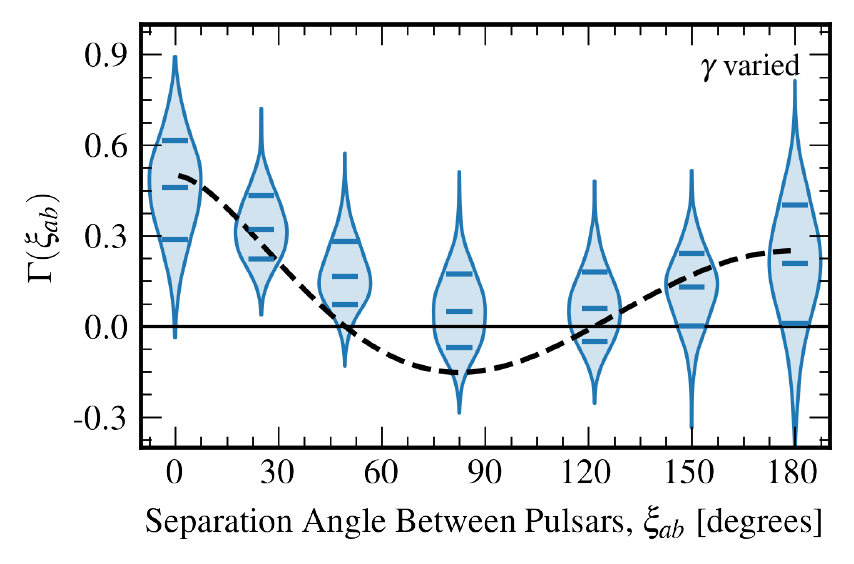}		
	\end{tabular} \vspace{-6pt}
	\caption{Summary of the main Bayesian and optimal-statistic analyses presented in this paper, which establish multiple lines of evidence for the presence of Hellings--Downs correlations in the 15-year NANOGrav data set. Throughout we refer to the $68.3\%$, $95.4\%$, and $99.7\%$ regions of distributions as $1/2/3\sigma$ regions, even in two dimensions.
	(a): Bayesian ``free-spectrum'' analysis, showing posteriors (gray violins) of independent variance parameters for a Hellings--Downs-correlated stochastic process at frequencies $i/T$, with $T$ the total data set time span.
	The blue represents the posterior median and $1/2\sigma$ posterior bands\textsuperscript{a} for a power-law model; the dashed black line corresponds to a $\gamma=13/3$ (SMBHB-like) power-law, plotted with the median posterior amplitude. See \S\ref{sec:bayes} for more details.
	(b): Posterior probability distribution of GWB amplitude and spectral exponent in a HD power-law model, showing $1/2/3\sigma$ credible regions. 
	The value $\gamma_\mathrm{GWB}=13/3$ (dashed black line) is included in the $99\%$ credible region.
	The amplitude is referenced to $f_\mathrm{ref}=1\,\mathrm{yr}^{-1}$ (blue) and $0.1\,\mathrm{yr}^{-1}$ (orange).
	The dashed blue and orange curves in the $\log_{10} A_\mathrm{GWB}$ subpanel shows its marginal posterior density for a $\gamma = 13/3$ model, with $f_\mathrm{ref}=1\,\mathrm{yr}^{-1}$ and $f_\mathrm{ref}=0.1\,\mathrm{yr}^{-1}$, respectively. See \S\ref{sec:bayes} for more details.
	(c): Angular-separation--binned inter-pulsar correlations, measured from 2,211 distinct pairings in our 67-pulsar array using the frequentist optimal statistic, assuming maximum-a-posteriori pulsar noise parameters and $\gamma=13/3$ common-process amplitude from a Bayesian inference analysis. The bin widths are chosen so that each includes approximately the same number of pulsar pairs, and central bin locations avoid zeros of the Hellings--Downs curve. This binned reconstruction accounts for correlations between pulsar pairs \citep{romano+2021,2022arXiv220807230A}.
	The dashed black line shows the Hellings--Downs correlation pattern, and the binned points are normalized by the amplitude of the $\gamma=13/3$ common process to be on the same scale. Note that we do not employ binning of inter-pulsar correlations in our detection statistics; this panel serves as a visual consistency check only. See \S\ref{sec:optimal} for more frequentist results.
	(d): Bayesian reconstruction of normalized inter-pulsar correlations, modeled as a cubic spline within a variable-exponent power-law model. The violins plot the marginal posterior densities (plus median and 68\% credible values) of the correlations at the knots. The knot positions are fixed, and are chosen on the basis of features of the Hellings--Downs curve (also shown as a dashed black line for reference): they include the maximum and minimum angular separations, the two zero crossings of the Hellings--Downs curve, and the position of minimum correlation.
	See \S\ref{sec:bayes} for more details. 
	\label{fig:spectrum_correlations_plot}}
\vspace{12pt}
\end{figure*}

Almost a century had to elapse between Einstein’s prediction of gravitational waves (GWs, \citealt{einstein1916approximative}) and their measurement from a coalescing binary of stellar-mass black holes \citep{2016PhRvL.116f1102A}. However, their existence had been confirmed in the late 1970s through measurements of the orbital decay of the Hulse--Taylor binary pulsar \citep{1975ApJ...195L..51H,1979Natur.277..437T}. Today, pulsars are again at the forefront of the quest to detect GWs, this time from binary systems of central galactic black holes.

Black holes with masses of $10^5$--$10^{10} \, M_\odot$
exist at the center of most galaxies and are closely correlated with the global properties of the host, suggesting a symbiotic evolution \citep{1998AJ....115.2285M,2013ApJ...764..184M}. 
Galaxy mergers are the main drivers of hierarchical structure formation over cosmic time \citep{blumenthal84} and lead to the formation of close massive--black-hole binaries long after the mergers \citep{bbr80,mm03}.
The most massive of these (supermassive black-hole binaries, SMBHBs, with masses $10^8$--$10^{10}\,M_\odot$)
emit GWs with slowly evolving frequencies, contributing to a noise-like broadband signal in the nHz range (the GW background, GWB; \citealt{rr95,jb03,wl03,shm+04,2014ApJ...789..156M,stc+19}).
If all contributing SMBHBs evolve purely by loss of circular orbital energy to gravitational radiation, the resultant GWB spectrum is well described by a simple $f^{-2/3}$ characteristic-strain power law \citep{p01}. However, GWB signals that are not produced by populations of inspiraling black holes may also lie within the nHz band; these include primordial GWs from inflation, scalar-induced GWs, and GW signals from multiple processes arising due to cosmological phase transitions, such as collisions of bubbles of the post-transition vacuum state, sound waves, turbulence, and the decay of any defects such as cosmic strings or domain walls that may have formed (see, e.g., \citealt{guzzetti2016gravitational,2018CQGra..35p3001C,2021Univ....7..398D}, and references therein).

The detection of nHz GWs follows the template outlined by \citet{1956AcPP...15..389P,2009GReGr..41.1215P}, whereby we time the propagation of light to measure modulations in the distance between freely falling reference masses.
\citet{1975GReGr...6..439E} derived the GW response of electromagnetic signals traveling between Earth and distant spacecraft, sparking interest in low-frequency GW detection.
\citet{saz78} and \citet{det79} described nHz GW detection using Galactic pulsars and (effectively) the solar system barycenter as references, relying on the regularity of pulsar emission and planetary motions to highlight GW effects.
The fact that pulsars are such accurate clocks enables precise measurements of their rotational, astrometric, and binary parameters (and more) from the times-of-arrival of their pulses, which are used to develop ever-refining end-to-end \emph{timing models}.
\citet{hd83} made the crucial suggestion that the correlations between the time-of-arrival perturbations of multiple pulsars could reveal a GW signal buried in pulsar noise; \citet{romani1989timing} and \citet{fb90} proposed that a \emph{pulsar timing array} (PTA) of highly stable millisecond pulsars \citep{1982Natur.300..615B} could be used to search for a GWB.
Nevertheless, the first multi-pulsar, long-term GWB limits were obtained by analyzing millisecond-pulsar residuals independently, rather than as an array \citep{1990PhRvL..65..285S,1994ApJ...428..713K}.

From a statistical-inference standpoint, the problem of detecting nHz GWs in PTA data is analogous to GW searches with terrestrial and future space-borne experiments, in which the propagation of light between reference masses is modeled with physical and phenomenological descriptions of signal and noise processes. It is distinguished by the irregular observation times, which encourage a time- rather than Fourier-domain formulation, and by noise sources (intrinsic pulsar noise, interstellar-medium--induced radio-frequency--dependent fluctuations, and timing-model errors) that are correlated on timescales common to the GWs of interest. This requires the joint estimation of GW signals and noise, which is similar to the kinds of global fitting procedures already used in terrestrial GW experiments, and proposed for space-borne experiments. 
GW analysts have therefore converged on a Bayesian framework that represents all noise sources as Gaussian processes \citep{2009MNRAS.395.1005V,2014PhRvD..90j4012V}, and relies on model comparison (i.e., Bayes factors, which are ratios of fully marginalized likelihoods) to define detection (see, e.g., \citealt{taylor2021nanohertz}).
This Bayesian approach is nevertheless complemented by null hypothesis testing, using a frequentist detection statistic\footnote{See \citet{2006ApJ...653.1571J} for an early example of a cross-correlation statistic for PTA GWB detection.} (the ``optimal statistic'' of \citealt{abc+2009,dfg+13,ccs+2015}) averaged over Bayesian posteriors of the noise parameters \citep{vite18}.

The GWB---rather than GW signals from individually resolved binary systems---is likely to become the first nHz source accessible to PTA observations \citep{rsg2015}. Because of its stochastic nature, the GWB cannot be identified as a distinctive phase-coherent signal in the way of individual compact-binary-coalescence GWs.
Rather, as PTA data sets grow in extent and sensitivity one expects to first observe the GWB as excess low-frequency residual power of consistent amplitude and spectral shape across multiple pulsars \citep{romano+2021,astro4cast}.
An observation following this behavior was reported in $2020$ (\citealt{abb+20}, henceforth \citetalias{abb+20}) 
for the 12.5-year data set collected by
the North American Nanohertz Observatory for Gravitational waves \citep[NANOGrav,][]{m13,ransom+19}, and then confirmed \citep{ppta_dr2_gwb,epta_dr2_gwb} by the Parkes Pulsar Timing Array \citep[PPTA,][]{2013PASA...30...17M} and the European Pulsar Timing Array \citep[EPTA,][]{dcl+16}, following many years of null results and steadily decreasing upper limits on the GWB amplitude. A combined International Pulsar Timing Array \citep[IPTA,][]{pdd+19} data release consisting of older data sets from the constituent PTAs also confirmed this observation \citep{2022MNRAS.510.4873A}.
Nevertheless, the finding of excess power cannot be attributed to a GWB origin merely by the consistency of amplitude and spectral shape, which could arise from intrinsic pulsar processes of similar magnitude \citep{gts+22,zhs+22}, or from a common systematic noise such as clock errors \citep{thk+2016}. 
Instead, definitive proof of GW origin is sought by establishing the presence of phase-coherent inter-pulsar correlations with the characteristic spatial pattern derived by Hellings and Downs (\citeyear[henceforth HD]{hd83}): for an isotropic GWB, the correlation between the GW-induced timing delays observed at Earth for any pair of pulsars is a universal, quasi-quadrupolar function of their angular separation in the sky. Even though this correlation pattern is modified if there is anisotropy in the GWB---which may be the case for a GWB generated by a SMBHB population \citep{2013PhRvD..88f2005M,2013PhRvD..88h4001T,Cornish:2013aba,2014PhRvD..90f2011M,2017NatAs...1..886M,2017ApJ...835...21R}---the HD template is effective for detecting even anisotropic GWBs in all but the most extreme scenarios \citep{Cornish:2013aba,cs16,2020PhRvD.102h4039T,2022ApJ...941..119B,2023PhRvD.107d3018A}.

In this letter we present multiple lines of evidence for an excess low-frequency signal with HD correlations in the NANOGrav 15-year data set (\autoref{fig:spectrum_correlations_plot}). Our key results are as follows. The Bayes factor between an HD-correlated, power-law GWB model and a spatially uncorrelated common-spectrum power-law model ranges from 200 to 1,000, depending on modeling choices (\autoref{fig:bfs}). The noise-marginalized optimal statistic, which is constructed to be selectively sensitive to HD-correlated power, achieves a signal-to-noise ratio of $\sim$~5 (\autoref{fig:background} and \autoref{fig:optstat}).
We calibrated these detection statistics
by removing correlations from the 15-year data set 
using the phase-shift technique, 
which removes inter-pulsar correlations by adding random phase shifts 
to the Fourier components of the common process \citep{tlb+17}. We find false-alarm probabilities of $p = 10^{-3}$ and $p = 5 \times 10^{-5}$ 
for the observed Bayes factor and optimal statistic, respectively (see \autoref{fig:background}).

For our fiducial power-law model ($f^{-2/3}$ for characteristic strain and $f^{-13/3}$ for timing residuals) and a log-uniform amplitude prior, the Bayesian posterior of GWB amplitude at the customary reference frequency 1 $\mathrm{yr}^{-1}$ 
is $\Agw=\AgwFixedGammaNinetyPercentConfidence$
(median with 90\% credible interval), which is compatible with current astrophysical estimates for the GWB from SMBHBs \citep[e.g.,][]{stc+19,aaa+23_smbhb}. This corresponds to a total integrated energy density of $\Omega_{\textrm{gw}} = \OmegaGWFixedGammaNinetyPercentConfidence$ or $\rho_{\textrm{gw}}=\RhoGWFixedGammaNinetyPercentConfidence$ (assuming $H_0 = 70 \; \mathrm{km}/\mathrm{s}/\mathrm{Mpc}$) in our sensitive frequency band.
For a more general model of the timing-residual power spectral density with variable power-law exponent $-\gamma$, we find $\Agw=\AgwVariableGammaNinetyPercentConfidence$, and $\gamma=\gammagwVariableGammaNinetyPercentConfidence$. 
See panel (b) of \autoref{fig:spectrum_correlations_plot} for $\Agw$ and $\gamma$ posteriors. 
The posterior for $\gamma$ is consistent with the value of $13/3$ 
predicted for a population of SMBHBs evolving by GW emission, although smaller values of $\gamma$ are preferred; 
however, the recovered posteriors are consistent with predictions from astrophysical models (see \citealt{aaa+23_smbhb}). 
We also note that, unlike our detection statistics (which are calibrated under our modeling assumptions), 
the estimation of $\gamma$ is very sensitive to minor details in the data model of a few pulsars.

The rest of this paper is organized as follows. We briefly describe our data set and data model in \S\ref{sec:data}. Our main results are discussed in detail in \S\ref{sec:bayes} and \S\ref{sec:optimal}; they are supported by a variety of robustness and validation studies, including a spectral analysis of the excess signal (\S\ref{subsec:spectral}), a correlation analysis that finds no significant evidence for additional spatially correlated processes (\S\ref{subsec:correlation}), and cross-validation studies with single-telescope data sets and leave-one-pulsar-out techniques (\S\ref{subsec:cross-validation}).
In the past two years we have performed an end-to-end review of the NANOGrav experiment, to identify and mitigate possible sources of systematic error or data set contamination: our improvements and considerations are partly described in a set of companion papers: on the NANOGrav statistical analysis as implemented in software \citep{code_review}, on the 15-year data set (\citealt{aaa+23}, hereafter \citetalias{aaa+23}), and on pulsar models (\citealt{aaa+23_noise}, hereafter \citetalias{aaa+23_noise}).
More companion papers address the possible SMBHB \citep{aaa+23_smbhb} and cosmological \citep{aaa+23_cosmo} interpretations of our results, with several more GW searches and signal studies in preparation. 
We look forward to the cross-validation analysis that will become possible with the independent data sets collected by other IPTA members.

\section{The 15-year Data Set and Data Model}
\label{sec:data}

\newcommand{\tresid}[0]{\bm{\delta t}}
\newcommand{\avec}[0]{\bm{a}}
\newcommand{\cvec}[0]{\bm{c}}
\newcommand{\rvec}[0]{\bm{r}}
\newcommand{\fmat}[0]{\bm{F}}
\newcommand{\cmat}[0]{\bm{C}}
\newcommand{\mmat}[0]{\bm{M}}
\newcommand{\dxivec}[0]{\bm{\delta\xi}}
\newcommand{\etavec}[0]{\bm{\eta}}
\newcommand{\epsilonvec}[0]{\bm{\epsilon}}

The NANOGrav 15-year data set\footnote{While the time between the first and last observations we analyze is 16.03 years, this data set is named ``15-year data set'' since no single pulsar exceeds 16 years of observation; we will use this nomenclature despite the discrepancy.}
\citepalias{aaa+23} contains observations of 68 pulsars obtained between 
July 2004 and August 2020 with the Arecibo Observatory (Arecibo), the Green Bank Telescope (GBT), and the Very Large Array (VLA), augmenting the 12.5-year data set \citep{aab+20,12yr_wideband} with $2.9$ years of timing data for the 47 pulsars in the previous data set, and with 21 new pulsars\footnote{The data set is available at \href{http://data.nanograv.org}{data.nanograv.org} with the code used to process it.}.
For this paper we analyze narrowband times of arrival (TOAs), which are computed separately for sub-bands of each receiver, and focus on the 67 pulsars with a timing baseline $\geq 3$\,years.
We adopt the TT(BIPM2019) timescale and the JPL DE440 ephemeris \citep{2021AJ....161..105P}, which improves Jupiter's orbit with ranging and VLBI observations of the Juno spacecraft.
Uncertainties in the Jovian orbit impacted NANOGrav's 11-year GWB search \citep{abb+18b,2020ApJ...893..112V}, but they are now negligible.

For each pulsar, we fit the TOAs to a timing model that includes pulsar spin period, spin period derivative, sky location, proper motion, and parallax. 
While not all pulsars have measurable parallax and proper motion, we always include these parameters because they induce delays with the same frequencies for all pulsars ($f=0.5 \, \mathrm{yr}^{-1}$ for parallax and $f=\mathrm{yr}^{-1}$ plus a linear envelope for proper motion), so there is a risk that a parallax or proper motion signal could be misidentified as a GW signal.
Fitting for these parameters in all pulsars reduces our sensitivity to GWs at those frequencies; 
however, this effect is minimal for GWB searches 
since these frequencies are much higher than the frequencies at which we expect the GWB to be significant. 
For binary pulsars, the timing model includes also five orbital elements for binary pulsars and additional non-Keplerian parameters when these improve the fit as determined by an $F$ test. 
We fit variations in dispersion measure as a piecewise constant ``DMX'' function \citep{abb+15,jml+2017}.
The individual analysis of each pulsar provides best-fit estimates of the timing residuals $\tresid$, of white measurement noise, and of intrinsic red noise, modeled as a power law \citep{cordes2013, lcc+2017, jml+2017}.\footnote{Throughout the paper we use ``red noise'' to describe noise whose power spectrum decreases with increasing frequency.}
White measurement noise is described by three parameters: a linear scaling of TOA uncertainties (``EFAC''), 
white noise added to the TOA uncertainties in quadrature (``EQUAD''), and 
noise common to all sub-bands at the same epoch (``ECORR''), 
with independent parameters for every receiver/backend combination (see \citetalias{aaa+23_noise}). 
We summarize white noise by its maximum \textit{a posteriori} (MAP) covariance $\cmat$.
See App.\ \ref{app:data_set_details} for more details of our instruments, observations, and data-reduction pipeline: a complete discussion of the data set can be found in \citetalias{aaa+23}. 

In our Bayesian GWB analysis, we model $\tresid$ as a finite Gaussian process consisting of time-correlated fluctuations that include intrinsic red pulsar noise and (potentially) a GW signal, along with timing-model uncertainties \citep{2009MNRAS.395.1005V, 2014PhRvD..90j4012V, taylor2021nanohertz}. The red noise is modeled with Fourier basis $\fmat$ and amplitudes $\cvec$ \citep{lentati+2013}. All Fourier bases (the columns of $\fmat$) are sines and cosines computed on the TOAs with frequencies $f_i = i/T$, where $T= 16.03$ yr is the TOA extent. The timing-model uncertainties are modeled with design-matrix basis $\mmat$ and coefficients $\epsilonvec$. The single-pulsar log likelihood is then
\begin{align}
    \ln p(\tresid | \cvec, \epsilonvec) = -\frac{1}{2}\left[\rvec^T\cmat^{-1}\rvec + \ln \mathrm{det}\left(2\pi\cmat \right) \right],
    \label{eq:likelihood}
\end{align}
with
\begin{align}
    \rvec = \tresid - \fmat\cvec - \mmat \epsilonvec.
\end{align}
The prior for the $\epsilonvec$ is taken to be uniform with infinite extent, so the posterior is driven entirely by the likelihood. The set of the $\{\cvec\}$ for all pulsars take a joint normal prior with zero mean and covariance
\begin{align}
    \langle c_{ai} c_{bj}\rangle = \delta_{ij} \left(
    \delta_{ab}\varphi_{ai} + \Phi_{ab,i}
    \right);\label{eq:gp_covariance}
\end{align}
here $a,b$ range over pulsars and $i,j$ over Fourier components; $\delta_{ij}$ is Kronecker's delta.
The term $\varphi_{ai}$ describes the spectrum of intrinsic red noise in pulsar $a$, while $\Phi_{ab,i}$ describes processes with common spectrum across all pulsars and (potentially) phase-coherent inter-pulsar correlations.
The $\{\cvec\}$ prior ties together the single-pulsar likelihoods (\autoref{eq:likelihood}) into a joint posterior, $p(\cvec, \epsilonvec, \etavec | \tresid)\propto p(\tresid | \cvec, \epsilonvec)p(\cvec,\epsilonvec|\etavec)p(\etavec)$, where we have dropped subscripts to denote the concatenation of vectors for all pulsars, and where $\etavec$ denotes all the hyperparameters (such as red-noise and GWB power spectrum amplitudes) that determine the covariances.
We marginalize over $\cvec$ and $\epsilonvec$ analytically, and use Markov chain Monte Carlo techniques (see App.\ \ref{sec:bayesapp}) to estimate $p(\etavec | \tresid)$ for different models of the intrinsic red noise and common spectrum.

The data-model variants adopted in this paper all share this probabilistic setup, but differ in the structure and parametrization of $\Phi_{ab,i}$.
For a model with intrinsic red noise only (henceforth \modelirn), $\Phi_{ab,i} = 0$; for common-spectrum spatially-uncorrelated red noise (\modelcurn), $\Phi_{ab,i} = \delta_{ab} \Phi_{\mathrm{CURN},i}$; for an isotropic GWB with Hellings--Downs correlations (\modelhd), $\Phi_{ab,i} = \Gamma(\xi_{ab}) \Phi_{\mathrm{HD},i}$, with $\Gamma$ the Hellings--Downs function of pulsar angular separations $\xi_{ab}$ 
\begin{align}
\Gamma(\xi_{ab}) &= \frac{3}{2}x\ln(x) - \frac{1}{4}x + \frac{1}{2} + \frac{1}{2}\delta_{ab},\\    
x&=\frac{1 - \cos \xi_{ab}}{2}.
\end{align}
In \citetalias{abb+20} we established strong Bayesian evidence for \modelcurn\ over \modelirn; finding that \modelhd\ is preferred over \modelcurn\ would point to the GWB origin of the common-spectrum signal. We also investigate other spatial correlation patterns, e.g., monopole or dipole, introduced in \S\ref{subsec:correlation}.

Throughout this paper, we set the spectral components $\varphi_{ai}$ of intrinsic pulsar noise (which have units of $\mathrm{s}^2$, as appropriate for the variance of timing residuals) to a power law,
\begin{align}
    \varphi_{ai} = \frac{A_a^2}{12\pi^2} \frac{1}{T} \! \left(\frac{f_i}{f_{\textrm{ref}}}\right)^{\!-\gamma_a}f_\textrm{ref}^{-3},
    \label{eq:powerlaw}
\end{align}
introducing two dimensionless hyperparameters for each pulsar: the intrinsic-noise amplitude $A_a$ and spectral index $\gamma_a$. We use log-uniform and uniform priors, respectively, on these hyperparameters; their bounds  are described in App.\ \ref{sec:bayesapp}. More sophisticated intrinsic-noise models are discussed in \S\ref{subsec:dm_models} and \citetalias{aaa+23_noise}.
In models \modelcurngamma\ and \modelhdgamma, the common spectra $\Phi_{\mathrm{CURN},i}$ and $\Phi_{\mathrm{HD},i}$ follow \autoref{eq:powerlaw}, 
\begin{align}
    \Phi_{\mathrm{CURN},i} &= \frac{A_{\mathrm{CURN}}^2}{12\pi^2} \frac{1}{T} \! \left(\frac{f_i}{f_{\textrm{ref}}}\right)^{\!-\gamma_{\mathrm{CURN}}} \!\!\! f_\textrm{ref}^{-3}, \\
     \Phi_{\mathrm{HD},i} &= \frac{A_{\mathrm{HD}}^2}{12\pi^2} \frac{1}{T} \! \left(\frac{f_i}{f_{\textrm{ref}}}\right)^{\!-\gamma_\mathrm{HD}} \!\!\! f_\textrm{ref}^{-3},
\end{align}
introducing hyperparameters $A_\mathrm{CURN}, \gamma_\mathrm{CURN}$ and $A_\mathrm{HD}, \gamma_\mathrm{HD}$ respectively.
However, we set $\gamma_\mathrm{HD} = 13/3$ for the GWB from a stationary ensemble of inspiraling binaries, and refer to that fiducial model as \modelhdgw.
For specific ``free spectrum'' studies we will instead model the individual $\Phi_{\mathrm{CURN},i}$ or $\Phi_{\mathrm{HD},i}$ elements, and refer to models \modelcurnfree\ and \modelhdfree.
Throughout this article we use frequencies $f_i = i/T$ with $i = 1\mbox{--}30$ for intrinsic noise ($f = 2\mbox{--}59\;\mathrm{nHz}$), covering a frequency range over which pulsar noise transitions from red-noise--dominated to white--noise-dominated. 
For common-spectrum noise, we limit the frequency range 
in order to reduce correlations with excess white noise at higher frequencies. 
Following \citetalias{abb+20}, we fit a \modelcurngamma\ model enhanced with a power-law break to our data, 
and limit frequencies to the MAP break frequencies ($i = 1\mbox{--}14$ or $f = 2\mbox{--}28\;\mathrm{nHz}$; see App.\ \ref{sec:appbroken}).

\section{Bayesian analysis}
\label{sec:bayes}

\begin{figure*}[t]
\begin{center}
    \includegraphics[width=0.9\textwidth]{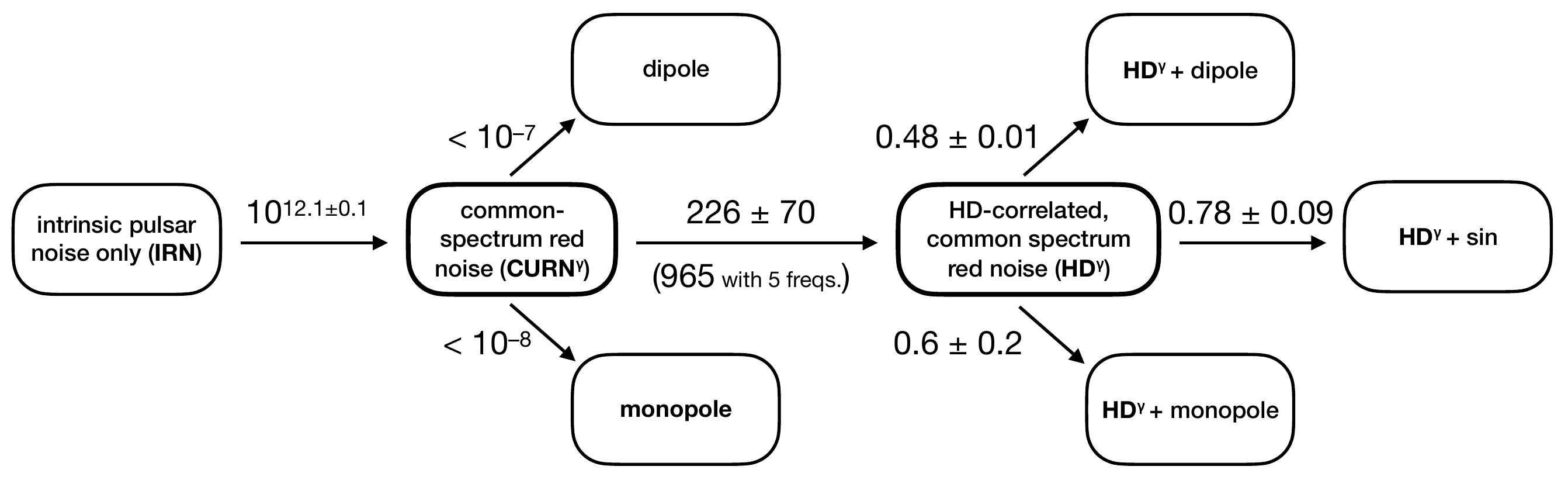} 
\end{center} \vspace{-18pt}
    \caption{Bayes factors between models of correlated red noise in the NANOGrav 15-year data set (see \S\ref{subsec:correlation} and App.~\ref{sec:bayesapp}). All models feature variable-$\gamma$ power laws. \modelcurngamma\ is vastly favored over \modelirn\ (i.e., we find very strong evidence for common-spectrum excess noise over pulsar intrinsic red-noise alone); \modelhdgamma\ is favored over \modelcurngamma\ (i.e., we find positive evidence for Hellings--Downs correlations in the common-spectrum process); dipole and monopole processes are strongly disfavored with respect to \modelcurngamma; adding correlated processes to \modelhdgamma\ is disfavored. While the interpretation of ``raw'' Bayes factors is somewhat subjective, they can be given a statistical significance within the hypothesis-testing framework by computing their background distributions and deriving the $p$-values of the observed factors, e.g., \autoref{fig:background}.
    \label{fig:bfs}}
\end{figure*}

When fit to the 15-year data set, the \modelcurngamma\ and \modelhdgamma\ models agree on the presence of a loud time-correlated stochastic signal with common amplitude and spectrum across pulsars\footnote{See App.\ \ref{sec:bayesapp} for details about our Bayesian methods, including the calculation of Bayes factors.}.
The joint $A_\mathrm{HD}$--$\,\gamma_\mathrm{HD}$ Bayesian posterior is shown in panel (b) of \autoref{fig:spectrum_correlations_plot}, with 1-D marginal posteriors in the horizontal and vertical subplots.
The posterior medians and 5--95\% quantiles are $A_\mathrm{HD} = \AgwVariableGammaNinetyPercentConfidence$ and $\gamma_\mathrm{HD} = \gammagwVariableGammaNinetyPercentConfidence$.
The thicker curve in the vertical subplot is the $A_\mathrm{HD}$ posterior for the \modelhdgw\ model, for which $A_{\mathrm{HD},13/3} = \AgwFixedGammaNinetyPercentConfidence$.
These amplitudes are compatible with astrophysical expectations of a GWB from inspiraling SMBHBs (see \S\ref{sec:discussion}).
The $A_\mathrm{HD}$ posterior has essentially no support below $10^{-15}$.

The strong $A_\mathrm{HD}$--$\,\gamma_\mathrm{HD}$ correlation is an artifact of using the conventional frequency $f_\mathrm{ref} = 1\,\mathrm{yr}^{-1}$ in \autoref{eq:powerlaw}, and it largely disappears when $f_\mathrm{ref}$ is moved to the band of greatest PTA sensitivity; see the dashed contours in panel (b) of \autoref{fig:spectrum_correlations_plot} for $f_\mathrm{ref} = (10 \, \mathrm{yr})^{-1}$.
The $\gamma_\mathrm{HD}$ posterior is in moderate tension with the theoretical universal binary-inspiral value $\gamma_\mathrm{HD}=13/3$, which lies at the $99$\% credible boundary: 
smaller values of $\gamma_\mathrm{HD}$ could be an indication that astrophysical effects, 
such as stellar scattering and gas dynamics, play a role in the evolution of SMBHBs emitting GWs in this frequency range 
(see \S\ref{sec:discussion} and \citealt{aaa+23_smbhb}). 
This highlights the importance of measuring this parameter. 
Furthermore, its estimation is sensitive to details in the modeling of intrinsic red noise and of interstellar-medium timing delays in a few pulsars (see the analysis in \S\ref{subsec:spectral}).
Notably, in the 12.5-year data set $\gamma_\mathrm{HD}=13/3$ was recovered at $\sim1\sigma$ below the median \citepalias{abb+20}; this anomaly is reversed in the newer data set.
It is likely that more expansive data sets or more sophisticated chromatic noise models, e.g., next generation Gaussian process models such as in \S\ref{subsec:dm_models} \citep{goncharov+21a,Chalumeau+2022,leg+2018}, will be needed to infer the presence of possible systematic errors in $\gamma_\mathrm{HD}$. 

\begin{figure*}
\begin{center}
    \includegraphics[width=0.95\textwidth]{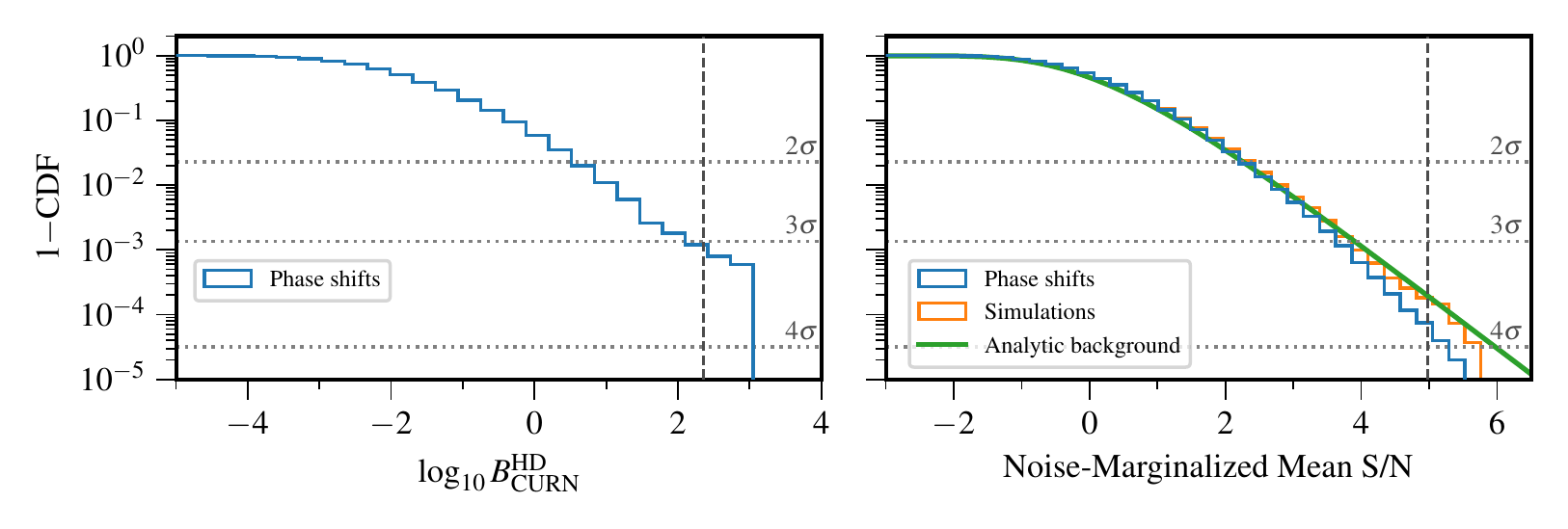} 
\end{center} \vspace{-18pt}
    \caption{Empirical background distribution of \modelhdgamma-to-\modelcurngamma\ Bayes factor (left, see \S\ref{sec:bayes}) and noise-marginalized optimal statistic (right, see \S\ref{sec:optimal}), as computed by the phase-shift technique \citep{tlb+17} to remove inter-pulsar correlations. We only compute 5,000 Bayesian phase shifts, compared to 400,000 optimal statistic phase shifts, 
because of the huge computational resources needed to perform the Bayesian analyses. For the optimal statistic, we also compute the background distribution using 27,000 simulations (orange line) and compare to an analytic calculation (green line). Dotted lines indicate Gaussian-equivalent 2$\sigma$, 3$\sigma$, and 4$\sigma$ thresholds. The dashed vertical lines indicate the values of the detection statistics for the unshifted data sets. For the Bayesian analyses, we find $p=10^{-3}$ (approx.\ $3\sigma$); for the optimal statistic analyses, we find $p = 5 \times 10^{-5}$--$1.9 \times 10^{-4}$ (approx.\ $3.5$--$4\sigma$). 
    \label{fig:background}}
\end{figure*}

Our Bayesian analysis provides evidence that the common-spectrum signal includes Hellings--Downs inter-pulsar correlations.
Specifically, the Bayes factor between the \modelhdgamma\ and \modelcurngamma\ models ranges from 200 (when 14 Fourier frequencies are included in $\Phi_i$) to 1,000 (when 5 frequencies are included, as in \citetalias{abb+20}). Results are similar for \modelhdgw\ vs.\ \modelcurngw.
\autoref{fig:bfs} recapitulates Bayes factors between a variety of models, including some with the alternative spatial-correlation structures discussed in \S\ref{subsec:correlation}. The very peaked $A_\mathrm{HD}$ posterior in panel (b) of \autoref{fig:spectrum_correlations_plot}, significantly separated from smaller amplitudes, supports the very large Bayes factor between \modelirn\ and \modelcurngamma. The 15-year data set favors \modelhdgamma\ over \modelcurngamma, and over models with monopolar or dipolar correlations, and it is inconclusive about, i.e., gives roughly even odds for, the presence of spatially correlated signals in addition to \modelhdgamma.

We can also regard the \modelhdgamma\ vs.\ \modelcurngamma\ Bayes factor as a detection statistic in a hypothesis-testing framework, and derive the $p$-value of the observed Bayes factor with respect to its empirical distribution under the \modelcurngamma\ model. 
We do so by computing Bayes factors on 5,000 bootstrapped data sets where inter-pulsar spatial correlations are removed by introducing random \textit{phase shifts}, drawn from a uniform distribution from 0 to $2\pi$, to the common-process Fourier components~\citep{tlb+17}. 
This procedure alters inter-pulsar correlations to have a mean of zero, while leaving the amplitudes of intrinsic pulsar noise and CURN unchanged, thus providing a way to test the null hypothesis that no inter-pulsar correlations are present.
The resulting background distribution of Bayes factors is shown in the left panel of \autoref{fig:background}---they exceed the observed value in five of the 5,000 phase shifts ($p = 10^{-3}$). 
We also performed sky scramble analyses~\citep{cs16}, which remove the dependence of inter-pulsar spatial correlations 
on the angular separations between the pulsars by attributing random sky positions to the pulsars. 
Sky scrambles generate a background distribution for which inter-pulsar correlations are present in the data but they are independent of the pulsars' angular separations: for this distribution, we find $p = 1.6 \times 10^{-3}$. 
A detailed discussion of sky scrambles and the results of these analyses can be found in App.\ \ref{sec:appskyscrambles}.

As in \citetalias{abb+20}, we also carried out a minimally modeled Bayesian reconstruction of the inter-pulsar correlation pattern, using spline interpolation over seven spline-knot positions. 
The choice of seven spline-knot positions is based on features of the Hellings--Downs pattern: two correspond to the maximum and minimum angular separations ($0^\circ$ and $180^\circ$, respectively), two are chosen to be at the theoretical zero crossings of the Hellings--Downs pattern ($49.2^\circ$ and $121.8^\circ$), one is at the theoretical minimum ($82.5^\circ$), 
and the final two are between the end points and zero crossings ($25^\circ$ and $150^\circ$) to allow additional flexibility in the fit.
Panel (d) of \autoref{fig:spectrum_correlations_plot} shows the marginal $1$-D posterior densities at these spline-knot positions for a power-law varied-exponent model. The reconstruction is consistent with the overplotted Hellings--Downs pattern; furthermore, the joint $2$-D marginal posterior densities for the knots, not shown in panel (d) of \autoref{fig:spectrum_correlations_plot}, at the HD zero-crossings is consistent with $(0,0)$ within $1\sigma$ credibility.

\section{Optimal statistic analysis}
\label{sec:optimal}

We complement our Bayesian search with a frequentist analysis using the \textit{optimal statistic} \citep{abc+2009,dfg+13,ccs+2015}, a summary statistic designed to measure correlated excess power in PTA residuals. 
(Note that there is no accepted definition of ``optimal statistic'' in modern statistical usage, but the term has become established in the PTA literature to refer to this specific method, so we use it for this reason.)
It is enlightening to describe the optimal statistic as a weighted average of the inter-pulsar correlation coefficients
\begin{equation}
\rho_{ab} = \frac{\tresid_a^T \mathbf{P}_a^{-1} \tilde{\mathbf{\Phi}}_{ab} \mathbf{P}_b^{-1} \tresid_b}{\mathrm{Tr} \, \mathbf{P}_a^{-1} \tilde{\mathbf{\Phi}}_{ab} \mathbf{P}_b^{-1} \tilde{\mathbf{\Phi}}_{ba}} \,,
\label{eq:rhoab}
\end{equation}
where $\tresid_a^T$ are the residuals of pulsar $a$, and $\mathbf{P}_a = \left\langle \tresid_a \tresid_a^T \right\rangle$ is their total auto-covariance matrix. The cross-covariance matrix $\tilde{\mathbf{\Phi}}_{ab}$ encodes the spectrum of the HD-correlated signal, normalized so that $\mathbf{\Phi}_{ab} = A^2 \Gamma(\xi_{ab}) \tilde{\mathbf{\Phi}}_{ab}$ (see \citealt{ptr2022}), and where elements of $\mathbf{\Phi}_{ab}$ are given by \autoref{eq:gp_covariance}.
Indeed, the $\rho_{ab}$ have expectation value $A^2 \Gamma(\xi_{ab})$, but their variance
$\sigma_{ab}^2 = (\mathrm{Tr} \, \mathbf{P}_a^{-1} \tilde{\mathbf{\Phi}}_{ab} \mathbf{P}_b^{-1} \tilde{\mathbf{\Phi}}_{ba})^{-1} + O(A^4)$
is too large to use them directly as estimators.
Thus we assemble the optimal statistic as the variance-weighted, $\Gamma$-template-matched average of the $\rho_{ab}$,
\begin{equation}
\hat{A}^2 = \frac{\sum_{a > b} \rho_{ab} \Gamma(\xi_{ab}) / \sigma_{ab}^2}{\sum_{a > b} \Gamma^2(\xi_{ab}) / \sigma_{ab}^2}.
\label{eq:os}
\end{equation}
This equation represents the optimal estimator of the HD amplitude $A^2$; it can also be interpreted as the best-fit $\hat{A}^2$ obtained by least-squares--fitting the $\rho_{ab}$ to the Hellings--Downs model $\hat{A}^2 \, \Gamma(\xi_{ab})$.
Because $\hat{A}^2$ is a function of intrinsic--red-noise and common-process hyperparameters through the $\mathbf{P}_a$, we use the results of an initial Bayesian-inference run to refer the statistic to MAP hyperparameters, or to marginalize it over their posteriors.
As discussed in \citet{vite18}, we obtain more accurate values of the amplitude by this marginalization.

\begin{figure}[t]
\begin{center}
    \includegraphics[width=\columnwidth]{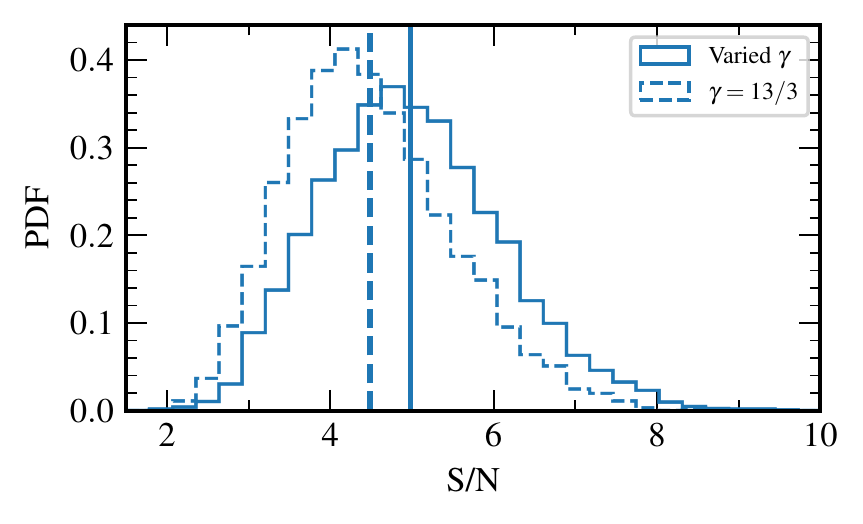}
\end{center}
	\vspace{-0.25in}
    \caption{
		Optimal statistic S/N for HD correlations, distributed over \modelcurngamma\  (solid lines) and \modelcurngw\ (dashed lines) noise-parameter posteriors. The vertical lines indicate the mean S/Ns. We find S/Ns of $5 \pm 1$ and $4 \pm 1$ 
		for \modelcurngamma\ and \modelcurngw, respectively.
		}
    \label{fig:optstat}
\end{figure}

To search for inter-pulsar correlations using the optimal statistic, we evaluate the frequency (the $p$-value) with which an uncorrelated common-spectrum process with parameters estimated from our data set would yield $\hat{A}^2$ greater than we observe.
In the absence of a signal, the expectation value of $\hat{A}^2$ is zero, and its distribution is approximately normal. Thus we divide the observed $\hat{A}^2$ by its standard deviation to define a formal signal-to-noise ratio 
\begin{equation}
\mathrm{S/N} = \frac{\sum_{a > b} \rho_{ab} \Gamma(\xi_{ab}) / \sigma_{ab}^2}{\left[\sum_{a > b} \Gamma^2(\xi_{ab}) / \sigma_{ab}^2\right]^{1/2}} \,.
\label{eq:ossnr}
\end{equation}
\autoref{fig:optstat} shows the distribution of this S/N over \modelcurngamma\ and \modelcurngw\
 noise-parameter posteriors, with S/Ns of $5 \pm 1$ and $4 \pm 1$, respectively (means $\pm$ standard deviations across noise-parameter posteriors). We use 14 frequency components to model the signal: the dependence on the number of frequency components is very weak.

Because the distribution of $\hat{A}^2$ is only approximately normal \citep{hazboun+2023GX2}, the S/N of \autoref{eq:ossnr} does not map analytically to a $p$-value, and it cannot be interpreted as a ``sigma'' level. 
Instead, optimal-statistic $p$-values can be computed empirically by removing inter-pulsar correlations from the 15-year data set with phase shifts \citep{tlb+17}.
We draw random phase offsets from 0 to $2\pi$ for the common-process Fourier components, which is equivalent 
to making uniform draws from the background distribution of CURN, and ask how often a random choice of phase offsets 
produces a HD-correlated signal.
The right panel of \autoref{fig:background} shows the distribution of noise-marginalized S/N over 400,000 phase shifts. 
There are 19 phase shifts with noise-marginalized S/N greater than observed, with $p = 5 \times 10^{-5}$.
We compare the phase-shift distribution with backgrounds obtained by 
simulation (right panel of \autoref{fig:background}, orange line) and analytic calculation (green line).
For the former, we simulate 27,000 \modelcurngamma\ realizations using MAP hyperparameters from the 15-yr data and compute the optimal-statistic S/N for each; for the latter, we evaluate the generalized $\chi^2$ distribution \citep{hazboun+2023GX2} with median \modelcurngamma\ hyperparameters.
Although neither method includes the marginalization over noise-parameter posteriors, we find good agreement with phase shifts, with $p = 1.8 \times 10^{-4}$ 
from simulations, and $p = 1.9 \times 10^{-4}$ 
from the analytic calculation. 
Finally, we use sky scrambles to compute the $p$-value for the null hypothesis that inter-pulsar correlations are present, 
but they have no dependence on the angular separation between the pulsars, for which we find $p < 10^{-4}$ (see App.~\ref{sec:appskyscrambles}).

Averaging the cross-correlations $\rho_{ab}$ in angular-separation bins with equal numbers of pulsar pairs reveals the Hellings--Downs pattern, as shown in panel (c) of \autoref{fig:spectrum_correlations_plot} for 15 bins. The $\rho_{ab}$ were evaluated with MAP \modelcurngw\ noise parameters. 
The black dashed curve traces the expected correlations for an HD-correlated background with the MAP amplitude; the vertical error bars display the expected 1$\sigma$ spreads of the binned cross-correlations, accounting for the $\langle \rho_{ab}\rho_{cd} \rangle$ covariances induced by the HD-correlated process \citep{romano+2021,2022arXiv220807230A}. (Neglecting those covariances yields 20--40\% smaller spreads. Note that they are \emph{not} included in $p$-value estimates because those are calculated under the null hypothesis of no spatially correlated process.)

Although each draw from the noise-parameter posterior would generate a slightly different plot, as would different binnings, the quality of the fit seen in \autoref{fig:spectrum_correlations_plot} provides a visual indication that the excess low-frequency power in the 15-year data set harbors HD correlations. The $\chi^2$ for this $15$-bin reconstruction with respect to the Hellings--Downs curve is $8.1$, where we account for $\rho_{ab}$ covariance in constructing the bins, and the covariance between bins in constructing the $\chi^2$~\citep{2022arXiv220807230A}. This corresponds to a $p$-value of 0.75, calculated using simulations based on the \modelhdgamma~model, or 0.92 if one assumes this value follows a canonical $\chi^2$ with 15 degrees of freedom.
These $p$-values are representative of what we find with different binnings: we find $p > 0.3$ when using eight to 20 bins 
(assuming a canonical $\chi^2$ distribution).

\section{Checks and validation}
\label{sec:validation}

Prior to analyzing the 15-year data set, we extensively reviewed our data collection and analysis procedures, methods, and tools, in an effort to eliminate contamination from systematic effects and human error.
Furthermore, the results presented in \S\ref{sec:bayes} and \S\ref{sec:optimal} are supported by a variety of consistency checks and auxiliary studies. In this section we present those that offer evidence for or against the presence of HD correlations, reveal anomalies, or otherwise highlight features of note in the data: alternative DM modeling (\S\ref{subsec:dm_models}), the spectral content (\S\ref{subsec:spectral}) and correlation pattern (\S\ref{subsec:correlation}) of the excess-noise signal, as well as the consistency of our findings across data set ``slices,'' pulsars, and telescopes (\S\ref{subsec:cross-validation}).

\subsection{Alternative DM models}
\label{subsec:dm_models}

In this paper and in previous GW searches (e.g., \citetalias{abb+20}), 
we model fluctuations in the DM using DMX parameters \citepalias[a piecewise-constant representation, see][]{aaa+23}. 
Adopting this DM model as the standard makes it easier to directly compare the results here 
to those in \citetalias{abb+20}. 
An alternative model where DM variations are modeled as a Fourier-domain Gaussian process, DMGP, has been used by \citet{2022MNRAS.510.4873A}, \citet{epta_dr2_gwb}, and \citet{ppta_dr2_gwb}.
The Fourier coefficients follow a power law similar to those of intrinsic and common-spectrum red noise, but their basis vectors include a $\nu^{-2}$ radio-frequency dependence, and the component frequencies $f_i = i/T$ range through $i = 1\mbox{--}100$.
Under the DMGP model we also include a deterministic solar-wind model \citep{2022ApJ...929...39H} and the two chromatic events in PSR J1713$+$0747 reported in \citet{leg+2018} which are modeled as deterministic exponential dips with the chromatic index quantifying the radio-frequency dependence of the dips left as a free parameter. If these chromatic events are not modeled, they raise estimated white noise \citep{hazboun:2020a}. A detailed discussion of chromatic noise effects can be found in \citetalias{aaa+23_noise}.

\begin{figure}[t]
\begin{center}
    \includegraphics[width=0.9\columnwidth]{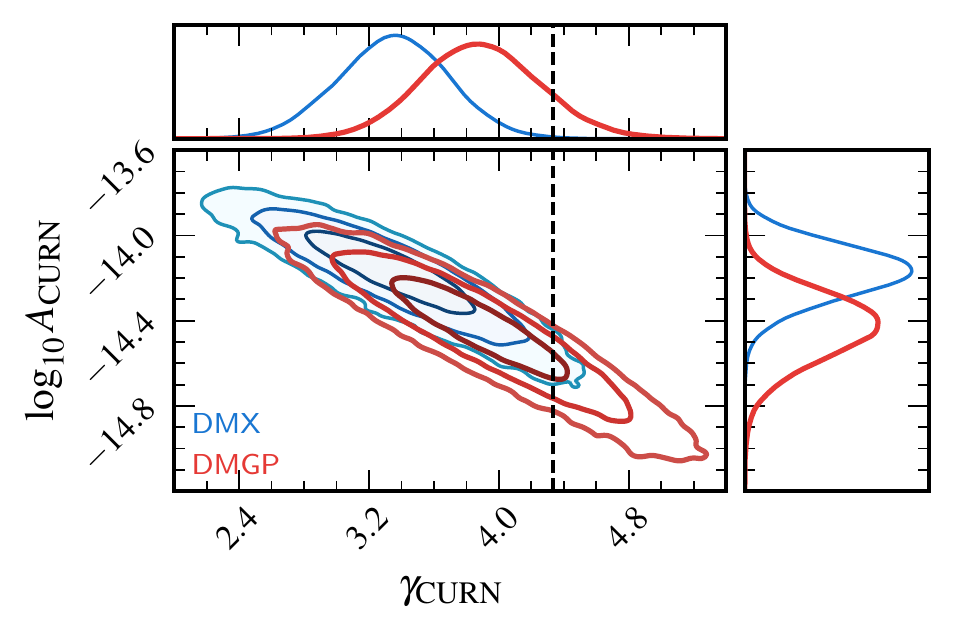}
\end{center}
	\vspace{-18pt}
    \caption{\modelcurngamma\ posterior distributions using DMGP (red) and DMX (blue) to model DM variations. The dashed line marks $\gamma_\mathrm{CURN} = 13/3$. While the posteriors are broadly consistent, DMGP shifts the $\gamma_\mathrm{CURN}$ posterior to higher values, making it more consistent with $\gamma_\mathrm{CURN}=13/3$.}
    \label{fig:dmgp}
\end{figure}

Using the DMGP model in place of DMX has minimal effects on nearly all pulsars in the array.
Only PSRs J1713$+$0747 and J1600$-$3053 show notable differences in their recovered intrinsic-noise parameters.
However, DMGP does affect the parameter estimation of common red noise, as seen in \autoref{fig:dmgp}, shifting the posterior for $\gamma$ to higher values that are more consistent with $13/3$. Despite this, we still recover HD correlations at the same significance as when we use DMX to model fluctuations in the DM, implying that the evidence reported for the presence of correlations in this work is independent of the choice of DM noise modeling.

\subsection{Spectral analysis}
\label{subsec:spectral}

Adopting power-law spectra for \modelcurn\ and \modelhd\ is a useful simplification that reduces the number of fit parameters and yields more informative constraints; furthermore, it is expedient to identify \modelhdgw\ with the hypothesis that we are observing the GWB from SMBHBs.
Nevertheless, the standard $\gamma = 13/3$ power law for GW inspirals may be altered by astrophysical processes such as stellar and gas friction in nuclei (see, e.g., \citealt{mm05} for a review), by appreciable eccentricity in SMBHB orbits \citep{en07}, and by low-number SMBHB statistics \citep{2008MNRAS.390..192S}.
\modelhdgamma\ parameter recovery may also be biased if intrinsic pulsar noise is not modeled well by a power law.
Indeed, our data show hints of a discrepancy from the idealized \modelhdgw\ model: the $\gamma_\mathrm{HD}$ posterior in panel (b) of \autoref{fig:spectrum_correlations_plot} favors slopes much shallower than 13/3, and the \modelhdgamma-to-\modelcurngamma\ Bayes factor drops from 1,000 to 200 when Fourier components at more than five frequencies are included in the model.
\begin{figure*}[t]
\begin{center}
    \includegraphics[width=\textwidth]{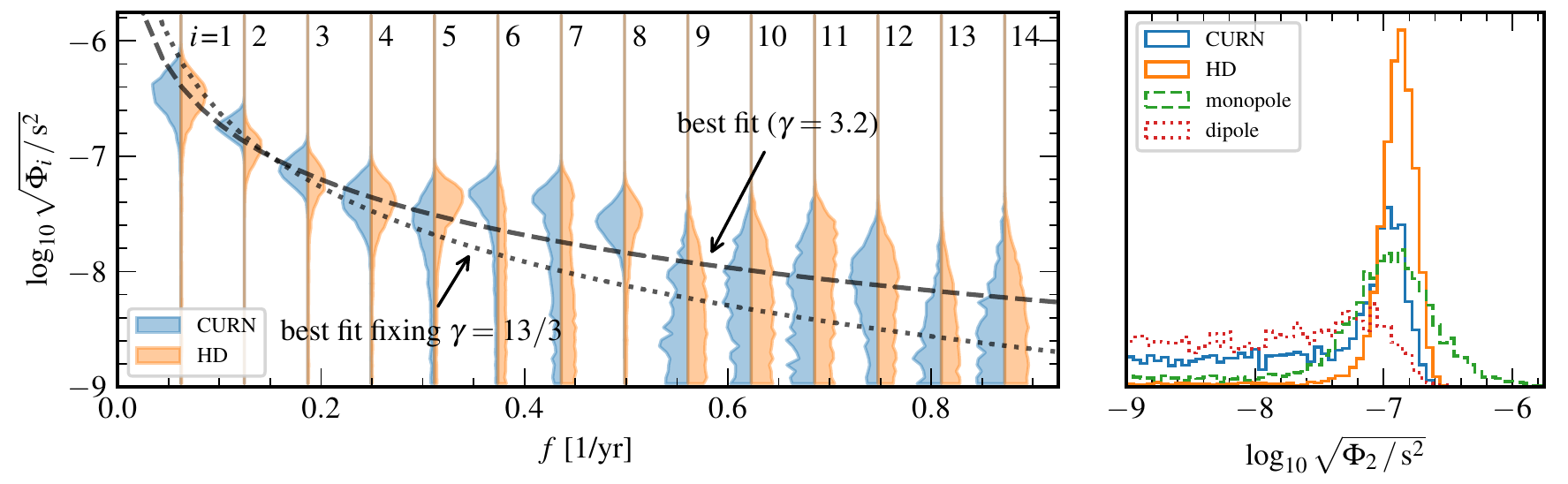}
\end{center}
	\vspace{-18pt}
    \caption{\textbf{Left}: Posteriors of Fourier component variance $\Phi_i$ for the \modelcurnfree\ (left) and \modelhdfree\ (right) models (see \S\ref{sec:data}), plotted at their corresponding frequencies $f_i = i/T$ with $T$ the 16.03-yr extent of the data set.
    Excess power is observed in bins 1--8 (somewhat marginally in bin 6); Hellings--Downs-correlated power in bins 1--5 and 8.
    The dashed line plots the best-fit power law, which has $\gamma \simeq 3.2$ (as in panel (d) of \autoref{fig:spectrum_correlations_plot}); the fit is pushed to lower $\gamma$ by bins 1 and 8.
    The dotted line plots the best-fit power law when $\gamma$ is fixed to 13/3; it overshoots in bin 1 and undershoots in bin 8.
    \textbf{Right}: Posteriors of variance $\Phi_2$ in Fourier bin 2 ($f_2 = 3.95$ nHz) in a
    \modelcurnfree\ + \modelhdfree\ + \modelmonofree\ + \modeldipolefree\
	model, showing evidence of a quasi-monochromatic monopole process (dashed). No monopole or dipole power is observed in all other bins of that joint model, with $\Phi_{\mathrm{CURN},i}$ and $\Phi_{\mathrm{HD},i}$ posteriors consistent with the left panel.
\label{fig:periodogram}}
\end{figure*}

We examine the spectral content of the 15-year data set using the \modelcurnfree\ and \modelhdfree\ models, which are parametrized by the variances of the Fourier components at each frequency.
Their marginal posteriors are shown in the left panel of \autoref{fig:periodogram}, where bin number $i$ corresponds to $f_i = i/T$, with $T = 16.03$ yr the extent of the data set.
For the purpose of illustration, we overlay best-fit power laws that thread the posteriors in a way similar to the factorized PTA likelihood of \citet{2022PhRvD.105h4049T} and \citet{2023arXiv230315442L}.

We deem excess power, either uncorrelated for \modelcurnfree\ or correlated for \modelhdfree,
to be observed in a bin when the support of the posterior is concentrated away from the lowest amplitudes.
No power of either kind is observed above $f_8$, consistent with the presence of a floor of white measurement noise.
Furthermore, no correlated power is observed in bins 6 and 7, where a power-law model would expect a smooth continuation of the trend of bins 1--5 (cf.\ the dashed fit of \autoref{fig:periodogram}): this may explain the drop in the Bayes factor.
However, correlated power reappears in bin 8, pushing the fit toward shallower slopes.
Indeed, repeating the fit by omitting subsets of the bins suggests that the low recovered $\gamma_\mathrm{HD}$ is due mostly to bin 8 and to the lower-than-expected correlated power found in bin 1. Obviously, excluding those bins leads to higher $\gamma_\mathrm{HD}$ estimates.

To explore deviations from a pure power law that may arise from statistical fluctuations of the astrophysical background or from unmodeled systematics (perhaps related to the timing model), 
in App.\ \ref{sec:apptprocess} we relax the normal $c_k$ prior (cf.\ \autoref{eq:gp_covariance}) to a multivariate Student's $t$-distribution that is more accepting of mild outliers.
The resulting estimate of $\gamma_\mathrm{CURN}$ peaks at a higher value and is broader than in \modelcurngamma, 
with posterior medians and 5-95\% quantiles of $\gamma_\mathrm{CURN} = 3.5^{+1.0}_{-1.0}$.

Similarly, spectral turnovers due to interactions between SMBHBs and their environments can result in reduced GWB power at lower frequencies, which might explain the slightly lower correlated power in bin 1. We investigate this hypothesis
in App.\ \ref{sec:appturnover} using the turnover spectrum
of \cite{scm2015}. For this \modelcurnturnover\ model, the 15-year data favor a spectral bend below 10 nHz (near $f_5$), but the Bayes factor against the standard \modelhdgamma\ is inconclusive.

Future data sets with longer time spans and the comparison of our data set with those of other PTAs should help clarify the astrophysical or systematic origin of these possible spectral features.

\subsection{Alternative correlation patterns}
\label{subsec:correlation}

Sources other than GWs can produce inter-pulsar residual correlations with spatial patterns other than HD.
For example, errors in the solar-system ephemerides create time-dependent Roemer delays with dipolar correlations \citep{r19,2020ApJ...893..112V}, 
and errors in the correction of telescope time to an inertial timescale \citep{2012MNRAS.427.2780H,2020MNRAS.491.5951H} create an identical time-dependent delay for all pulsars (i.e., a delay with monopolar correlations).

\citet{grtm14} showed that, for a pulsar array distributed uniformly across the sky,
HD correlations can be decomposed as
\begin{align}
	\Gamma_{\mathrm{HD},{ab}} &= \sum_{l=0}^{\infty}g_l\,P_l(\cos\xi_{ab}), \nonumber\\
	g_0 = 0,\; g_1 = 0,\; g_l &= \frac{3}{2}(2l+1)\frac{(l-2)!}{(l+2)!}\,\,\, \mathrm{for}\,\,\,l\geq 2,
	\label{eq:hdlegendre}
\end{align}
where the $P_l(\cos\xi_{ab})$ are Legendre polynomials of order $l$ evaluated at the pulsar angular separation $\xi_{ab}$. In other words, a HD-correlated signal should have no power at $l=0$ or $l=1$.

\begin{figure}[t]
\begin{center}
    \includegraphics[width=\columnwidth]{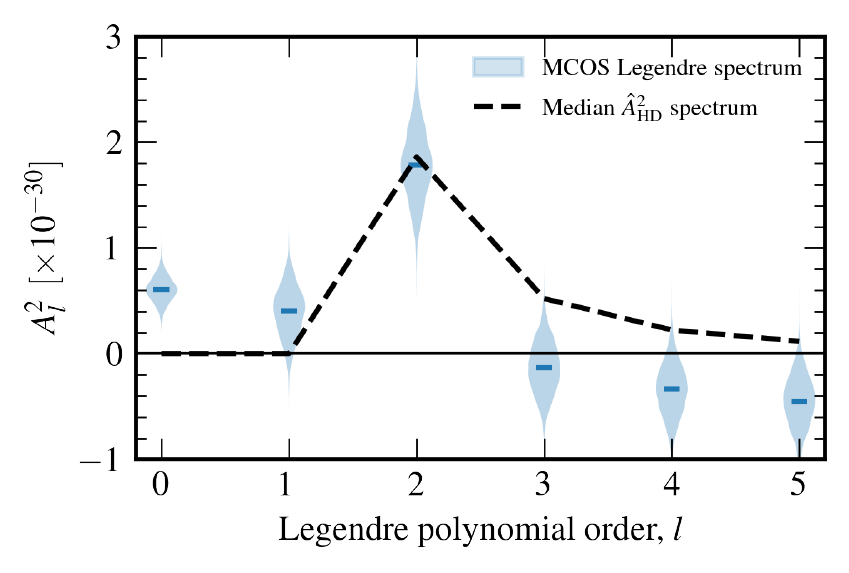}
\end{center}
	\vspace{-18pt}
    \caption{Multiple-component optimal statistic for a Legendre polynomial basis (\autoref{eq:hdlegendre}) with with $l_\mathrm{max}=5$.
    The violin plots show the distributions of the normalized Legendre coefficients $A_l^2 = A^2 g_l$ over \modelcurngamma\ noise-parameter posteriors.
    The black dashed line shows the Legendre spectrum of a pure-HD signal, with the median posterior $\hat{A}^2_\mathrm{HD}$.}
    \label{fig:optstat_legendre_vg}
\end{figure}

We can perform a frequentist generic correlation search using Legendre polynomials\footnote{A Bayesian method for fitting correlations using Legendre polynomials can be found in \citet{nay+23}.} with the 
multiple-component optimal statistic (MCOS; \citealt{sv2023})---a generalized statistic that allows multiple correlation patterns to be fit simultaneously to the correlation coefficients $\rho_{ab}$. 
\autoref{fig:optstat_legendre_vg} shows the constraints on $A_l^2 = A^2 g_l$ obtained by fitting the correlations $\rho_{ab}$ to this Legendre series using the MCOS and marginalizing over \modelcurngamma\ noise-parameter posteriors.
The quadrupolar structure of the data is evident, along with a small but significant monopolar contribution. 

The same feature from the Legendre decomposition appears if we use the MCOS to search for multiple correlations simultaneously: 
a multiple regression analysis favors models that contain both significant HD and monopole correlations (see App.~\ref{sec:appmcos}). From simulations of 15-year-like data sets (see App.~\ref{sec:appastrosims}), we find a $p$-value of 0.11 (approx.~$2\sigma$) for observing a monopole at this significance or higher with a pure-HD injection of amplitude similar to what we observe. We also perform a model-checking study to assess whether the observed monopole is consistent with the \modelhdgw\ model (see App.~\ref{sec:appmodelsims}), and we find a $p$-value of 0.11 for producing an apparent monopole when the signal is purely \modelhdgw.
Thus, we conclude that it is possible for a HD-correlated signal to appear to have monopole correlations 
in an optimal statistic analysis at this significance level.

In contrast, Bayesian searches for additional correlations do not find evidence 
of additional monopole- or dipole-correlated red noise processes: 
as shown in \autoref{fig:bfs}, the Bayes factors for these processes are $\sim1$. 
We also perform a general Bayesian search for correlations using a 
\modelcurnfree\ + \modelhdfree\ + \modelmonofree\ + \modeldipolefree\
model, which allows for independent uncorrelated and correlated components at every frequency bin. 
We note that this analysis is more flexible than the ones described above, which assume a power-law power spectral density. 
We find no significant dipole-correlated power at any frequency, and 
we find monopole-correlated power only in the second frequency bin ($f_2 = 3.95$ nHz); 
posteriors of variance for that bin are shown in the right panel of \autoref{fig:periodogram}.

Motivated by this finding, we perform a search for \modelhdgamma\ + \modelsinusoid,  
which includes a deterministic sinusoidal delay (applied to all pulsars alike, as appropriate for a monopole) 
with free frequency, amplitude, and phase. 
The sinusoid's posteriors match the free-spectral analysis in frequency and amplitude; 
however, the Bayes factor between \modelhdgamma\ + \modelsinusoid\ and \modelhdgamma\, calculated using two methods~\citep{hee15,HourihaneMeyers2022}, is only $\sim 1$, 
so the signal cannot be considered statistically significant. 
Astrophysically motivated searches for sources that produce sinusoidal 
or sinusoid-like delays in the residuals, 
such as an individual SMBHB or perturbations to the local gravitational field induced by 
fuzzy dark matter \citep{2014JCAP...02..019K}, also yield Bayes factors $\sim1$. 
Thus we conclude that there is some evidence of additional power at 3.95 nHz with monopole correlations; 
however, the significance in the Bayesian analyses is low, 
while the optimal-statistic S/N could be produced by a HD-correlated signal. 
Therefore, we cannot definitively say whether the signal is present, or determine the source. 
We note that performing an MCOS analysis after subtracting off realizations of a sinusoid 
using \modelhdgamma\ + \modelsinusoid\ posteriors reduces the 
$(\mathrm{S}/\mathrm{N})_\mathrm{monopole} \simeq 0$ 
while $(\mathrm{S}/\mathrm{N})_\mathrm{HD}$ remains unchanged, 
indicating that this single-frequency monopole-correlated signal is likely causing 
the nonzero monopole signal observed in the MCOS analysis.

Similar hints of a monopolar signal (though weaker) were found in the NANOGrav 12.5-year data set, unsurprisingly given that it is a subset of the current data set.
To exercise due diligence, we audited the correction of telescope time to GPS time at the Arecibo Observatory and at the Green Bank Telescope, and found nothing that could explain our observations.
The subsequent steps in the time-correction pipeline rely on very accurate atomic clocks and are unlikely to introduce considerable systematics \citep{petit2022}. 
An important test will be whether this signal persists in future data sets. 
If this monopolar feature is a truly an astrophysical signal, 
we would expect it to increase in significance as our data set grows. 
Comparisons with other PTAs and combined IPTA data sets will also provide crucial insight.

\subsection{Dropout and cross-validation}
\label{subsec:cross-validation}

The GWB is by its nature a signal affecting all of the pulsars in the PTA, although it may appear more significant in some based on their observing time span, noise properties, and on the particular realization of pulsar and Earth contributions \citep{2023MNRAS.518.1802S}. 
One way to assess the significance of the GWB in each pulsar is a Bayesian \emph{dropout analysis} \citep{aab+19,abb+20}, which introduces a binary parameter that turns on and off the common signal (or its inter-pulsar correlations) for a single pulsar, leaving all other pulsars unchanged. 
The Bayes factor associated with this parameter, also referred to as the ``dropout factor,'' describes how much each pulsar likes to ``participate'' in the common signal. 

\begin{figure}[t]
\begin{center}
    \includegraphics[width=\columnwidth]{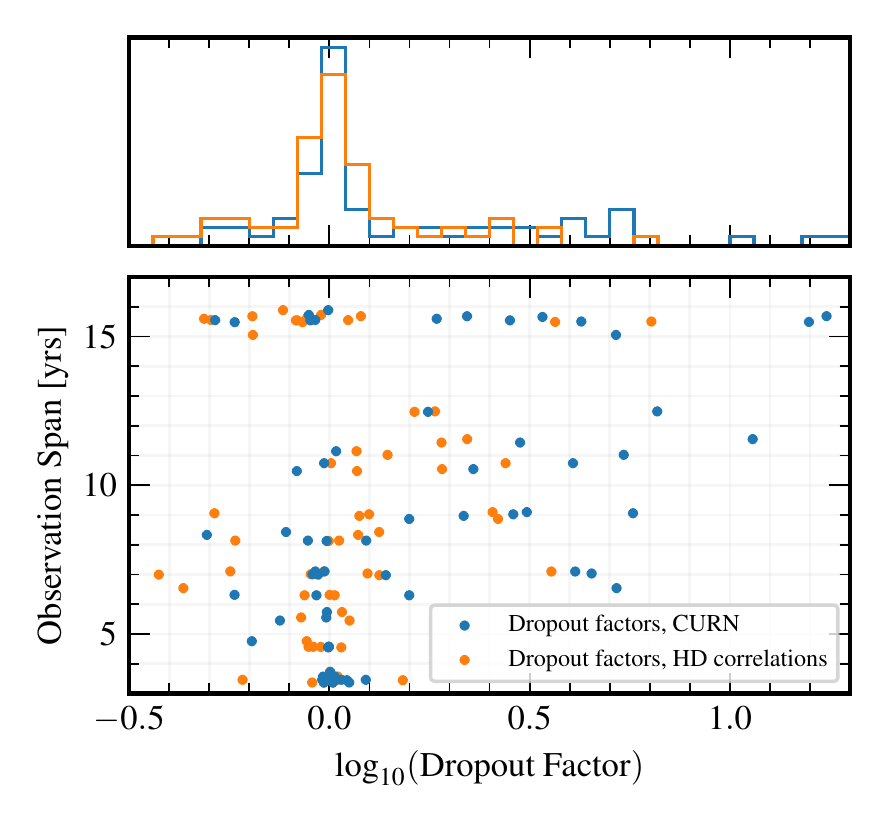}
\end{center}
\vspace{-12pt}
    \caption{Support for \modelcurngamma\ (blue) and \modelhdgamma\ correlations (orange) in each pulsar, as measured by a dropout analysis. 
    Dropout factors greater than 1 indicate support for the \modelcurngamma\ or \modelhdgamma\, while those less than 1 
    show that the pulsar disfavors it. We find significant spread in the dropout factors 
    among pulsars with long observation times, but overall more pulsars favor \modelcurngamma\ participation and \modelhdgamma\ correlations than disfavor them.}
    \label{fig:dropout}
\end{figure}

\autoref{fig:dropout} plots \modelcurngamma\ vs. \modelirn\ dropout factors for all 67 pulsars (blue).
We find positive dropout factors (i.e., dropout factors $>2$) for an uncorrelated common process in twenty pulsars, 
while only one has a dropout factor $<0.5$. 
For comparison, in the NANOGrav 12.5-year data set 
ten pulsars showed positive dropout factors for an uncorrelated common process, while three had negative dropout factors. 
We also show HD correlations vs. \modelcurngamma\ dropout factors (orange). 
For these, the uncorrelated common process is always present in all pulsars, 
but the cross-correlations for all pulsar pairs involving a given pulsar may be dropped from the likelihood. 
We find positive factors for HD correlations vs. \modelcurngamma\ in seven pulsars, while three are negative.
We expect more pulsars to have positive dropout factors for \modelcurngamma\ vs. \modelirn\ than for Hellings--Downs vs. \modelcurngamma\ because the Bayes factor comparing the first two models is significantly higher than the one comparing the second two models (see \autoref{fig:bfs}).
Negative dropout factors could be caused by noise fluctuations or they could be an indication that more advanced chromatic noise modeling is necessary~\citep{aab+20}. They could also be caused by the GWB itself, which induces both correlated and uncorrelated noise in the pulsars (the so-called ``Earth terms'' and ``pulsar terms''; \citealt{2018JPhCo...2j5002M}).

In addition to Bayes factors, the goodness-of-fit of probabilistic models can be evaluated by assessing their predictive performance \citep{gelman2013bayesian}.
Specifically, given that the GWB is correlated across pulsars, we can (partially) predict the timing residuals $\tresid_a$ of  pulsar $a$ from the residuals $\tresid_{-a}$ of all other pulsars by way of the ``leave-one-out'' \emph{posterior predictive likelihood} (PPL)
\begin{align}
p(\tresid_a | \tresid_{-a}) = \int d \bm{\theta}_a \, p(\tresid_a | \bm{\theta}_a) \, p(\bm{\theta}_a | \tresid_{-a}),
\end{align} 
where $\bm{\theta}_a$ are all the parameters and hyperparameters that affect pulsar $a$ in a given model.
As discussed in \citet{meyers+23}, we compare the predictive performance of \modelcurngw\ and \modelhdgw\ for each pulsar in turn by taking the ratio of the corresponding leave-one-out PPLs. These ratios are closely related to the dropout factors plotted in \autoref{fig:dropout}.
Multiplying the PPL ratios for all pulsars yields the \emph{pseudo Bayes factor} (PBF).
For the 15-year data set we find $\mathrm{PBF}_{15\,\mathrm{yr}} =$ 1,400 in favor of \modelhdgw\ over \modelcurngw. The PBF does not have a ``betting odds'' interpretation, but we obtain a crude estimate of its significance by building its background distribution on 40 \modelcurngw\ simulations with the MAP $\log_{10} A_\mathrm{CURN}$ inferred from the 15-year data set. For all simulations except one, the PBF favors the null hypothesis, and  $\log_{10} \mathrm{PBF}_{15\,\mathrm{yr}}$ is displaced by approx.\ three standard deviations from the mean $\log_{10} \mathrm{PBF}$.
\begin{figure}[t]
\begin{center}
    \includegraphics[width=\columnwidth]{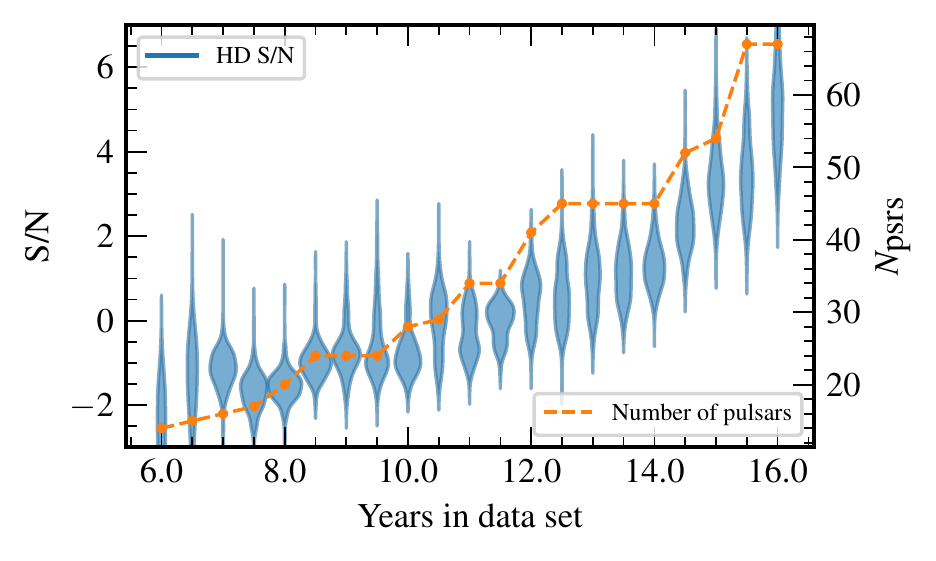}
\end{center}
\vspace{-12pt}
    \caption{S/N growth as a function of time and number of pulsars. As we move from left to right we add an additional six months of data at each step. New pulsars are added when they accumulate three years of data. The blue violin plot shows the distribution of the optimal statistic S/N over \modelcurngamma\ noise parameters. The dashed orange line shows the number of pulsars used for each time slice.}
     \label{fig:os_time_slice_forward}
\end{figure}

A different sort of cross-validation relies on evaluating the optimal statistic for temporal subsets of the data set, as in \citet{hazboun:2020a}.
In the regime where the lowest frequencies of our data are dominated by the GWB, the optimal statistic S/N should grow with the square root of the time span of the data and linearly 
with the number of pulsars in the array~\citep{Siemens:2013zla}; in this regime increasing the number of pulsars is the best way to boost PTA sensitivity to the GWB.
To verify that this is indeed the case, we analyze ``slices'' of the data set in six-month increments, starting from a six-year data set.
Once a new pulsar accumulates three years of data, we add it to the array.
We perform a separate Bayesian \modelcurngamma\ analysis for each slice and calculate the Hellings--Downs optimal statistic over the noise-parameter posterior.
In \autoref{fig:os_time_slice_forward}, we plot the S/N distributions against time span and the number of pulsars. As expected, we observe essentially 
monotonic growth associated with the increase in the number of pulsars.

\begin{figure}[t]
\begin{center}
    \includegraphics[width=0.9\columnwidth]{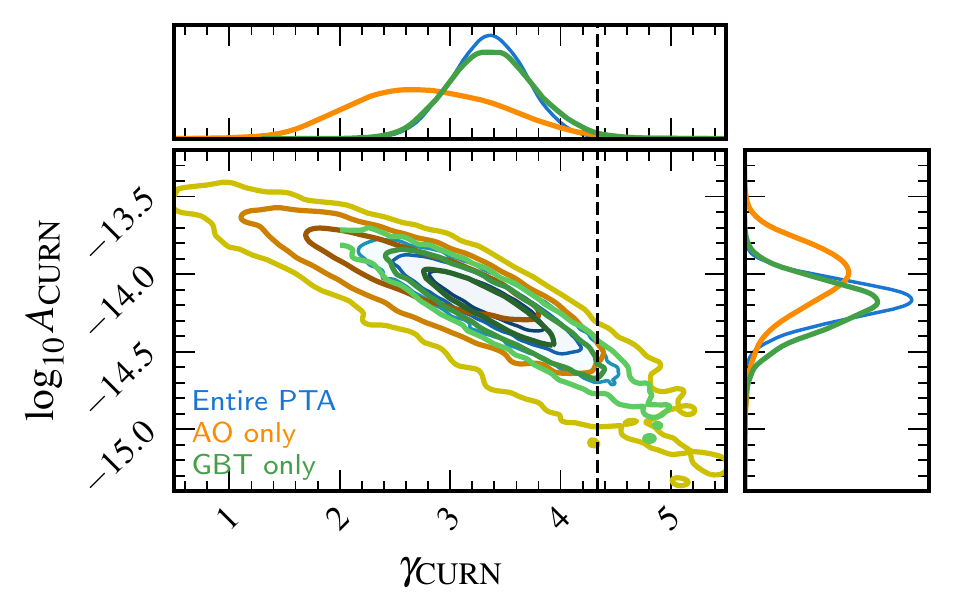}
\end{center}
	\vspace{-18pt}
    \caption{\modelcurngamma\ posterior distributions for Arecibo (orange) 
    and GBT (green) split-telescope data sets, and for the full data set (blue).
    The dashed line marks $\gamma_\mathrm{CURN} = 13/3$. 
    The posteriors for the split-telescope data sets are consistent with each other 
    and with the posteriors for the full data set.}
    \label{fig:telescope_psd}
\end{figure}

The signal should also be consistent between timing observations made with Arecibo and GBT.
To test this, we analyze the two split-telescope data sets (see App.\ \ref{app:data_set_details});
both show evidence of common-spectrum excess noise.
\autoref{fig:telescope_psd} shows Arecibo (orange) and GBT (green) \modelcurngamma\ posteriors, 
which are broadly consistent with each other and with full-data posteriors (blue). 
Arecibo yields $\log_{10}A = \logAgwVariableGammaAreciboSixtyEightPercent$ and $\gamma = \gammagwAreciboSixtyEightPercent$ (medians with 68\% credible intervals), 
while GBT yields $\log_{10}A = \logAgwVariableGammaGBTSixtyEightPercent$ and $\gamma = \gammagwGBTSixtyEightPercent$. 

The split-telescope data sets are significantly less sensitive to spatial correlations 
than the full data set, because they have fewer pulsars and therefore pulsar pairs 
(see \autoref{fig:ang_sep} of App.\ \ref{app:data_set_details}). 
Nevertheless, we can search them for spatial correlations using the optimal statistic.
We find a noise-marginalized Hellings--Downs S/N of 2.9 for Arecibo and 3.3 for GBT, 
consistent with the split-telescope data sets having about half the number of pulsars 
as the full data set. 
The S/Ns for Arecibo and GBT are comparable: 
while telescope sensitivity, observing cadence, 
and distribution of pulsars all affect GWB sensitivity, 
the dominant factor is the number of pulsars 
because the S/N scales linearly with the number of pulsars 
but only as $\propto (\sigma \sqrt{c})^{-1/\gamma}$, where $\sigma$ is the residual root-mean-squared, 
and $c$ is the observing cadence~\citep{Siemens:2013zla}. 
We also note that the distributions of angular separations probed by Arecibo and GBT are similar, 
although GBT observes more pulsar pairs with large angular separations (see \autoref{fig:ang_sep}).

\section{Discussion}
\label{sec:discussion}

In this letter we have reported on a search for an isotropic stochastic GWB in the 15-year NANOGrav data set.
A previous analysis of the 12.5-year NANOGrav data set found strong evidence for excess low-frequency noise with common spectral properties across the array, but inconclusive evidence for Hellings--Downs inter-pulsar correlations, which would point to the GW origin of the background.
By contrast, the 12.5-year data disfavored purely monopolar (clock-error--like) and dipolar (ephemeris-error--like) correlations.
Subsequent independent analyses by the PPTA and EPTA collaborations reported results consistent with ours \citep{ppta_dr2_gwb,epta_dr2_gwb}, as did the search of a combined data set  \citep{2022MNRAS.510.4873A}---a syzygy of tantalizing discoveries that portend the rise of low-frequency GW astronomy.

We analyzed timing data for 67 pulsars in the 15-year data set (those that span $> 3$ years), with a total time span of 16.03 years, and more than twice the pulsar pairs than in the 12.5-year data set.
The common-spectrum stochastic signal gains even greater significance and is detected in a larger number of pulsars.
For the first time, we find compelling evidence of Hellings--Downs inter-pulsar correlations, using both Bayesian and frequentist detection statistics (see \autoref{fig:spectrum_correlations_plot}), with false-alarm probabilities of $p=10^{-3}$ and $p=5 \times 10^{-5}$--$1.9\times10^{-4}$, respectively (see \autoref{fig:background}).

The significance of Hellings--Downs correlations increases as we increase the number of frequency components in the analysis up to five, indicating that the correlated signal extends over a range of frequencies. 
A detailed spectral analysis supports a power-law signal, but at least two frequency bins show deviations that may skew the determination of spectral slope (\autoref{fig:periodogram}).
These discrepancies may arise from astrophysical or systematic effects.
Furthermore, slope determination changes significantly using an alternative DM model (\autoref{fig:dmgp}). 
The study of spatial correlations with the optimal statistic confirms a Hellings--Downs quasi-quadrupolar pattern (\autoref{fig:optstat_legendre_vg} and panel c of \autoref{fig:spectrum_correlations_plot}), with some indications of an additional monopolar signal confined to a narrow frequency range near 4~nHz. 
However, the Bayesian evidence for this monopolar signal is inconclusive, and we could not ascribe it to any astrophysical or terrestrial source (e.g., an individual SMBHB or errors in the chain of timing corrections).

The GWB is a persistent signal that should increase in significance with number of pulsars and observing time span.
This is indeed what we observe by analyzing slices of the data set (see \autoref{fig:os_time_slice_forward}). 
Furthermore, the signal is present in multiple pulsars (\autoref{fig:dropout}), and can be found in independent single-telescope data sets (\autoref{fig:telescope_psd}). We are preparing a number of other papers searching the 15-year data set 
for stochastic and deterministic signals, including an all-sky, all-frequency search for GWs from individual circular SMBHBs. 
This search, together with the same analysis of the 12.5-year data set \citep{2023arXiv230103608A}, indicates that the spectrum and correlations we observe cannot be produced by an individual circular SMBHB.

If the Hellings--Downs-correlated signal is indeed an astrophysical GWB, its origin remains indeterminate.
Among the many possible sources in the PTA frequency band, numerous studies have focused on the unresolved background from a population of close-separation SMBHBs.
The SMBHB population is a direct byproduct of hierarchical structure formation, which is driven by galaxy mergers \citep[e.g.,][]{blumenthal84}. In a post-merger galaxy, the SMBHs sink to the center of the common merger remnant through dynamical interactions with their astrophysical environment, eventually leading to the formation of a binary \citep{bbr80}. 
GW emission from a SMBHB at nHz frequencies is quasi-monochromatic because the binaries evolve very slowly.
Under the assumption of purely GW-driven binary evolution, the expected characteristic-strain spectrum is $\propto f^{-2/3}$ (or $f^{-13/3}$ for pulsar-timing residuals).

The GWB spectrum may also feature a low-frequency turnover induced by the dynamical interactions of binaries with their astrophysical environment (e.g., with stars or gas, see \citealt{2002ApJ...567L...9A,shm+04,mm05}) or possibly by non-negligible orbital eccentricities persisting to small separations \citep{en07}. 
We find little support for a low-frequency turnover in our data (see App.\ \ref{sec:appturnover}).

\begin{figure}
	\begin{center}
	\includegraphics[width=\columnwidth]{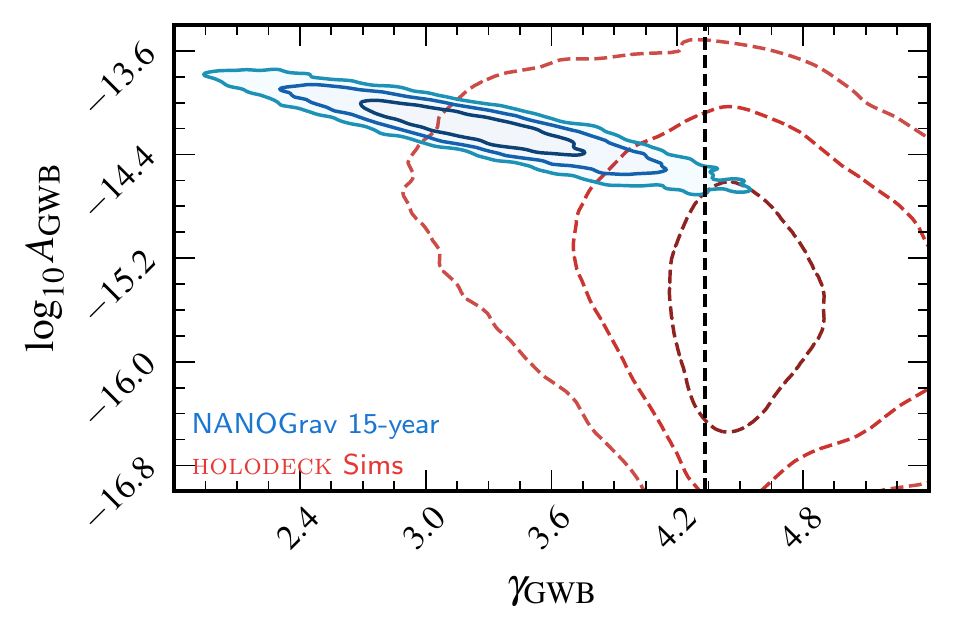}
	\end{center}
	\vspace{-0.25in}
	\caption{Posteriors of \modelhdgamma\ amplitude (for $f_\mathrm{ref}=1\,\mathrm{yr}^{-1}$) and spectral slope for the 15-year data set (blue), compared to power-law fits to simulated GWB spectra (red, dashed) from a population of SMBHBs generated by \textsc{holodeck} \citep{holodeck} under the assumption of purely GW-driven binary evolution, and narrowly distributed model parameters based on galaxy merger-observations.
We show $1/2/3\sigma$ regions, and the dashed line indicates $\gamma = 13/3$.
The broad contours confirm that population variance can lead to a significant spread of spectral characteristics.}
	\label{fig:astro_compare}
\end{figure}

The GWB amplitude is determined primarily by SMBH masses and by the occurrence rate of close binaries, which in turn depends on the galaxy merger rate, the occupation fraction of SMBHs, and the binary evolution timescale; population models predict amplitudes ranging over more than an order of magnitude \citep{rr95,wl03,jb03,2014ApJ...789..156M,2013MNRAS.433L...1S}, under a variety of assumptions.
\autoref{fig:astro_compare} displays a comparison of \modelhdgamma\ parameter posteriors with power-law spectral fits from an observationally constrained semi-analytic model of the SMBHB population constructed with the \textsc{holodeck} package \citep{holodeck}. This particular set of SMBHB populations assumes purely GW-driven binary evolution, and uses relatively narrow distributions of model parameters based on literature constraints from galaxy-merger observations \citep[see, e.g.,][]{Tomczak+2014}.
While the amplitude recovered in our analysis is consistent with models derived directly from our understanding of SMBH and galaxy evolution, it is toward the upper end of predictions implying a combination of relatively high SMBH masses and binary fractions. A detailed discussion of the GWB from SMBHBs in light of our results is given in \citet{aaa+23_smbhb}.

In addition to SMBHBs, more exotic cosmological sources such as inflation, cosmic strings, phase transitions, domain walls, and curvature-induced GWs can also produce detectable GWBs in the nHz range \citep[see, e.g.,][and references therein]{guzzetti2016gravitational,2018CQGra..35p3001C}.
Similarities in the spectral shapes of cosmological and astrophysical signals make it challenging to determine the origin of the background from its spectral characterization \citep{kpm+22}.
The question could be settled by the detection of signals from individual loud SMBHBs or by the observation of spatial anisotropies, since the anisotropies expected from SMBHBs are orders of magnitude larger than those produced by most cosmological sources \citep{2018CQGra..35p3001C,2022JCAP...11..009B}.
We discuss these models in the context of our results in \citet{aaa+23_cosmo}.

The EPTA and Indian Pulsar Timing Array (InPTA; \citealt{InPTA}), PPTA, and Chinese Pulsar Timing Array (CPTA; \citealt{CPTA}) collaborations have also recently searched their most recent data for signatures of a gravitational-wave background \citep{epta23,ppta23,cpta23}, and an upcoming IPTA paper will compare the results of these searches. The IPTA's forthcoming Data Release 3 will combine the NANOGrav $15$-year data set with observations from the EPTA, PPTA, and InPTA collaborations, comprising about 80 pulsars with time spans up to 24 years, and offering significantly greater sensitivity to spatial correlations and spectral characteristics than single-PTA data sets. Future PTA observation campaigns will improve our understanding of this signal and of its astrophysical and cosmological interpretation. 
Longer data sets will tighten spectral constraints on the GWB, clarifying its origin \citep{astro4cast}. Greater numbers of pulsars will allow us to probe anisotropy in the GWB \citep{ptr2022} and its polarization structure \citep[see, e.g.,][and references therein]{2021ApJ...923L..22A}. The observation of a stochastic signal with spatial correlations in PTA data, suggesting a GWB origin, expands the horizon of GW astronomy with a new Galaxy-scale observatory sensitive to the most massive black-hole systems in the Universe and to exotic cosmological processes.

\acknowledgments

\emph{Author contributions.}
An alphabetical-order author list was used for this paper in recognition of the fact that a large, decade timescale project such as NANOGrav is necessarily the result of the work of many people. All authors contributed to the activities of the NANOGrav collaboration leading to the work presented here, and reviewed the manuscript, text, and figures prior to the paper's submission. 
Additional specific contributions to this paper are as follows.
G.A., A.A., A.M.A., Z.A., P.T.B., P.R.B., H.T.C., K.C., M.E.D., P.B.D., T.D., E.C.F., W.F., E.F., G.E.F., N.G., P.A.G., J.G., D.C.G., J.S.H., R.J.J., M.L.J., D.L.K., M.K., M.T.L., D.R.L., J.L., R.S.L., A.M., M.A.M., N.M., B.W.M., C.N., D.J.N., T.T.P., B.B.P.P., N.S.P., H.A.R., S.M.R., P.S.R., A.S., C.S., B.J.S., I.H.S., K.S., A.S., J.K.S., and H.M.W.~developed the 15-year data set through a combination of 
observations, arrival time calculations, data checks and refinements, 
and timing model development and analysis; 
additional specific contributions to the data set are summarized in \citetalias{aaa+23}.
%
S.R.T.~and S.J.V.~led the search and coordinated the writing of this paper. 
P.T.B., J.S.H., P.M.M., N.S.P., J.S., S.R.T., M.V., and S.J.V.~wrote the paper, made the figures, and generated the bibliography. 
K.P.I., N.L., N.S.P., X.S., J.S., J.P.S., S.R.T., and S.J.V.~performed analyses and developed new techniques on a preliminary 14-year data set.
P.T.B., B.B., B.D.C., S.C., B.D., G.E.F., K.G., A.D.J., A.R.K., L.Z.K., N.L., W.G.L., N.S.P., S.C.S., L.S., X.S., J.S., J.P.S., S.R.T., C.U., M.V., S.J.V., Q.W., and C.A.W.~analyzed preliminary versions of the 15-year data set.
J.S.H., J.G., N.L., N.S.P., J.P.S., and J.K.S.~performed noise analyses on the data set.
P.T.B., B.B., R.B., R.C., M.C., N.J.C., D.D., H.D., G.E.F., K.A.G., S.H., A.D.J., N.L., W.G.L., P.M.M., P.P., N.S.P., S.C.S., X.S., J.S., J.P.S., J.T., S.R.T., M.V., S.J.V., and C.A.W.~performed the Bayesian and frequentist analyses presented in this paper. 
K.G., J.S.H., P.M.M., J.D.R., and S.J.V.~computed the background distribution of our detection statistics.
P.R.B., P.M.M., N.S.P., and M.V.~performed simulations that were used to compute false alarm probabilities.
A.M.A., P.T.B., B.B., B.D., S.C., J.A.E., G.E.F., J.S.H., A.D.J., A.R.K., N.L., M.T.L., K.D.O., T.T.P., N.S.P., J.S., J.P.S., J.K.S., S.R.T., M.V.,  and S.J.V.~contributed to the development of our code.
P.T.B., N.J.C., J.S.H., A.D.J., T.B.L., P.M.M., J.D.R., and M.V.~performed an internal code review.
K.C., C.C., J.A.E., J.S.H., D.R.M., M.A.Mc., C.M.F.M., K.D.O., N.S.P., S.R.T., M.V., R.v.H., and S.J.V.~were part of a NANOGrav Detection Study Group.
M.A.M.~and P.N.~served on the IPTA Detection Committee. K.C., J.S.H., T.J.W.L., P.M.M., N.S.P., J.D.R., X.S., S.R.T., S.J.V., and C.A.W.~responded to the Detection Checklist from the IPTA Detection Committee.

\emph{Acknowledgments.}
The NANOGrav collaboration receives support from National Science Foundation (NSF) Physics Frontiers Center award numbers 1430284 and 2020265, the Gordon and Betty Moore Foundation, NSF AccelNet award number 2114721, an NSERC Discovery Grant, and CIFAR. 
The Arecibo Observatory is a facility of the NSF operated under cooperative agreement (AST-1744119) by the University of Central Florida (UCF) in alliance with Universidad Ana G. M{\'e}ndez (UAGM) and Yang Enterprises (YEI), Inc. The Green Bank Observatory is a facility of the NSF operated under cooperative agreement by Associated Universities, Inc. The National Radio Astronomy Observatory is a facility of the NSF operated under cooperative agreement by Associated Universities, Inc. 
This work used the Extreme Science and Engineering Discovery Environment (XSEDE), which is supported by National Science Foundation grant number ACI-1548562. Specifically, it used the Bridges-2 system, which is supported by NSF award number ACI-1928147, at the Pittsburgh Supercomputing Center (PSC). This work was conducted using the Thorny Flat HPC Cluster at West Virginia University (WVU), which is funded in part by National Science Foundation (NSF) Major Research Instrumentation Program (MRI) Award number 1726534, and West Virginia University. This work was also conducted in part using the resources of the Advanced Computing Center for Research and Education (ACCRE) at Vanderbilt University, Nashville, TN. This work was facilitated through the use of advanced computational, storage, and networking infrastructure provided by the Hyak supercomputer system at the University of Washington. This research was supported in part through computational resources and services provided by Advanced Research Computing at the University of Michigan, Ann Arbor.
NANOGrav is part of the International Pulsar Timing Array (IPTA); we would like to thank our IPTA colleagues for their feedback on this paper. We thank members of the IPTA Detection Committee for developing the IPTA Detection Checklist. We thank Bruce Allen for useful feedback. We thank Valentina Di Marco and Eric Thrane for uncovering a bug in the sky scramble code. We thank Jolien Creighton and Leo Stein for helpful conversations about background estimation.
L.B. acknowledges support from the National Science Foundation under award AST-1909933 and from the Research Corporation for Science Advancement under Cottrell Scholar Award No. 27553.
P.R.B. is supported by the Science and Technology Facilities Council, grant number ST/W000946/1.
S.B. gratefully acknowledges the support of a Sloan Fellowship, and the support of NSF under award \#1815664.
The work of R.B., R.C., D.D., N.La., X.S., J.P.S., and J.T. is partly supported by the George and Hannah Bolinger Memorial Fund in the College of Science at Oregon State University.
M.C., P.P., and S.R.T. acknowledge support from NSF AST-2007993.
M.C. and N.S.P. were supported by the Vanderbilt Initiative in Data Intensive Astrophysics (VIDA) Fellowship.
K.Ch., A.D.J., and M.V. acknowledge support from the Caltech and Jet Propulsion Laboratory President's and Director's Research and Development Fund.
K.Ch. and A.D.J. acknowledge support from the Sloan Foundation.
Support for this work was provided by the NSF through the Grote Reber Fellowship Program administered by Associated Universities, Inc./National Radio Astronomy Observatory.
Support for H.T.C. is provided by NASA through the NASA Hubble Fellowship Program grant \#HST-HF2-51453.001 awarded by the Space Telescope Science Institute, which is operated by the Association of Universities for Research in Astronomy, Inc., for NASA, under contract NAS5-26555.
K.Cr. is supported by a UBC Four Year Fellowship (6456).
M.E.D. acknowledges support from the Naval Research Laboratory by NASA under contract S-15633Y.
T.D. and M.T.L. are supported by an NSF Astronomy and Astrophysics Grant (AAG) award number 2009468.
E.C.F. is supported by NASA under award number 80GSFC21M0002.
G.E.F., S.C.S., and S.J.V. are supported by NSF award PHY-2011772.
K.A.G. and S.R.T. acknowledge support from an NSF CAREER award \#2146016.
The Flatiron Institute is supported by the Simons Foundation.
S.H. is supported by the National Science Foundation Graduate Research Fellowship under Grant No. DGE-1745301.
N.La. acknowledges the support from Larry W. Martin and Joyce B. O'Neill Endowed Fellowship in the College of Science at Oregon State University.
Part of this research was carried out at the Jet Propulsion Laboratory, California Institute of Technology, under a contract with the National Aeronautics and Space Administration (80NM0018D0004).
D.R.L. and M.A.Mc. are supported by NSF \#1458952.
M.A.Mc. is supported by NSF \#2009425.
C.M.F.M. was supported in part by the National Science Foundation under Grants No. NSF PHY-1748958 and AST-2106552.
A.Mi. is supported by the Deutsche Forschungsgemeinschaft under Germany's Excellence Strategy - EXC 2121 Quantum Universe - 390833306.
P.N. acknowledges support from the BHI, funded by grants from the John Templeton Foundation and the Gordon and Betty Moore Foundation.
The Dunlap Institute is funded by an endowment established by the David Dunlap family and the University of Toronto.
K.D.O. was supported in part by NSF Grant No. 2207267.
T.T.P. acknowledges support from the Extragalactic Astrophysics Research Group at E\"{o}tv\"{o}s Lor\'{a}nd University, funded by the E\"{o}tv\"{o}s Lor\'{a}nd Research Network (ELKH), which was used during the development of this research.
S.M.R. and I.H.S. are CIFAR Fellows.
Portions of this work performed at NRL were supported by ONR 6.1 basic research funding.
J.D.R. also acknowledges support from start-up funds from Texas Tech University.
J.S. is supported by an NSF Astronomy and Astrophysics Postdoctoral Fellowship under award AST-2202388, and acknowledges previous support by the NSF under award 1847938.
C.U. acknowledges support from BGU (Kreitman fellowship), and the Council for Higher Education and Israel Academy of Sciences and Humanities (Excellence fellowship).
C.A.W. acknowledges support from CIERA, the Adler Planetarium, and the Brinson Foundation through a CIERA-Adler postdoctoral fellowship.
O.Y. is supported by the National Science Foundation Graduate Research Fellowship under Grant No. DGE-2139292.

\textit{Dedication.} This paper is dedicated to the memory of Donald Backer:  a pioneer in pulsar timing arrays, a term he coined; a discoverer of the first millisecond pulsar; a master developer of pulsar timing instrumentation; a founding member of NANOGrav; and a friend and mentor to many of us.

\facilities{Arecibo, GBT, VLA}

\software{\texttt{acor}, \texttt{astropy} \citep{2022ApJ...935..167A}, \texttt{ceffyl} \citep{2023arXiv230315442L}, \texttt{chainconsumer} \citep{Hinton2016}, \texttt{ENTERPRISE} \citep{enterprise}, \texttt{enterprise\_extensions} \citep{enterprise_ext}, \texttt{hasasia} \citep{hasasia}, \texttt{holodeck} \citep{holodeck}, \texttt{Jupyter} \citep{Kluyver2016jupyter}, \texttt{libstempo} \citep{libstempo}, \texttt{matplotlib} \citep{matplotlib}, \texttt{numpy} \citep{harris2020array}, \texttt{PINT} \citep{2021ApJ...911...45L}, \texttt{PTMCMC} \citep{ptmcmc}, \texttt{scipy} \citep{2020SciPy-NMeth}, \texttt{Tempo2} \citep{tempo2}}


\vspace{12pt}

\bibliographystyle{aasjournal}
\bibliography{apjjabb,bib}

\begin{thebibliography}{}
\expandafter\ifx\csname natexlab\endcsname\relax\def\natexlab#1{#1}\fi
\providecommand{\url}[1]{\href{#1}{#1}}
\providecommand{\dodoi}[1]{doi:~\href{http://doi.org/#1}{\nolinkurl{#1}}}
\providecommand{\doeprint}[1]{\href{http://ascl.net/#1}{\nolinkurl{http://ascl.net/#1}}}
\providecommand{\doarXiv}[1]{\href{https://arxiv.org/abs/#1}{\nolinkurl{https://arxiv.org/abs/#1}}}

\bibitem[{{Abbott} {et~al.}(2016){Abbott}, {Abbott}, {Abbott}, {Abernathy},
  {Acernese}, {Ackley}, {Adams}, {Adams}, {Addesso}, {Adhikari}, {Adya},
  {Affeldt}, {Agathos}, {Agatsuma}, {Aggarwal}, {Aguiar}, {Aiello}, {Ain},
  {Ajith}, {Allen}, {Allocca}, {Altin}, {Anderson}, {Anderson}, {Arai},
  {Arain}, {Araya}, {Arceneaux}, {Areeda}, {Arnaud}, {Arun}, {Ascenzi},
  {Ashton}, {Ast}, {Aston}, {Astone}, {Aufmuth}, {Aulbert}, {Babak}, {Bacon},
  {Bader}, {Baker}, {Baldaccini}, {Ballardin}, {Ballmer}, {Barayoga},
  {Barclay}, {Barish}, {Barker}, {Barone}, {Barr}, {Barsotti}, {Barsuglia},
  {Barta}, {Bartlett}, {Barton}, {Bartos}, {Bassiri}, {Basti}, {Batch},
  {Baune}, {Bavigadda}, {Bazzan}, {Behnke}, {Bejger}, {Belczynski}, {Bell},
  {Bell}, {Berger}, {Bergman}, {Bergmann}, {Berry}, {Bersanetti}, {Bertolini},
  {Betzwieser}, {Bhagwat}, {Bhandare}, {Bilenko}, {Billingsley}, {Birch},
  {Birney}, {Birnholtz}, {Biscans}, {Bisht}, {Bitossi}, {Biwer}, {Bizouard},
  {Blackburn}, {Blair}, {Blair}, {Blair}, {Bloemen}, {Bock}, {Bodiya}, {Boer},
  {Bogaert}, {Bogan}, {Bohe}, {Bojtos}, {Bond}, {Bondu}, {Bonnand}, {Boom},
  {Bork}, {Boschi}, {Bose}, {Bouffanais}, {Bozzi}, {Bradaschia}, {Brady},
  {Braginsky}, {Branchesi}, {Brau}, {Briant}, {Brillet}, {Brinkmann},
  {Brisson}, {Brockill}, {Brooks}, {Brown}, {Brown}, {Brown}, {Buchanan},
  {Buikema}, {Bulik}, {Bulten}, {Buonanno}, {Buskulic}, {Buy}, {Byer},
  {Cabero}, {Cadonati}, {Cagnoli}, {Cahillane}, {Bustillo}, {Callister},
  {Calloni}, {Camp}, {Cannon}, {Cao}, {Capano}, {Capocasa}, {Carbognani},
  {Caride}, {Casanueva Diaz}, {Casentini}, {Caudill}, {Cavagli{\`a}},
  {Cavalier}, {Cavalieri}, {Cella}, {Cepeda}, {Baiardi}, {Cerretani},
  {Cesarini}, {Chakraborty}, {Chalermsongsak}, {Chamberlin}, {Chan}, {Chao},
  {Charlton}, {Chassande-Mottin}, {Chen}, {Chen}, {Cheng}, {Chincarini},
  {Chiummo}, {Cho}, {Cho}, {Chow}, {Christensen}, {Chu}, {Chua}, {Chung},
  {Ciani}, {Clara}, {Clark}, {Cleva}, {Coccia}, {Cohadon}, {Colla}, {Collette},
  {Cominsky}, {Constancio}, {Conte}, {Conti}, {Cook}, {Corbitt}, {Cornish},
  {Corsi}, {Cortese}, {Costa}, {Coughlin}, {Coughlin}, {Coulon}, {Countryman},
  {Couvares}, {Cowan}, {Coward}, {Cowart}, {Coyne}, {Coyne}, {Craig},
  {Creighton}, {Creighton}, {Cripe}, {Crowder}, {Cruise}, {Cumming},
  {Cunningham}, {Cuoco}, {Dal Canton}, {Danilishin}, {D'Antonio}, {Danzmann},
  {Darman}, {Da Silva Costa}, {Dattilo}, {Dave}, {Daveloza}, {Davier},
  {Davies}, {Daw}, {Day}, {De}, {DeBra}, {Debreczeni}, {Degallaix}, {De
  Laurentis}, {Del{\'e}glise}, {Del Pozzo}, {Denker}, {Dent}, {Dereli},
  {Dergachev}, {DeRosa}, {De Rosa}, {DeSalvo}, {Dhurandhar}, {D{\'\i}az}, {Di
  Fiore}, {Di Giovanni}, {Di Lieto}, {Di Pace}, {Di Palma}, {Di Virgilio},
  {Dojcinoski}, {Dolique}, {Donovan}, {Dooley}, {Doravari}, {Douglas},
  {Downes}, {Drago}, {Drever}, {Driggers}, {Du}, {Ducrot}, {Dwyer}, {Edo},
  {Edwards}, {Effler}, {Eggenstein}, {Ehrens}, {Eichholz}, {Eikenberry},
  {Engels}, {Essick}, {Etzel}, {Evans}, {Evans}, {Everett}, {Factourovich},
  {Fafone}, {Fair}, {Fairhurst}, {Fan}, {Fang}, {Farinon}, {Farr}, {Farr},
  {Favata}, {Fays}, {Fehrmann}, {Fejer}, {Feldbaum}, {Ferrante}, {Ferreira},
  {Ferrini}, {Fidecaro}, {Finn}, {Fiori}, {Fiorucci}, {Fisher}, {Flaminio},
  {Fletcher}, {Fong}, {Fournier}, {Franco}, {Frasca}, {Frasconi}, {Frede},
  {Frei}, {Freise}, {Frey}, {Frey}, {Fricke}, {Fritschel}, {Frolov}, {Fulda},
  {Fyffe}, {Gabbard}, {Gair}, {Gammaitoni}, {Gaonkar}, {Garufi}, {Gatto},
  {Gaur}, {Gehrels}, {Gemme}, {Gendre}, {Genin}, {Gennai}, {George}, {Gergely},
  {Germain}, {Ghosh}, {Ghosh}, {Ghosh}, {Giaime}, {Giardina}, {Giazotto},
  {Gill}, {Glaefke}, {Gleason}, {Goetz}, {Goetz}, {Gondan}, {Gonz{\'a}lez},
  {Castro}, {Gopakumar}, {Gordon}, {Gorodetsky}, {Gossan}, {Gosselin},
  {Gouaty}, {Graef}, {Graff}, {Granata}, {Grant}, {Gras}, {Gray}, {Greco},
  {Green}, {Greenhalgh}, {Groot}, {Grote}, {Grunewald}, {Guidi}, {Guo},
  {Gupta}, {Gupta}, {Gushwa}, {Gustafson}, {Gustafson}, {Hacker}, {Hall},
  {Hall}, {Hammond}, {Haney}, {Hanke}, {Hanks}, {Hanna}, {Hannam}, {Hanson},
  {Hardwick}, {Harms}, {Harry}, {Harry}, {Hart}, {Hartman}, {Haster},
  {Haughian}, {Healy}, {Heefner}, {Heidmann}, {Heintze}, {Heinzel}, {Heitmann},
  {Hello}, {Hemming}, {Hendry}, {Heng}, {Hennig}, {Heptonstall}, {Heurs},
  {Hild}, {Hoak}, {Hodge}, {Hofman}, {Hollitt}, {Holt}, {Holz}, {Hopkins},
  {Hosken}, {Hough}, {Houston}, {Howell}, {Hu}, {Huang}, {Huerta}, {Huet},
  {Hughey}, {Husa}, {Huttner}, {Huynh-Dinh}, {Idrisy}, {Indik}, {Ingram},
  {Inta}, {Isa}, {Isac}, {Isi}, {Islas}, {Isogai}, {Iyer}, {Izumi}, {Jacobson},
  {Jacqmin}, {Jang}, {Jani}, {Jaranowski}, {Jawahar}, {Jim{\'e}nez-Forteza},
  {Johnson}, {Johnson-McDaniel}, {Jones}, {Jones}, {Jonker}, {Ju}, {Haris},
  {Kalaghatgi}, {Kalogera}, {Kandhasamy}, {Kang}, {Kanner}, {Karki},
  {Kasprzack}, {Katsavounidis}, {Katzman}, {Kaufer}, {Kaur}, {Kawabe},
  {Kawazoe}, {K{\'e}f{\'e}lian}, {Kehl}, {Keitel}, {Kelley}, {Kells},
  {Kennedy}, {Keppel}, {Key}, {Khalaidovski}, {Khalili}, {Khan}, {Khan},
  {Khan}, {Khazanov}, {Kijbunchoo}, {Kim}, {Kim}, {Kim}, {Kim}, {Kim}, {Kim},
  {King}, {King}, {Kinzel}, {Kissel}, {Kleybolte}, {Klimenko}, {Koehlenbeck},
  {Kokeyama}, {Koley}, {Kondrashov}, {Kontos}, {Koranda}, {Korobko}, {Korth},
  {Kowalska}, {Kozak}, {Kringel}, {Krishnan}, {Kr{\'o}lak}, {Krueger}, {Kuehn},
  {Kumar}, {Kumar}, {Kuo}, {Kutynia}, {Kwee}, {Lackey}, {Landry}, {Lange},
  {Lantz}, {Lasky}, {Lazzarini}, {Lazzaro}, {Leaci}, {Leavey}, {Lebigot},
  {Lee}, {Lee}, {Lee}, {Lee}, {Lenon}, {Leonardi}, {Leong}, {Leroy},
  {Letendre}, {Levin}, {Levine}, {Li}, {Libson}, {Littenberg}, {Lockerbie},
  {Logue}, {Lombardi}, {London}, {Lord}, {Lorenzini}, {Loriette}, {Lormand},
  {Losurdo}, {Lough}, {Lousto}, {Lovelace}, {L{\"u}ck}, {Lundgren}, {Luo},
  {Lynch}, {Ma}, {MacDonald}, {Machenschalk}, {MacInnis}, {Macleod},
  {Maga{\~n}a-Sandoval}, {Magee}, {Mageswaran}, {Majorana}, {Maksimovic},
  {Malvezzi}, {Man}, {Mandel}, {Mandic}, {Mangano}, {Mansell}, {Manske},
  {Mantovani}, {Marchesoni}, {Marion}, {M{\'a}rka}, {M{\'a}rka}, {Markosyan},
  {Maros}, {Martelli}, {Martellini}, {Martin}, {Martin}, {Martynov}, {Marx},
  {Mason}, {Masserot}, {Massinger}, {Masso-Reid}, {Matichard}, {Matone},
  {Mavalvala}, {Mazumder}, {Mazzolo}, {McCarthy}, {McClelland}, {McCormick},
  {McGuire}, {McIntyre}, {McIver}, {McManus}, {McWilliams}, {Meacher},
  {Meadors}, {Meidam}, {Melatos}, {Mendell}, {Mendoza-Gandara}, {Mercer},
  {Merilh}, {Merzougui}, {Meshkov}, {Messenger}, {Messick}, {Meyers},
  {Mezzani}, {Miao}, {Michel}, {Middleton}, {Mikhailov}, {Milano}, {Miller},
  {Millhouse}, {Minenkov}, {Ming}, {Mirshekari}, {Mishra}, {Mitra},
  {Mitrofanov}, {Mitselmakher}, {Mittleman}, {Moggi}, {Mohan}, {Mohapatra},
  {Montani}, {Moore}, {Moore}, {Moraru}, {Moreno}, {Morriss}, {Mossavi},
  {Mours}, {Mow-Lowry}, {Mueller}, {Mueller}, {Muir}, {Mukherjee}, {Mukherjee},
  {Mukherjee}, {Mukund}, {Mullavey}, {Munch}, {Murphy}, {Murray}, {Mytidis},
  {Nardecchia}, {Naticchioni}, {Nayak}, {Necula}, {Nedkova}, {Nelemans},
  {Neri}, {Neunzert}, {Newton}, {Nguyen}, {Nielsen}, {Nissanke}, {Nitz},
  {Nocera}, {Nolting}, {Normandin}, {Nuttall}, {Oberling}, {Ochsner}, {O'Dell},
  {Oelker}, {Ogin}, {Oh}, {Oh}, {Ohme}, {Oliver}, {Oppermann}, {Oram},
  {O'Reilly}, {O'Shaughnessy}, {Ott}, {Ottaway}, {Ottens}, {Overmier}, {Owen},
  {Pai}, {Pai}, {Palamos}, {Palashov}, {Palomba}, {Pal-Singh}, {Pan}, {Pan},
  {Pankow}, {Pannarale}, {Pant}, {Paoletti}, {Paoli}, {Papa}, {Paris},
  {Parker}, {Pascucci}, {Pasqualetti}, {Passaquieti}, {Passuello},
  {Patricelli}, {Patrick}, {Pearlstone}, {Pedraza}, {Pedurand}, {Pekowsky},
  {Pele}, {Penn}, {Perreca}, {Pfeiffer}, {Phelps}, {Piccinni}, {Pichot},
  {Pickenpack}, {Piergiovanni}, {Pierro}, {Pillant}, {Pinard}, {Pinto},
  {Pitkin}, {Poeld}, {Poggiani}, {Popolizio}, {Post}, {Powell}, {Prasad},
  {Predoi}, {Premachandra}, {Prestegard}, {Price}, {Prijatelj}, {Principe},
  {Privitera}, {Prix}, {Prodi}, {Prokhorov}, {Puncken}, {Punturo}, {Puppo},
  {P{\"u}rrer}, {Qi}, {Qin}, {Quetschke}, {Quintero}, {Quitzow-James}, {Raab},
  {Rabeling}, {Radkins}, {Raffai}, {Raja}, {Rakhmanov}, {Ramet}, {Rapagnani},
  {Raymond}, {Razzano}, {Re}, {Read}, {Reed}, {Regimbau}, {Rei}, {Reid},
  {Reitze}, {Rew}, {Reyes}, {Ricci}, {Riles}, {Robertson}, {Robie}, {Robinet},
  {Rocchi}, {Rolland}, {Rollins}, {Roma}, {Romano}, {Romano}, {Romanov},
  {Romie}, {Rosi{\'n}ska}, {Rowan}, {R{\"u}diger}, {Ruggi}, {Ryan}, {Sachdev},
  {Sadecki}, {Sadeghian}, {Salconi}, {Saleem}, {Salemi}, {Samajdar}, {Sammut},
  {Sampson}, {Sanchez}, {Sandberg}, {Sandeen}, {Sanders}, {Sanders},
  {Sassolas}, {Sathyaprakash}, {Saulson}, {Sauter}, {Savage}, {Sawadsky},
  {Schale}, {Schilling}, {Schmidt}, {Schmidt}, {Schnabel}, {Schofield},
  {Sch{\"o}nbeck}, {Schreiber}, {Schuette}, {Schutz}, {Scott}, {Scott},
  {Sellers}, {Sengupta}, {Sentenac}, {Sequino}, {Sergeev}, {Serna},
  {Setyawati}, {Sevigny}, {Shaddock}, {Shaffer}, {Shah}, {Shahriar}, {Shaltev},
  {Shao}, {Shapiro}, {Shawhan}, {Sheperd}, {Shoemaker}, {Shoemaker}, {Siellez},
  {Siemens}, {Sigg}, {Silva}, {Simakov}, {Singer}, {Singer}, {Singh}, {Singh},
  {Singhal}, {Sintes}, {Slagmolen}, {Smith}, {Smith}, {Smith}, {Smith}, {Son},
  {Sorazu}, {Sorrentino}, {Souradeep}, {Srivastava}, {Staley}, {Steinke},
  {Steinlechner}, {Steinlechner}, {Steinmeyer}, {Stephens}, {Stevenson},
  {Stone}, {Strain}, {Straniero}, {Stratta}, {Strauss}, {Strigin}, {Sturani},
  {Stuver}, {Summerscales}, {Sun}, {Sutton}, {Swinkels}, {Szczepa{\'n}czyk},
  {Tacca}, {Talukder}, {Tanner}, {T{\'a}pai}, {Tarabrin}, {Taracchini},
  {Taylor}, {Theeg}, {Thirugnanasambandam}, {Thomas}, {Thomas}, {Thomas},
  {Thorne}, {Thorne}, {Thrane}, {Tiwari}, {Tiwari}, {Tokmakov}, {Tomlinson},
  {Tonelli}, {Torres}, {Torrie}, {T{\"o}yr{\"a}}, {Travasso}, {Traylor},
  {Trifir{\`o}}, {Tringali}, {Trozzo}, {Tse}, {Turconi}, {Tuyenbayev},
  {Ugolini}, {Unnikrishnan}, {Urban}, {Usman}, {Vahlbruch}, {Vajente},
  {Valdes}, {Vallisneri}, {van Bakel}, {van Beuzekom}, {van den Brand}, {Van
  Den Broeck}, {Vander-Hyde}, {van der Schaaf}, {van Heijningen}, {van Veggel},
  {Vardaro}, {Vass}, {Vas{\'u}th}, {Vaulin}, {Vecchio}, {Vedovato}, {Veitch},
  {Veitch}, {Venkateswara}, {Verkindt}, {Vetrano}, {Vicer{\'e}}, {Vinciguerra},
  {Vine}, {Vinet}, {Vitale}, {Vo}, {Vocca}, {Vorvick}, {Voss}, {Vousden},
  {Vyatchanin}, {Wade}, {Wade}, {Wade}, {Waldman}, {Walker}, {Wallace},
  {Walsh}, {Wang}, {Wang}, {Wang}, {Wang}, {Wang}, {Ward}, {Ward}, {Warner},
  {Was}, {Weaver}, {Wei}, {Weinert}, {Weinstein}, {Weiss}, {Welborn}, {Wen},
  {We{\ss}els}, {Westphal}, {Wette}, {Whelan}, {Whitcomb}, {White}, {Whiting},
  {Wiesner}, {Wilkinson}, {Willems}, {Williams}, {Williams}, {Williamson},
  {Willis}, {Willke}, {Wimmer}, {Winkelmann}, {Winkler}, {Wipf}, {Wiseman},
  {Wittel}, {Woan}, {Worden}, {Wright}, {Wu}, {Yablon}, {Yakushin}, {Yam},
  {Yamamoto}, {Yancey}, {Yap}, {Yu}, {Yvert}, {Zadro{\.Z}ny}, {Zangrando},
  {Zanolin}, {Zendri}, {Zevin}, {Zhang}, {Zhang}, {Zhang}, {Zhang}, {Zhao},
  {Zhou}, {Zhou}, {Zhu}, {Zucker}, {Zuraw}, {Zweizig}, {LIGO Scientific
  Collaboration}, \& {Virgo Collaboration}}]{2016PhRvL.116f1102A}
{Abbott}, B.~P., {Abbott}, R., {Abbott}, T.~D., {et~al.} 2016, \prl, 116,
  061102, \dodoi{10.1103/PhysRevLett.116.061102}

\bibitem[{{Afzal} {et~al.}(2023)}]{aaa+23_cosmo}
{Afzal}, A., {et~al.} 2023, in preparation, \dodoi{10.3847/2041-8213/acdc91}

\bibitem[{{Agazie} {et~al.}(2023{\natexlab{a}})}]{aaa+23_smbhb}
{Agazie}, G., {et~al.} 2023{\natexlab{a}}, in preparation

\bibitem[{{Agazie} {et~al.}(2023{\natexlab{b}})}]{aaa+23}
---. 2023{\natexlab{b}}, in preparation, \dodoi{10.3847/2041-8213/acda9a}

\bibitem[{{Agazie} {et~al.}(2023{\natexlab{c}})}]{aaa+23_noise}
---. 2023{\natexlab{c}}, in preparation, \dodoi{10.3847/2041-8213/acda88}

\bibitem[{{Aggarwal} {et~al.}(2019){Aggarwal}, {Arzoumanian}, {Baker},
  {Brazier}, {Brinson}, {Brook}, {Burke-Spolaor}, {Chatterjee}, {Cordes},
  {Cornish}, {Crawford}, {Crowter}, {Cromartie}, {DeCesar}, {Demorest},
  {Dolch}, {Ellis}, {Ferdman}, {Ferrara}, {Fonseca}, {Garver-Daniels},
  {Gentile}, {Hazboun}, {Holgado}, {Huerta}, {Islo}, {Jennings}, {Jones},
  {Jones}, {Kaiser}, {Kaplan}, {Kelley}, {Key}, {Lam}, {Lazio}, {Levin},
  {Lorimer}, {Luo}, {Lynch}, {Madison}, {McLaughlin}, {McWilliams},
  {Mingarelli}, {Ng}, {Nice}, {Pennucci}, {Pol}, {Ransom}, {Ray}, {Siemens},
  {Simon}, {Spiewak}, {Stairs}, {Stinebring}, {Stovall}, {Swiggum}, {Taylor},
  {Turner}, {Vallisneri}, {van Haasteren}, {Vigeland }, {Witt}, {Zhu}, \&
  {NANOGrav Collaboration}}]{aab+19}
{Aggarwal}, K., {Arzoumanian}, Z., {Baker}, P.~T., {et~al.} 2019, \apj, 880,
  116, \dodoi{10.3847/1538-4357/ab2236}

\bibitem[{Akaike(1998)}]{Akaike1998}
Akaike, H. 1998, Information Theory and an Extension of the Maximum Likelihood
  Principle, ed. E.~Parzen, K.~Tanabe, \& G.~Kitagawa (New York, NY: Springer
  New York), 199--213, \dodoi{10.1007/978-1-4612-1694-0_15}

\bibitem[{{Alam} {et~al.}(2021{\natexlab{a}}){Alam}, {Arzoumanian}, {Baker},
  {Blumer}, {Bohler}, {Brazier}, {Brook}, {Burke-Spolaor}, {Caballero},
  {Camuccio}, {Chamberlain}, {Chatterjee}, {Cordes}, {Cornish}, {Crawford},
  {Cromartie}, {Decesar}, {Demorest}, {Dolch}, {Ellis}, {Ferdman}, {Ferrara},
  {Fiore}, {Fonseca}, {Garcia}, {Garver-Daniels}, {Gentile}, {Good},
  {Gusdorff}, {Halmrast}, {Hazboun}, {Islo}, {Jennings}, {Jessup}, {Jones},
  {Kaiser}, {Kaplan}, {Kelley}, {Key}, {Lam}, {Lazio}, {Lorimer}, {Luo},
  {Lynch}, {Madison}, {Maraccini}, {McLaughlin}, {Mingarelli}, {Ng}, {Nguyen},
  {Nice}, {Pennucci}, {Pol}, {Ramette}, {Ransom}, {Ray}, {Shapiro-Albert},
  {Siemens}, {Simon}, {Spiewak}, {Stairs}, {Stinebring}, {Stovall}, {Swiggum},
  {Taylor}, {Tripepi}, {Vallisneri}, {Vigeland}, {Witt}, {Zhu}, \& {Nanograv
  Collaboration}}]{aab+20}
{Alam}, M.~F., {Arzoumanian}, Z., {Baker}, P.~T., {et~al.} 2021{\natexlab{a}},
  \apjs, 252, 4, \dodoi{10.3847/1538-4365/abc6a0}

\bibitem[{{Alam} {et~al.}(2021{\natexlab{b}}){Alam}, {Arzoumanian}, {Baker},
  {Blumer}, {Bohler}, {Brazier}, {Brook}, {Burke-Spolaor}, {Caballero},
  {Camuccio}, {Chamberlain}, {Chatterjee}, {Cordes}, {Cornish}, {Crawford},
  {Cromartie}, {Decesar}, {Demorest}, {Dolch}, {Ellis}, {Ferdman}, {Ferrara},
  {Fiore}, {Fonseca}, {Garcia}, {Garver-Daniels}, {Gentile}, {Good},
  {Gusdorff}, {Halmrast}, {Hazboun}, {Islo}, {Jennings}, {Jessup}, {Jones},
  {Kaiser}, {Kaplan}, {Kelley}, {Key}, {Lam}, {Lazio}, {Lorimer}, {Luo},
  {Lynch}, {Madison}, {Maraccini}, {McLaughlin}, {Mingarelli}, {Ng}, {Nguyen},
  {Nice}, {Pennucci}, {Pol}, {Ramette}, {Ransom}, {Ray}, {Shapiro-Albert},
  {Siemens}, {Simon}, {Spiewak}, {Stairs}, {Stinebring}, {Stovall}, {Swiggum},
  {Taylor}, {Tripepi}, {Vallisneri}, {Vigeland}, {Witt}, {Zhu}, \& {Nanograv
  Collaboration}}]{12yr_wideband}
---. 2021{\natexlab{b}}, \apjs, 252, 5, \dodoi{10.3847/1538-4365/abc6a1}

\bibitem[{{Allen}(2023)}]{2023PhRvD.107d3018A}
{Allen}, B. 2023, \prd, 107, 043018, \dodoi{10.1103/PhysRevD.107.043018}

\bibitem[{{Allen} \& {Romano}(2022)}]{2022arXiv220807230A}
{Allen}, B., \& {Romano}, J.~D. 2022, arXiv e-prints, arXiv:2208.07230,
  \dodoi{10.48550/arXiv.2208.07230}

\bibitem[{{Anholm} {et~al.}(2009){Anholm}, {Ballmer}, {Creighton}, {Price}, \&
  {Siemens}}]{abc+2009}
{Anholm}, M., {Ballmer}, S., {Creighton}, J.~D.~E., {Price}, L.~R., \&
  {Siemens}, X. 2009, \prd, 79, 084030, \dodoi{10.1103/PhysRevD.79.084030}

\bibitem[{{Antoniadis} {et~al.}(2022){Antoniadis}, {Arzoumanian}, {Babak},
  {Bailes}, {Bak Nielsen}, {Baker}, {Bassa}, {B{\'e}csy}, {Berthereau},
  {Bonetti}, {Brazier}, {Brook}, {Burgay}, {Burke-Spolaor}, {Caballero},
  {Casey-Clyde}, {Chalumeau}, {Champion}, {Charisi}, {Chatterjee}, {Chen},
  {Cognard}, {Cordes}, {Cornish}, {Crawford}, {Cromartie}, {Crowter}, {Dai},
  {DeCesar}, {Demorest}, {Desvignes}, {Dolch}, {Drachler}, {Falxa}, {Ferrara},
  {Fiore}, {Fonseca}, {Gair}, {Garver-Daniels}, {Goncharov}, {Good}, {Graikou},
  {Guillemot}, {Guo}, {Hazboun}, {Hobbs}, {Hu}, {Islo}, {Janssen}, {Jennings},
  {Johnson}, {Jones}, {Kaiser}, {Kaplan}, {Karuppusamy}, {Keith}, {Kelley},
  {Kerr}, {Key}, {Kramer}, {Lam}, {Lamb}, {Lazio}, {Lee}, {Lentati}, {Liu},
  {Luo}, {Lynch}, {Lyne}, {Madison}, {Main}, {Manchester}, {McEwen}, {McKee},
  {McLaughlin}, {Mickaliger}, {Mingarelli}, {Ng}, {Nice}, {Os{\l}owski},
  {Parthasarathy}, {Pennucci}, {Perera}, {Perrodin}, {Petiteau}, {Pol},
  {Porayko}, {Possenti}, {Ransom}, {Ray}, {Reardon}, {Russell}, {Samajdar},
  {Sampson}, {Sanidas}, {Sarkissian}, {Schmitz}, {Schult}, {Sesana},
  {Shaifullah}, {Shannon}, {Shapiro-Albert}, {Siemens}, {Simon}, {Smith},
  {Speri}, {Spiewak}, {Stairs}, {Stappers}, {Stinebring}, {Swiggum}, {Taylor},
  {Theureau}, {Tiburzi}, {Vallisneri}, {van der Wateren}, {Vecchio},
  {Verbiest}, {Vigeland}, {Wahl}, {Wang}, {Wang}, {Wang}, {Witt}, {Zhang}, \&
  {Zhu}}]{2022MNRAS.510.4873A}
{Antoniadis}, J., {Arzoumanian}, Z., {Babak}, S., {et~al.} 2022, \mnras, 510,
  4873, \dodoi{10.1093/mnras/stab3418}

\bibitem[{{Antoniadis} {et~al.}(2023)}]{epta23}
{Antoniadis}, J., {et~al.} 2023, in preparation

\bibitem[{{Armitage} \& {Natarajan}(2002)}]{2002ApJ...567L...9A}
{Armitage}, P.~J., \& {Natarajan}, P. 2002, \apjl, 567, L9,
  \dodoi{10.1086/339770}

\bibitem[{Arzoumanian {et~al.}(2015)Arzoumanian, Brazier, Burke-Spolaor,
  Chamberlin, Chatterjee, Christy, Cordes, Cornish, Crowter, Demorest, Dolch,
  Ellis, Ferdman, Fonseca, Garver-Daniels, Gonzalez, Jenet, Jones, Jones,
  Kaspi, Koop, Lam, Lazio, Levin, Lommen, Lorimer, Luo, Lynch, Madison,
  McLaughlin, Mcwilliams, Nice, Palliyaguru, Pennucci, Ransom, Siemens, Stairs,
  Stinebring, Stovall, Swiggum, Vallisneri, van Haasteren, Wang, \&
  Zhu}]{abb+15}
Arzoumanian, Z., Brazier, A., Burke-Spolaor, S., {et~al.} 2015, \apj, 813, 65,
  \dodoi{10.1088/0004-637X/813/1/65}

\bibitem[{Arzoumanian {et~al.}(2016)Arzoumanian, Brazier, Burke-Spolaor,
  Chamberlin, Chatterjee, Christy, Cordes, Cornish, Crowter, Demorest, Deng,
  Dolch, Ellis, Ferdman, Fonseca, Garver-Daniels, Gonzalez, Jenet, Jones,
  Jones, Kaspi, Koop, Lam, Lazio, Levin, Lommen, Lorimer, Luo, Lynch, Madison,
  Mclaughlin, Mcwilliams, Mingarelli, Nice, Palliyaguru, Pennucci, Ransom,
  Sampson, Sanidas, Sesana, Siemens, Simon, Stairs, Stinebring, Stovall,
  Swiggum, Taylor, Vallisneri, van Haasteren, Wang, Zhu, \&
  Collaboration}]{abb+16}
---. 2016, \apj, 821, 13, \dodoi{10.3847/0004-637X/821/1/13}

\bibitem[{{Arzoumanian} {et~al.}(2018){Arzoumanian}, {Baker}, {Brazier},
  {Burke-Spolaor}, {Chamberlin}, {Chatterjee}, {Christy}, {Cordes}, {Cornish},
  {Crawford}, {Thankful Cromartie}, {Crowter}, {DeCesar}, {Demorest}, {Dolch},
  {Ellis}, {Ferdman}, {Ferrara}, {Folkner}, {Fonseca}, {Garver-Daniels},
  {Gentile}, {Haas}, {Hazboun}, {Huerta}, {Islo}, {Jones}, {Jones}, {Kaplan},
  {Kaspi}, {Lam}, {Lazio}, {Levin}, {Lommen}, {Lorimer}, {Luo}, {Lynch},
  {Madison}, {McLaughlin}, {McWilliams}, {Mingarelli}, {Ng}, {Nice}, {Park},
  {Pennucci}, {Pol}, {Ransom}, {Ray}, {Rasskazov}, {Siemens}, {Simon},
  {Spiewak}, {Stairs}, {Stinebring}, {Stovall}, {Swiggum}, {Taylor},
  {Vallisneri}, {van Haasteren}, {Vigeland}, {Zhu}, \& {NANOGrav
  Collaboration}}]{abb+18b}
{Arzoumanian}, Z., {Baker}, P.~T., {Brazier}, A., {et~al.} 2018, \apj, 859, 47,
  \dodoi{10.3847/1538-4357/aabd3b}

\bibitem[{Arzoumanian {et~al.}(2020)Arzoumanian, Baker, Blumer, B{\'{e} }csy,
  Brazier, Brook, Burke-Spolaor, Chatterjee, Chen, Cordes, Cornish, Crawford,
  Cromartie, DeCesar, Demorest, Dolch, Ellis, Ferrara, Fiore, Fonseca,
  Garver-Daniels, Gentile, Good, Hazboun, Holgado, Islo, Jennings, Jones,
  Kaiser, Kaplan, Kelley, Key, Laal, Lam, Lazio, Lorimer, Luo, Lynch, Madison,
  McLaughlin, Mingarelli, Ng, Nice, Pennucci, Pol, Ransom, Ray, Shapiro-Albert,
  Siemens, Simon, Spiewak, Stairs, Stinebring, Stovall, Sun, Swiggum, Taylor,
  Turner, Vallisneri, Vigeland, \& Witt}]{abb+20}
Arzoumanian, Z., Baker, P.~T., Blumer, H., {et~al.} 2020, The Astrophysical
  Journal Letters, 905, L34, \dodoi{10.3847/2041-8213/abd401}

\bibitem[{{Arzoumanian} {et~al.}(2021){Arzoumanian}, {Baker}, {Blumer},
  {B{\'e}csy}, {Brazier}, {Brook}, {Burke-Spolaor}, {Charisi}, {Chatterjee},
  {Chen}, {Cordes}, {Cornish}, {Crawford}, {Cromartie}, {Decesar}, {Degan},
  {Demorest}, {Dolch}, {Drachler}, {Ellis}, {Ferrara}, {Fiore}, {Fonseca},
  {Garver-Daniels}, {Gentile}, {Good}, {Hazboun}, {Holgado}, {Islo},
  {Jennings}, {Jones}, {Kaiser}, {Kaplan}, {Kelley}, {Key}, {Laal}, {Lam}, {W.
  Lazio}, {Lorimer}, {Liu}, {Luo}, {Lynch}, {Madison}, {McEwen}, {McLaughlin},
  {Mingarelli}, {Ng}, {Nice}, {Olum}, {Pennucci}, {Pol}, {Ransom}, {Ray},
  {Romano}, {Sardesai}, {Shapiro-Albert}, {Siemens}, {Simon}, {Siwek},
  {Spiewak}, {Stairs}, {Stinebring}, {Stovall}, {Sun}, {Swiggum}, {Taylor},
  {Turner}, {Vallisneri}, {Vigeland}, {Wahl}, {Witt}, \& {NANOGRAV
  Collaboration}}]{2021ApJ...923L..22A}
{Arzoumanian}, Z., {Baker}, P.~T., {Blumer}, H., {et~al.} 2021, \apjl, 923,
  L22, \dodoi{10.3847/2041-8213/ac401c}

\bibitem[{{Arzoumanian} {et~al.}(2023){Arzoumanian}, {Baker}, {Blecha},
  {Blumer}, {Brazier}, {Brook}, {Burke-Spolaor}, {B{\'e}csy}, {Casey-Clyde},
  {Charisi}, {Chatterjee}, {Chen}, {Cordes}, {Cornish}, {Crawford},
  {Cromartie}, {DeCesar}, {Demorest}, {Dolch}, {Drachler}, {Ellis}, {Ferrara},
  {Fiore}, {Fonseca}, {Freedman}, {Garver-Daniels}, {Gentile}, {Glaser},
  {Good}, {G{\"u}ltekin}, {Hazboun}, {Jennings}, {Johnson}, {Jones}, {Kaiser},
  {Kaplan}, {Kelley}, {Shapiro Key}, {Laal}, {Lam}, {Lamb}, {Lazio},
  {Lewandowska}, {Liu}, {Lorimer}, {Luo}, {Lynch}, {Madison}, {McEwen},
  {McLaughlin}, {Mingarelli}, {Ng}, {Nice}, {Ocker}, {Olum}, {Pennucci}, {Pol},
  {Ransom}, {Ray}, {Romano}, {Shapiro-Albert}, {Siemens}, {Simon}, {Siwek},
  {Spiewak}, {Stairs}, {Stinebring}, {Stovall}, {Swiggum}, {Sydnor}, {Taylor},
  {Turner}, {Vallisneri}, {Vigeland}, {Wahl}, {Walsh}, {Witt}, \&
  {Young}}]{2023arXiv230103608A}
{Arzoumanian}, Z., {Baker}, P.~T., {Blecha}, L., {et~al.} 2023, arXiv e-prints,
  arXiv:2301.03608, \dodoi{10.48550/arXiv.2301.03608}

\bibitem[{{Astropy Collaboration} {et~al.}(2022){Astropy Collaboration},
  {Price-Whelan}, {Lim}, {Earl}, {Starkman}, {Bradley}, {Shupe}, {Patil},
  {Corrales}, {Brasseur}, {N{\"o}the}, {Donath}, {Tollerud}, {Morris},
  {Ginsburg}, {Vaher}, {Weaver}, {Tocknell}, {Jamieson}, {van Kerkwijk},
  {Robitaille}, {Merry}, {Bachetti}, {G{\"u}nther}, {Aldcroft},
  {Alvarado-Montes}, {Archibald}, {B{\'o}di}, {Bapat}, {Barentsen},
  {Baz{\'a}n}, {Biswas}, {Boquien}, {Burke}, {Cara}, {Cara}, {Conroy},
  {Conseil}, {Craig}, {Cross}, {Cruz}, {D'Eugenio}, {Dencheva}, {Devillepoix},
  {Dietrich}, {Eigenbrot}, {Erben}, {Ferreira}, {Foreman-Mackey}, {Fox},
  {Freij}, {Garg}, {Geda}, {Glattly}, {Gondhalekar}, {Gordon}, {Grant},
  {Greenfield}, {Groener}, {Guest}, {Gurovich}, {Handberg}, {Hart},
  {Hatfield-Dodds}, {Homeier}, {Hosseinzadeh}, {Jenness}, {Jones}, {Joseph},
  {Kalmbach}, {Karamehmetoglu}, {Ka{\l}uszy{\'n}ski}, {Kelley}, {Kern},
  {Kerzendorf}, {Koch}, {Kulumani}, {Lee}, {Ly}, {Ma}, {MacBride}, {Maljaars},
  {Muna}, {Murphy}, {Norman}, {O'Steen}, {Oman}, {Pacifici}, {Pascual},
  {Pascual-Granado}, {Patil}, {Perren}, {Pickering}, {Rastogi}, {Roulston},
  {Ryan}, {Rykoff}, {Sabater}, {Sakurikar}, {Salgado}, {Sanghi}, {Saunders},
  {Savchenko}, {Schwardt}, {Seifert-Eckert}, {Shih}, {Jain}, {Shukla}, {Sick},
  {Simpson}, {Singanamalla}, {Singer}, {Singhal}, {Sinha}, {Sip{\H{o}}cz},
  {Spitler}, {Stansby}, {Streicher}, {{\v{S}}umak}, {Swinbank}, {Taranu},
  {Tewary}, {Tremblay}, {de Val-Borro}, {Van Kooten}, {Vasovi{\'c}}, {Verma},
  {de Miranda Cardoso}, {Williams}, {Wilson}, {Winkel}, {Wood-Vasey}, {Xue},
  {Yoachim}, {Zhang}, {Zonca}, \& {Astropy Project
  Contributors}}]{2022ApJ...935..167A}
{Astropy Collaboration}, {Price-Whelan}, A.~M., {Lim}, P.~L., {et~al.} 2022,
  \apj, 935, 167, \dodoi{10.3847/1538-4357/ac7c74}

\bibitem[{{Backer} {et~al.}(1982){Backer}, {Kulkarni}, {Heiles}, {Davis}, \&
  {Goss}}]{1982Natur.300..615B}
{Backer}, D.~C., {Kulkarni}, S.~R., {Heiles}, C., {Davis}, M.~M., \& {Goss},
  W.~M. 1982, \nat, 300, 615, \dodoi{10.1038/300615a0}

\bibitem[{{Bartolo} {et~al.}(2022){Bartolo}, {Bertacca}, {Caldwell},
  {Contaldi}, {Cusin}, {De Luca}, {Dimastrogiovanni}, {Fasiello}, {Figueroa},
  {Franciolini}, {Jenkins}, {Peloso}, {Pieroni}, {Renzini}, {Ricciardone},
  {Riotto}, {Sakellariadou}, {Sorbo}, {Tasinato}, {Torrado}, {Clesse},
  {Kuroyanagi}, \& {LISA Cosmology Working Group}}]{2022JCAP...11..009B}
{Bartolo}, N., {Bertacca}, D., {Caldwell}, R., {et~al.} 2022, \jcap, 2022, 009,
  \dodoi{10.1088/1475-7516/2022/11/009}

\bibitem[{{B{\'e}csy} {et~al.}(2022){B{\'e}csy}, {Cornish}, \&
  {Kelley}}]{2022ApJ...941..119B}
{B{\'e}csy}, B., {Cornish}, N.~J., \& {Kelley}, L.~Z. 2022, \apj, 941, 119,
  \dodoi{10.3847/1538-4357/aca1b2}

\bibitem[{{Begelman} {et~al.}(1980){Begelman}, {Blandford}, \& {Rees}}]{bbr80}
{Begelman}, M.~C., {Blandford}, R.~D., \& {Rees}, M.~J. 1980, Nature, 287, 307,
  \dodoi{10.1038/287307a0}

\bibitem[{{Blumenthal} {et~al.}(1984){Blumenthal}, {Faber}, {Primack}, \&
  {Rees}}]{blumenthal84}
{Blumenthal}, G.~R., {Faber}, S.~M., {Primack}, J.~R., \& {Rees}, M.~J. 1984,
  \nat, 311, 517, \dodoi{10.1038/311517a0}

\bibitem[{{Burke-Spolaor} {et~al.}(2019){Burke-Spolaor}, {Taylor}, {Charisi},
  {Dolch}, {Hazboun}, {Holgado}, {Kelley}, {Lazio}, {Madison}, {McMann},
  {Mingarelli}, {Rasskazov}, {Siemens}, {Simon}, \& {Smith}}]{stc+19}
{Burke-Spolaor}, S., {Taylor}, S.~R., {Charisi}, M., {et~al.} 2019, \aapr, 27,
  5, \dodoi{10.1007/s00159-019-0115-7}

\bibitem[{{Caprini} \& {Figueroa}(2018)}]{2018CQGra..35p3001C}
{Caprini}, C., \& {Figueroa}, D.~G. 2018, Classical and Quantum Gravity, 35,
  163001, \dodoi{10.1088/1361-6382/aac608}

\bibitem[{Carlin \& Chib(1995)}]{cc95}
Carlin, B.~P., \& Chib, S. 1995, Journal of the Royal Statistical Society.
  Series B (Methodological), 57, 473.
\newblock \url{http://www.jstor.org/stable/2346151}

\bibitem[{{Chalumeau} {et~al.}(2022){Chalumeau}, {Babak}, {Petiteau}, {Chen},
  {Samajdar}, {Caballero}, {Theureau}, {Guillemot}, {Desvignes},
  {Parthasarathy}, {Liu}, {Shaifullah}, {Hu}, {van der Wateren}, {Antoniadis},
  {Bak Nielsen}, {Bassa}, {Berthereau}, {Burgay}, {Champion}, {Cognard},
  {Falxa}, {Ferdman}, {Freire}, {Gair}, {Graikou}, {Guo}, {Jang}, {Janssen},
  {Karuppusamy}, {Keith}, {Kramer}, {Lee}, {Liu}, {Lyne}, {Main}, {McKee},
  {Mickaliger}, {Perera}, {Perrodin}, {Porayko}, {Possenti}, {Sanidas},
  {Sesana}, {Speri}, {Stappers}, {Tiburzi}, {Vecchio}, {Verbiest}, {Wang},
  {Wang}, \& {Xu}}]{Chalumeau+2022}
{Chalumeau}, A., {Babak}, S., {Petiteau}, A., {et~al.} 2022, \mnras, 509, 5538,
  \dodoi{10.1093/mnras/stab3283}

\bibitem[{{Chamberlin} {et~al.}(2015){Chamberlin}, {Creighton}, {Siemens},
  {Demorest}, {Ellis}, {Price}, \& {Romano}}]{ccs+2015}
{Chamberlin}, S.~J., {Creighton}, J.~D.~E., {Siemens}, X., {et~al.} 2015, \prd,
  91, 044048, \dodoi{10.1103/PhysRevD.91.044048}

\bibitem[{{Chen} {et~al.}(2021){Chen}, {Caballero}, {Guo}, {Chalumeau}, {Liu},
  {Shaifullah}, {Lee}, {Babak}, {Desvignes}, {Parthasarathy}, {Hu}, {van der
  Wateren}, {Antoniadis}, {Bak Nielsen}, {Bassa}, {Berthereau}, {Burgay},
  {Champion}, {Cognard}, {Falxa}, {Ferdman}, {Freire}, {Gair}, {Graikou},
  {Guillemot}, {Jang}, {Janssen}, {Karuppusamy}, {Keith}, {Kramer}, {Liu},
  {Lyne}, {Main}, {McKee}, {Mickaliger}, {Perera}, {Perrodin}, {Petiteau},
  {Porayko}, {Possenti}, {Samajdar}, {Sanidas}, {Sesana}, {Speri}, {Stappers},
  {Theureau}, {Tiburzi}, {Vecchio}, {Verbiest}, {Wang}, {Wang}, \&
  {Xu}}]{epta_dr2_gwb}
{Chen}, S., {Caballero}, R.~N., {Guo}, Y.~J., {et~al.} 2021, \mnras, 508, 4970,
  \dodoi{10.1093/mnras/stab2833}

\bibitem[{Cordes(2013)}]{cordes2013}
Cordes, J.~M. 2013, Classical and Quantum Gravity, 30, 224002,
  \dodoi{10.1088/0264-9381/30/22/224002}

\bibitem[{Cornish \& Littenberg(2015)}]{cornish2015bayeswave}
Cornish, N.~J., \& Littenberg, T.~B. 2015, Classical and Quantum Gravity, 32,
  135012

\bibitem[{{Cornish} \& {Sampson}(2016)}]{cs16}
{Cornish}, N.~J., \& {Sampson}, L. 2016, \prd, 93, 104047,
  \dodoi{10.1103/PhysRevD.93.104047}

\bibitem[{Cornish \& Sesana(2013)}]{Cornish:2013aba}
Cornish, N.~J., \& Sesana, A. 2013, Class. Quant. Grav., 30, 224005,
  \dodoi{10.1088/0264-9381/30/22/224005}

\bibitem[{{Demorest}(2007)}]{Demorest2007}
{Demorest}, P.~B. 2007, PhD thesis, University of California, Berkeley

\bibitem[{{Demorest} {et~al.}(2013){Demorest}, {Ferdman}, {Gonzalez}, {Nice},
  {Ransom}, {Stairs}, {Arzoumanian}, {Brazier}, {Burke-Spolaor}, {Chamberlin},
  {Cordes}, {Ellis}, {Finn}, {Freire}, {Giampanis}, {Jenet}, {Kaspi}, {Lazio},
  {Lommen}, {McLaughlin}, {Palliyaguru}, {Perrodin}, {Shannon}, {Siemens},
  {Stinebring}, {Swiggum}, \& {Zhu}}]{dfg+13}
{Demorest}, P.~B., {Ferdman}, R.~D., {Gonzalez}, M.~E., {et~al.} 2013, ApJ,
  762, 94, \dodoi{10.1088/0004-637X/762/2/94}

\bibitem[{{Desvignes} {et~al.}(2016){Desvignes}, {Caballero}, {Lentati},
  {Verbiest}, {Champion}, {Stappers}, {Janssen}, {Lazarus}, {Os{\l}owski},
  {Babak}, {Bassa}, {Brem}, {Burgay}, {Cognard}, {Gair}, {Graikou},
  {Guillemot}, {Hessels}, {Jessner}, {Jordan}, {Karuppusamy}, {Kramer},
  {Lassus}, {Lazaridis}, {Lee}, {Liu}, {Lyne}, {McKee}, {Mingarelli},
  {Perrodin}, {Petiteau}, {Possenti}, {Purver}, {Rosado}, {Sanidas}, {Sesana},
  {Shaifullah}, {Smits}, {Taylor}, {Theureau}, {Tiburzi}, {van Haasteren}, \&
  {Vecchio}}]{dcl+16}
{Desvignes}, G., {Caballero}, R.~N., {Lentati}, L., {et~al.} 2016, \mnras, 458,
  3341, \dodoi{10.1093/mnras/stw483}

\bibitem[{{Detweiler}(1979)}]{det79}
{Detweiler}, S. 1979, \apj, 234, 1100, \dodoi{10.1086/157593}

\bibitem[{Dickey(1971)}]{dickey1971}
Dickey, J.~M. 1971, The Annals of Mathematical Statistics, 42, 204.
\newblock \url{http://www.jstor.org/stable/2958475}

\bibitem[{{Dom{\`e}nech}(2021)}]{2021Univ....7..398D}
{Dom{\`e}nech}, G. 2021, Universe, 7, 398, \dodoi{10.3390/universe7110398}

\bibitem[{{DuPlain} {et~al.}(2008){DuPlain}, {Ransom}, {Demorest}, {Brandt},
  {Ford}, \& {Shelton}}]{DuPlain2008}
{DuPlain}, R., {Ransom}, S., {Demorest}, P., {et~al.} 2008, in \procspie, Vol.
  7019, Advanced Software and Control for Astronomy II, 70191D,
  \dodoi{10.1117/12.790003}

\bibitem[{Einstein(1916)}]{einstein1916approximative}
Einstein, A. 1916, Sitzungsber. Preuss. Akad. Wiss. Berlin (Math. Phys.), 1916,
  1

\bibitem[{Ellis \& van Haasteren(2017)}]{ptmcmc}
Ellis, J., \& van Haasteren, R. 2017, jellis18/PTMCMCSampler: Official Release,
  \dodoi{10.5281/zenodo.1037579}

\bibitem[{{Ellis} {et~al.}(2019){Ellis}, {Vallisneri}, {Taylor}, \&
  {Baker}}]{enterprise}
{Ellis}, J.~A., {Vallisneri}, M., {Taylor}, S.~R., \& {Baker}, P.~T. 2019,
  {ENTERPRISE: Enhanced Numerical Toolbox Enabling a Robust PulsaR Inference
  SuitE}.
\newblock \doeprint{1912.015}

\bibitem[{{Enoki} \& {Nagashima}(2007)}]{en07}
{Enoki}, M., \& {Nagashima}, M. 2007, Progress of Theoretical Physics, 117,
  241, \dodoi{10.1143/PTP.117.241}

\bibitem[{{Estabrook} \& {Wahlquist}(1975)}]{1975GReGr...6..439E}
{Estabrook}, F.~B., \& {Wahlquist}, H.~D. 1975, General Relativity and
  Gravitation, 6, 439, \dodoi{10.1007/BF00762449}

\bibitem[{{Ford} {et~al.}(2010){Ford}, {Demorest}, \& {Ransom}}]{Ford2010}
{Ford}, J.~M., {Demorest}, P., \& {Ransom}, S. 2010, in \procspie, Vol. 7740,
  Software and Cyberinfrastructure for Astronomy, 77400A,
  \dodoi{10.1117/12.857666}

\bibitem[{{Foster} \& {Backer}(1990)}]{fb90}
{Foster}, R.~S., \& {Backer}, D.~C. 1990, \apj, 361, 300,
  \dodoi{10.1086/169195}

\bibitem[{Gair {et~al.}(2014)Gair, Romano, Taylor, \& Mingarelli}]{grtm14}
Gair, J., Romano, J.~D., Taylor, S., \& Mingarelli, C. M.~F. 2014, Phys. Rev.
  D, 90, 082001, \dodoi{10.1103/PhysRevD.90.082001}

\bibitem[{Gelman {et~al.}(2013)Gelman, Carlin, Stern, Dunson, Vehtari, \&
  Rubin}]{gelman2013bayesian}
Gelman, A., Carlin, J., Stern, H., {et~al.} 2013, Bayesian Data Analysis, Third
  Edition, Chapman \& Hall/CRC Texts in Statistical Science (Taylor \& Francis)

\bibitem[{Gelman \& Meng(1998)}]{gelman1998simulating}
Gelman, A., \& Meng, X.-L. 1998, Statistical science, 163,
  \dodoi{10.1214/ss/1028905934}

\bibitem[{Gelman {et~al.}(1996)Gelman, Meng, \& Stern}]{gelman_bayes_pvals}
Gelman, A., Meng, X.-L., \& Stern, H. 1996, Statistica Sinica, 6, 733.
\newblock \url{http://www.jstor.org/stable/24306036}

\bibitem[{Gelman \& Rubin(1992)}]{gelmanrubin1992}
Gelman, A., \& Rubin, D.~B. 1992, Statistical Science, 7, 457.
\newblock \url{http://www.jstor.org/stable/2246093}

\bibitem[{Godsill(2001)}]{g01}
Godsill, S.~J. 2001, Journal of Computational and Graphical Statistics, 10,
  230.
\newblock \url{http://www.jstor.org/stable/1391010}

\bibitem[{{Goncharov} {et~al.}(2021{\natexlab{a}}){Goncharov}, {Shannon},
  {Reardon}, {Hobbs}, {Zic}, {Bailes}, {Cury{\l}o}, {Dai}, {Kerr}, {Lower},
  {Manchester}, {Mandow}, {Middleton}, {Miles}, {Parthasarathy}, {Thrane},
  {Thyagarajan}, {Xue}, {Zhu}, {Cameron}, {Feng}, {Luo}, {Russell},
  {Sarkissian}, {Spiewak}, {Wang}, {Wang}, {Zhang}, \& {Zhang}}]{ppta_dr2_gwb}
{Goncharov}, B., {Shannon}, R.~M., {Reardon}, D.~J., {et~al.}
  2021{\natexlab{a}}, \apjl, 917, L19, \dodoi{10.3847/2041-8213/ac17f4}

\bibitem[{{Goncharov} {et~al.}(2021{\natexlab{b}}){Goncharov}, {Reardon},
  {Shannon}, {Zhu}, {Thrane}, {Bailes}, {Bhat}, {Dai}, {Hobbs}, {Kerr},
  {Manchester}, {Os{\l}owski}, {Parthasarathy}, {Russell}, {Spiewak},
  {Thyagarajan}, \& {Wang}}]{goncharov+21a}
{Goncharov}, B., {Reardon}, D.~J., {Shannon}, R.~M., {et~al.}
  2021{\natexlab{b}}, \mnras, 502, 478, \dodoi{10.1093/mnras/staa3411}

\bibitem[{{Goncharov} {et~al.}(2022){Goncharov}, {Thrane}, {Shannon}, {Harms},
  {Bhat}, {Hobbs}, {Kerr}, {Manchester}, {Reardon}, {Russell}, {Zhu}, \&
  {Zic}}]{gts+22}
{Goncharov}, B., {Thrane}, E., {Shannon}, R.~M., {et~al.} 2022, \apjl, 932,
  L22, \dodoi{10.3847/2041-8213/ac76bb}

\bibitem[{Guzzetti {et~al.}(2016)Guzzetti, Bartolo, Liguori, \&
  Matarrese}]{guzzetti2016gravitational}
Guzzetti, M.~C., Bartolo, N., Liguori, M., \& Matarrese, S. 2016, La Rivista
  del Nuovo Cimento, 39, 399, \dodoi{10.1393/ncr/i2016-10127-1}

\bibitem[{Harris {et~al.}(2020)Harris, Millman, van~der Walt, Gommers,
  Virtanen, Cournapeau, Wieser, Taylor, Berg, Smith, Kern, Picus, Hoyer, van
  Kerkwijk, Brett, Haldane, del R{\'{i}}o, Wiebe, Peterson,
  G{\'{e}}rard-Marchant, Sheppard, Reddy, Weckesser, Abbasi, Gohlke, \&
  Oliphant}]{harris2020array}
Harris, C.~R., Millman, K.~J., van~der Walt, S.~J., {et~al.} 2020, Nature, 585,
  357, \dodoi{10.1038/s41586-020-2649-2}

\bibitem[{{Hazboun} {et~al.}(2023){Hazboun}, {Meyers}, {Romano}, {Siemens}, \&
  {Archibald}}]{hazboun+2023GX2}
{Hazboun}, J., {Meyers}, P.~M., {Romano}, J.~D., {Siemens}, X., \& {Archibald},
  A.~M. 2023, arXiv e-prints, arXiv:2305.01116.
\newblock \doarXiv{2305.01116}

\bibitem[{{Hazboun} {et~al.}(2019){Hazboun}, {Romano}, \& {Smith}}]{hasasia}
{Hazboun}, J., {Romano}, J., \& {Smith}, T. 2019, The Journal of Open Source
  Software, 4, 1775, \dodoi{10.21105/joss.01775}

\bibitem[{Hazboun {et~al.}(2019)Hazboun, Romano, \& Smith}]{Hazboun:2019vhv}
Hazboun, J.~S., Romano, J.~D., \& Smith, T.~L. 2019, Phys. Rev. D, 100, 104028,
  \dodoi{10.1103/PhysRevD.100.104028}

\bibitem[{{Hazboun} {et~al.}(2020){Hazboun}, {Simon}, {Taylor}, {Lam},
  {Vigeland}, {Islo}, {Key}, {Arzoumanian}, {Baker}, {Brazier}, {Brook},
  {Burke-Spolaor}, {Chatterjee}, {Cordes}, {Cornish}, {Crawford}, {Crowter},
  {Cromartie}, {DeCesar}, {Demorest}, {Dolch}, {Ellis}, {Ferdman}, {Ferrara},
  {Fonseca}, {Garver-Daniels}, {Gentile}, {Good}, {Holgado}, {Huerta},
  {Jennings}, {Jones}, {Jones}, {Kaiser}, {Kaplan}, {Kelley}, {Lazio}, {Levin},
  {Lommen}, {Lorimer}, {Luo}, {Lynch}, {Madison}, {McLaughlin}, {McWilliams},
  {Mingarelli}, {Ng}, {Nice}, {Pennucci}, {Pol}, {Ransom}, {Ray}, {Siemens},
  {Spiewak}, {Stairs}, {Stinebring}, {Stovall}, {Swiggum}, {Turner},
  {Vallisneri}, {van Haasteren}, {Witt}, \& {Zhu}}]{hazboun:2020a}
{Hazboun}, J.~S., {Simon}, J., {Taylor}, S.~R., {et~al.} 2020, \apj, 890, 108,
  \dodoi{10.3847/1538-4357/ab68db}

\bibitem[{{Hazboun} {et~al.}(2022){Hazboun}, {Simon}, {Madison}, {Arzoumanian},
  {Cromartie}, {Crowter}, {Decesar}, {Demorest}, {Dolch}, {Ellis}, {Ferdman},
  {Ferrara}, {Fonseca}, {Gentile}, {Jones}, {Jones}, {Lam}, {Levin}, {Lorimer},
  {Lynch}, {McLaughlin}, {Ng}, {Nice}, {Pennucci}, {Ransom}, {Ray}, {Spiewak},
  {Stairs}, {Stovall}, {Swiggum}, {Zhu}, \& {The Nanograv
  Collaboration}}]{2022ApJ...929...39H}
{Hazboun}, J.~S., {Simon}, J., {Madison}, D.~R., {et~al.} 2022, \apj, 929, 39,
  \dodoi{10.3847/1538-4357/ac5829}

\bibitem[{Heck {et~al.}(2019)Heck, Overstall, Gronau, \&
  Wagenmakers}]{heck2019quantifying}
Heck, D.~W., Overstall, A.~M., Gronau, Q.~F., \& Wagenmakers, E.-J. 2019,
  Statistics and Computing, 29, 631, \dodoi{10.1007/s11222-018-9828-0}

\bibitem[{Hee {et~al.}(2015)Hee, Handley, Hobson, \& Lasenby}]{hee15}
Hee, S., Handley, W.~J., Hobson, M.~P., \& Lasenby, A.~N. 2015, Monthly Notices
  of the Royal Astronomical Society, 455, 2461, \dodoi{10.1093/mnras/stv2217}

\bibitem[{{Hellings} \& {Downs}(1983)}]{hd83}
{Hellings}, R.~W., \& {Downs}, G.~S. 1983, ApJL, 265, L39,
  \dodoi{10.1086/183954}

\bibitem[{{Hinton}(2016)}]{Hinton2016}
{Hinton}, S.~R. 2016, The Journal of Open Source Software, 1, 00045,
  \dodoi{10.21105/joss.00045}

\bibitem[{{Hobbs} \& {Edwards}(2012)}]{tempo2}
{Hobbs}, G., \& {Edwards}, R. 2012, {Tempo2: Pulsar Timing Package}.
\newblock \doeprint{1210.015}

\bibitem[{{Hobbs} {et~al.}(2012){Hobbs}, {Coles}, {Manchester}, {Keith},
  {Shannon}, {Chen}, {Bailes}, {Bhat}, {Burke-Spolaor}, {Champion},
  {Chaudhary}, {Hotan}, {Khoo}, {Kocz}, {Levin}, {Oslowski}, {Preisig}, {Ravi},
  {Reynolds}, {Sarkissian}, {van Straten}, {Verbiest}, {Yardley}, \&
  {You}}]{2012MNRAS.427.2780H}
{Hobbs}, G., {Coles}, W., {Manchester}, R.~N., {et~al.} 2012, \mnras, 427,
  2780, \dodoi{10.1111/j.1365-2966.2012.21946.x}

\bibitem[{{Hobbs} {et~al.}(2020){Hobbs}, {Guo}, {Caballero}, {Coles}, {Lee},
  {Manchester}, {Reardon}, {Matsakis}, {Tong}, {Arzoumanian}, {Bailes},
  {Bassa}, {Bhat}, {Brazier}, {Burke-Spolaor}, {Champion}, {Chatterjee},
  {Cognard}, {Dai}, {Desvignes}, {Dolch}, {Ferdman}, {Graikou}, {Guillemot},
  {Janssen}, {Keith}, {Kerr}, {Kramer}, {Lam}, {Liu}, {Lyne}, {Lazio}, {Lynch},
  {McKee}, {McLaughlin}, {Mingarelli}, {Nice}, {Os{\l}owski}, {Pennucci},
  {Perera}, {Perrodin}, {Possenti}, {Russell}, {Sanidas}, {Sesana},
  {Shaifullah}, {Shannon}, {Simon}, {Spiewak}, {Stairs}, {Stappers}, {Swiggum},
  {Taylor}, {Theureau}, {Toomey}, {van Haasteren}, {Wang}, {Wang}, \&
  {Zhu}}]{2020MNRAS.491.5951H}
{Hobbs}, G., {Guo}, L., {Caballero}, R.~N., {et~al.} 2020, \mnras, 491, 5951,
  \dodoi{10.1093/mnras/stz3071}

\bibitem[{{Hourihane} {et~al.}(2023){Hourihane}, {Meyers}, {Johnson},
  {Chatziioannou}, \& {Vallisneri}}]{HourihaneMeyers2022}
{Hourihane}, S., {Meyers}, P., {Johnson}, A., {Chatziioannou}, K., \&
  {Vallisneri}, M. 2023, \prd, 107, 084045, \dodoi{10.1103/PhysRevD.107.084045}

\bibitem[{{Hulse} \& {Taylor}(1975)}]{1975ApJ...195L..51H}
{Hulse}, R.~A., \& {Taylor}, J.~H. 1975, \apjl, 195, L51,
  \dodoi{10.1086/181708}

\bibitem[{{Hunter}(2007)}]{matplotlib}
{Hunter}, J.~D. 2007, Computing in Science and Engineering, 9, 90,
  \dodoi{10.1109/MCSE.2007.55}

\bibitem[{{Jaffe} \& {Backer}(2003)}]{jb03}
{Jaffe}, A.~H., \& {Backer}, D.~C. 2003, Astrophysical Journal, 583, 616,
  \dodoi{10.1086/345443}

\bibitem[{{Jenet} {et~al.}(2006){Jenet}, {Hobbs}, {van Straten}, {Manchester},
  {Bailes}, {Verbiest}, {Edwards}, {Hotan}, {Sarkissian}, \&
  {Ord}}]{2006ApJ...653.1571J}
{Jenet}, F.~A., {Hobbs}, G.~B., {van Straten}, W., {et~al.} 2006, \apj, 653,
  1571, \dodoi{10.1086/508702}

\bibitem[{{Johnson} {et~al.}(2023)}]{code_review}
{Johnson}, A., {et~al.} 2023, in preparation

\bibitem[{{Jones} {et~al.}(2017){Jones}, {McLaughlin}, {Lam}, {Cordes},
  {Levin}, {Chatterjee}, {Arzoumanian}, {Crowter}, {Demorest}, {Dolch},
  {Ellis}, {Ferdman}, {Fonseca}, {Gonzalez}, {Jones}, {Lazio}, {Nice},
  {Pennucci}, {Ransom}, {Stinebring}, {Stairs}, {Stovall}, {Swiggum}, \&
  {Zhu}}]{jml+2017}
{Jones}, M.~L., {McLaughlin}, M.~A., {Lam}, M.~T., {et~al.} 2017, \apj, 841,
  125, \dodoi{10.3847/1538-4357/aa73df}

\bibitem[{{Joshi} {et~al.}(2018){Joshi}, {Arumugasamy}, {Bagchi},
  {Bandyopadhyay}, {Basu}, {Dhand a Batra}, {Bethapudi}, {Choudhary}, {De},
  {Dey}, {Gopakumar}, {Gupta}, {Krishnakumar}, {Maan}, {Manoharan}, {Naidu},
  {Nandi}, {Pathak}, {Surnis}, \& {Susobhanan}}]{InPTA}
{Joshi}, B.~C., {Arumugasamy}, P., {Bagchi}, M., {et~al.} 2018, Journal of
  Astrophysics and Astronomy, 39, 51, \dodoi{10.1007/s12036-018-9549-y}

\bibitem[{{Kaiser} {et~al.}(2022){Kaiser}, {Pol}, {McLaughlin}, {Chen},
  {Hazboun}, {Kelley}, {Simon}, {Taylor}, {Vigeland}, \& {Witt}}]{kpm+22}
{Kaiser}, A.~R., {Pol}, N.~S., {McLaughlin}, M.~A., {et~al.} 2022, \apj, 938,
  115, \dodoi{10.3847/1538-4357/ac86cc}

\bibitem[{{Kaspi} {et~al.}(1994){Kaspi}, {Taylor}, \&
  {Ryba}}]{1994ApJ...428..713K}
{Kaspi}, V.~M., {Taylor}, J.~H., \& {Ryba}, M.~F. 1994, \apj, 428, 713,
  \dodoi{10.1086/174280}

\bibitem[{{Kelley} {et~al.}(2023)}]{holodeck}
{Kelley}, L.~Z., {et~al.} 2023, in preparation

\bibitem[{{Khmelnitsky} \& {Rubakov}(2014)}]{2014JCAP...02..019K}
{Khmelnitsky}, A., \& {Rubakov}, V. 2014, \jcap, 2014, 019,
  \dodoi{10.1088/1475-7516/2014/02/019}

\bibitem[{Kluyver {et~al.}(2016)Kluyver, Ragan-Kelley, P{\'e}rez, Granger,
  Bussonnier, Frederic, Kelley, Hamrick, Grout, Corlay, Ivanov, Avila, Abdalla,
  \& Willing}]{Kluyver2016jupyter}
Kluyver, T., Ragan-Kelley, B., P{\'e}rez, F., {et~al.} 2016, in Positioning and
  Power in Academic Publishing: Players, Agents and Agendas, ed. F.~Loizides \&
  B.~Schmidt, IOS Press, 87 -- 90

\bibitem[{{Lam} {et~al.}(2017){Lam}, {Cordes}, {Chatterjee}, {Arzoumanian},
  {Crowter}, {Demorest}, {Dolch}, {Ellis}, {Ferdman}, {Fonseca}, {Gonzalez},
  {Jones}, {Jones}, {Levin}, {Madison}, {McLaughlin}, {Nice}, {Pennucci},
  {Ransom}, {Shannon}, {Siemens}, {Stairs}, {Stovall}, {Swiggum}, \&
  {Zhu}}]{lcc+2017}
{Lam}, M.~T., {Cordes}, J.~M., {Chatterjee}, S., {et~al.} 2017, \apj, 834, 35,
  \dodoi{10.3847/1538-4357/834/1/35}

\bibitem[{{Lam} {et~al.}(2018){Lam}, {Ellis}, {Grillo}, {Jones}, {Hazboun},
  {Brook}, {Turner}, {Chatterjee}, {Cordes}, {Lazio}, {DeCesar}, {Arzoumanian},
  {Blumer}, {Cromartie}, {Demorest}, {Dolch}, {Ferdman}, {Ferrara}, {Fonseca},
  {Garver-Daniels}, {Gentile}, {Gupta}, {Lorimer}, {Lynch}, {Madison},
  {McLaughlin}, {Ng}, {Nice}, {Pennucci}, {Ransom}, {Spiewak}, {Stairs},
  {Stinebring}, {Stovall}, {Swiggum}, {Vigeland}, \& {Zhu}}]{leg+2018}
{Lam}, M.~T., {Ellis}, J.~A., {Grillo}, G., {et~al.} 2018, \apj, 861, 132,
  \dodoi{10.3847/1538-4357/aac770}

\bibitem[{{Lamb} {et~al.}(2023){Lamb}, {Taylor}, \& {van
  Haasteren}}]{2023arXiv230315442L}
{Lamb}, W.~G., {Taylor}, S.~R., \& {van Haasteren}, R. 2023, arXiv e-prints,
  arXiv:2303.15442, \dodoi{10.48550/arXiv.2303.15442}

\bibitem[{{Lee}(2016)}]{CPTA}
{Lee}, K.~J. 2016, in Astronomical Society of the Pacific Conference Series,
  Vol. 502, Frontiers in Radio Astronomy and FAST Early Sciences Symposium
  2015, ed. L.~{Qain} \& D.~{Li}, 19

\bibitem[{{Lentati} {et~al.}(2013){Lentati}, {Alexander}, {Hobson}, {Taylor},
  {Gair}, {Balan}, \& {van Haasteren}}]{lentati+2013}
{Lentati}, L., {Alexander}, P., {Hobson}, M.~P., {et~al.} 2013, \prd, 87,
  104021, \dodoi{10.1103/PhysRevD.87.104021}

\bibitem[{{Luo} {et~al.}(2021){Luo}, {Ransom}, {Demorest}, {Ray}, {Archibald},
  {Kerr}, {Jennings}, {Bachetti}, {van Haasteren}, {Champagne}, {Colen},
  {Phillips}, {Zimmerman}, {Stovall}, {Lam}, \& {Jenet}}]{2021ApJ...911...45L}
{Luo}, J., {Ransom}, S., {Demorest}, P., {et~al.} 2021, \apj, 911, 45,
  \dodoi{10.3847/1538-4357/abe62f}

\bibitem[{{Magorrian} {et~al.}(1998){Magorrian}, {Tremaine}, {Richstone},
  {Bender}, {Bower}, {Dressler}, {Faber}, {Gebhardt}, {Green}, {Grillmair},
  {Kormendy}, \& {Lauer}}]{1998AJ....115.2285M}
{Magorrian}, J., {Tremaine}, S., {Richstone}, D., {et~al.} 1998, \aj, 115,
  2285, \dodoi{10.1086/300353}

\bibitem[{{Manchester} {et~al.}(2013){Manchester}, {Hobbs}, {Bailes}, {Coles},
  {van Straten}, {Keith}, {Shannon}, {Bhat}, {Brown}, {Burke-Spolaor},
  {Champion}, {Chaudhary}, {Edwards}, {Hampson}, {Hotan}, {Jameson}, {Jenet},
  {Kesteven}, {Khoo}, {Kocz}, {Maciesiak}, {Oslowski}, {Ravi}, {Reynolds},
  {Sarkissian}, {Verbiest}, {Wen}, {Wilson}, {Yardley}, {Yan}, \&
  {You}}]{2013PASA...30...17M}
{Manchester}, R.~N., {Hobbs}, G., {Bailes}, M., {et~al.} 2013, \pasa, 30, e017,
  \dodoi{10.1017/pasa.2012.017}

\bibitem[{{McConnell} \& {Ma}(2013)}]{2013ApJ...764..184M}
{McConnell}, N.~J., \& {Ma}, C.-P. 2013, \apj, 764, 184,
  \dodoi{10.1088/0004-637X/764/2/184}

\bibitem[{{McLaughlin}(2013)}]{m13}
{McLaughlin}, M.~A. 2013, Classical and Quantum Gravity, 30, 224008,
  \dodoi{10.1088/0264-9381/30/22/224008}

\bibitem[{{McWilliams} {et~al.}(2014){McWilliams}, {Ostriker}, \&
  {Pretorius}}]{2014ApJ...789..156M}
{McWilliams}, S.~T., {Ostriker}, J.~P., \& {Pretorius}, F. 2014, \apj, 789,
  156, \dodoi{10.1088/0004-637X/789/2/156}

\bibitem[{{Merritt} \& {Milosavljevi{\'c}}(2005)}]{mm05}
{Merritt}, D., \& {Milosavljevi{\'c}}, M. 2005, Living Reviews in Relativity,
  8, 8, \dodoi{10.12942/lrr-2005-8}

\bibitem[{{Meyers} {et~al.}(2023){Meyers}, {Chatziioannou}, {Vallisneri}, \&
  {Chua}}]{meyers+23}
{Meyers}, P.~M., {Chatziioannou}, K., {Vallisneri}, M., \& {Chua}, A. J.~K.
  2023, arXiv e-prints, arXiv:2306.05559, \dodoi{10.48550/arXiv.2306.05559}

\bibitem[{{Milosavljevi{\'c}} \& {Merritt}(2003)}]{mm03}
{Milosavljevi{\'c}}, M., \& {Merritt}, D. 2003, \apj, 596, 860,
  \dodoi{10.1086/378086}

\bibitem[{{Mingarelli} \& {Mingarelli}(2018)}]{2018JPhCo...2j5002M}
{Mingarelli}, C. M.~F., \& {Mingarelli}, A.~B. 2018, Journal of Physics
  Communications, 2, 105002, \dodoi{10.1088/2399-6528/aae06d}

\bibitem[{{Mingarelli} \& {Sidery}(2014)}]{2014PhRvD..90f2011M}
{Mingarelli}, C. M.~F., \& {Sidery}, T. 2014, \prd, 90, 062011,
  \dodoi{10.1103/PhysRevD.90.062011}

\bibitem[{{Mingarelli} {et~al.}(2013){Mingarelli}, {Sidery}, {Mandel}, \&
  {Vecchio}}]{2013PhRvD..88f2005M}
{Mingarelli}, C.~M.~F., {Sidery}, T., {Mandel}, I., \& {Vecchio}, A. 2013,
  \prd, 88, 062005, \dodoi{10.1103/PhysRevD.88.062005}

\bibitem[{{Mingarelli} {et~al.}(2017){Mingarelli}, {Lazio}, {Sesana}, {Greene},
  {Ellis}, {Ma}, {Croft}, {Burke-Spolaor}, \& {Taylor}}]{2017NatAs...1..886M}
{Mingarelli}, C. M.~F., {Lazio}, T. J.~W., {Sesana}, A., {et~al.} 2017, Nature
  Astronomy, 1, 886, \dodoi{10.1038/s41550-017-0299-6}

\bibitem[{{Nay} {et~al.}(2023){Nay}, {Boddy}, {Smith}, \&
  {Mingarelli}}]{nay+23}
{Nay}, J., {Boddy}, K.~K., {Smith}, T.~L., \& {Mingarelli}, C. M.~F. 2023,
  arXiv e-prints, arXiv:2306.06168, \dodoi{10.48550/arXiv.2306.06168}

\bibitem[{Ogata(1989)}]{ogata1989monte}
Ogata, Y. 1989, Numerische Mathematik, 55, 137, \dodoi{10.1007/BF01406511}

\bibitem[{{Park} {et~al.}(2021){Park}, {Folkner}, {Williams}, \&
  {Boggs}}]{2021AJ....161..105P}
{Park}, R.~S., {Folkner}, W.~M., {Williams}, J.~G., \& {Boggs}, D.~H. 2021,
  \aj, 161, 105, \dodoi{10.3847/1538-3881/abd414}

\bibitem[{{Perera} {et~al.}(2019){Perera}, {DeCesar}, {Demorest}, {Kerr},
  {Lentati}, {Nice}, {Os{\l}owski}, {Ransom}, {Keith}, {Arzoumanian}, {Bailes},
  {Baker}, {Bassa}, {Bhat}, {Brazier}, {Burgay}, {Burke-Spolaor}, {Caballero},
  {Champion}, {Chatterjee}, {Chen}, {Cognard}, {Cordes}, {Crowter}, {Dai},
  {Desvignes}, {Dolch}, {Ferdman}, {Ferrara}, {Fonseca}, {Goldstein},
  {Graikou}, {Guillemot}, {Hazboun}, {Hobbs}, {Hu}, {Islo}, {Janssen},
  {Karuppusamy}, {Kramer}, {Lam}, {Lee}, {Liu}, {Luo}, {Lyne}, {Manchester},
  {McKee}, {McLaughlin}, {Mingarelli}, {Parthasarathy}, {Pennucci}, {Perrodin},
  {Possenti}, {Reardon}, {Russell}, {Sanidas}, {Sesana}, {Shaifullah},
  {Shannon}, {Siemens}, {Simon}, {Spiewak}, {Stairs}, {Stappers}, {Swiggum},
  {Taylor}, {Theureau}, {Tiburzi}, {Vallisneri}, {Vecchio}, {Wang}, {Zhang},
  {Zhang}, {Zhu}, \& {Zhu}}]{pdd+19}
{Perera}, B.~B.~P., {DeCesar}, M.~E., {Demorest}, P.~B., {et~al.} 2019, \mnras,
  490, 4666, \dodoi{10.1093/mnras/stz2857}

\bibitem[{Petit(2022)}]{petit2022}
Petit, G. 2022, personal communication

\bibitem[{{Phinney}(2001)}]{p01}
{Phinney}, E.~S. 2001, arXiv e-prints, astro,
  \dodoi{10.48550/arXiv.astro-ph/0108028}

\bibitem[{{Pirani}(1956)}]{1956AcPP...15..389P}
{Pirani}, F.~A.~E. 1956, Acta Physica Polonica, 15, 389,
  \dodoi{10.1007/s10714-009-0787-9}

\bibitem[{{Pirani}(2009)}]{2009GReGr..41.1215P}
{Pirani}, F. A.~E. 2009, General Relativity and Gravitation, 41, 1215,
  \dodoi{10.1007/s10714-009-0787-9}

\bibitem[{{Pol} {et~al.}(2022){Pol}, {Taylor}, \& {Romano}}]{ptr2022}
{Pol}, N., {Taylor}, S.~R., \& {Romano}, J.~D. 2022, \apj, 940, 173,
  \dodoi{10.3847/1538-4357/ac9836}

\bibitem[{{Pol} {et~al.}(2021){Pol}, {Taylor}, {Kelley}, {Vigeland}, {Simon},
  {Chen}, {Arzoumanian}, {Baker}, {B{\'e}csy}, {Brazier}, {Brook},
  {Burke-Spolaor}, {Chatterjee}, {Cordes}, {Cornish}, {Crawford}, {Thankful
  Cromartie}, {Decesar}, {Demorest}, {Dolch}, {Ferrara}, {Fiore}, {Fonseca},
  {Garver-Daniels}, {Good}, {Hazboun}, {Jennings}, {Jones}, {Kaiser}, {Kaplan},
  {Shapiro Key}, {Lam}, {Lazio}, {Luo}, {Lynch}, {Madison}, {McEwen},
  {McLaughlin}, {Mingarelli}, {Ng}, {Nice}, {Pennucci}, {Ransom}, {Ray},
  {Shapiro-Albert}, {Siemens}, {Stairs}, {Stinebring}, {Swiggum}, {Vallisneri},
  {Wahl}, {Witt}, \& {Nanograv Collaboration}}]{astro4cast}
{Pol}, N.~S., {Taylor}, S.~R., {Kelley}, L.~Z., {et~al.} 2021, \apjl, 911, L34,
  \dodoi{10.3847/2041-8213/abf2c9}

\bibitem[{{Rajagopal} \& {Romani}(1995)}]{rr95}
{Rajagopal}, M., \& {Romani}, R.~W. 1995, \apj, 446, 543,
  \dodoi{10.1086/175813}

\bibitem[{{Ransom} {et~al.}(2019){Ransom}, {Brazier}, {Chatterjee}, {Cohen},
  {Cordes}, {DeCesar}, {Demorest}, {Hazboun}, {Lam}, {Lynch}, {McLaughlin},
  {Ransom}, {Siemens}, {Taylor}, \& {Vigeland}}]{ransom+19}
{Ransom}, S., {Brazier}, A., {Chatterjee}, S., {et~al.} 2019, in \baas,
  Vol.~51, 195.
\newblock \doarXiv{1908.05356}

\bibitem[{{Reardon} {et~al.}(2023)}]{ppta23}
{Reardon}, D.~J., {et~al.} 2023, in preparation

\bibitem[{{Roebber}(2019)}]{r19}
{Roebber}, E. 2019, \apj, 876, 55, \dodoi{10.3847/1538-4357/ab100e}

\bibitem[{{Roebber} \& {Holder}(2017)}]{2017ApJ...835...21R}
{Roebber}, E., \& {Holder}, G. 2017, \apj, 835, 21,
  \dodoi{10.3847/1538-4357/835/1/21}

\bibitem[{Romani(1989)}]{romani1989timing}
Romani, R.~W. 1989, in Timing Neutron Stars, ed. H.~{\"O}gelman \& E.~P.~J.
  Heuvel (Springer), 113--117

\bibitem[{{Romano} {et~al.}(2021){Romano}, {Hazboun}, {Siemens}, \&
  {Archibald}}]{romano+2021}
{Romano}, J.~D., {Hazboun}, J.~S., {Siemens}, X., \& {Archibald}, A.~M. 2021,
  \prd, 103, 063027, \dodoi{10.1103/PhysRevD.103.063027}

\bibitem[{{Rosado} {et~al.}(2015){Rosado}, {Sesana}, \& {Gair}}]{rsg2015}
{Rosado}, P.~A., {Sesana}, A., \& {Gair}, J. 2015, \mnras, 451, 2417,
  \dodoi{10.1093/mnras/stv1098}

\bibitem[{{Sampson} {et~al.}(2015){Sampson}, {Cornish}, \&
  {McWilliams}}]{scm2015}
{Sampson}, L., {Cornish}, N.~J., \& {McWilliams}, S.~T. 2015, \prd, 91, 084055,
  \dodoi{10.1103/PhysRevD.91.084055}

\bibitem[{{Sardesai} \& {Vigeland}(2023)}]{sv2023}
{Sardesai}, S.~C., \& {Vigeland}, S.~J. 2023, arXiv e-prints, arXiv:2303.09615.
\newblock \doarXiv{2303.09615}

\bibitem[{{Sazhin}(1978)}]{saz78}
{Sazhin}, M.~V. 1978, \sovast, 22, 36

\bibitem[{{Sesana}(2013)}]{2013MNRAS.433L...1S}
{Sesana}, A. 2013, \mnras, 433, L1, \dodoi{10.1093/mnrasl/slt034}

\bibitem[{{Sesana} {et~al.}(2004){Sesana}, {Haardt}, {Madau}, \&
  {Volonteri}}]{shm+04}
{Sesana}, A., {Haardt}, F., {Madau}, P., \& {Volonteri}, M. 2004, \apj, 611,
  623, \dodoi{10.1086/422185}

\bibitem[{{Sesana} {et~al.}(2008){Sesana}, {Vecchio}, \&
  {Colacino}}]{2008MNRAS.390..192S}
{Sesana}, A., {Vecchio}, A., \& {Colacino}, C.~N. 2008, \mnras, 390, 192,
  \dodoi{10.1111/j.1365-2966.2008.13682.x}

\bibitem[{Siemens {et~al.}(2013)Siemens, Ellis, Jenet, \&
  Romano}]{Siemens:2013zla}
Siemens, X., Ellis, J., Jenet, F., \& Romano, J.~D. 2013, Class. Quant. Grav.,
  30, 224015, \dodoi{10.1088/0264-9381/30/22/224015}

\bibitem[{{Speri} {et~al.}(2023){Speri}, {Porayko}, {Falxa}, {Chen}, {Gair},
  {Sesana}, \& {Taylor}}]{2023MNRAS.518.1802S}
{Speri}, L., {Porayko}, N.~K., {Falxa}, M., {et~al.} 2023, \mnras, 518, 1802,
  \dodoi{10.1093/mnras/stac3237}

\bibitem[{{Stinebring} {et~al.}(1990){Stinebring}, {Ryba}, {Taylor}, \&
  {Romani}}]{1990PhRvL..65..285S}
{Stinebring}, D.~R., {Ryba}, M.~F., {Taylor}, J.~H., \& {Romani}, R.~W. 1990,
  \prl, 65, 285, \dodoi{10.1103/PhysRevLett.65.285}

\bibitem[{{Taylor} {et~al.}(1979){Taylor}, {Fowler}, \&
  {McCulloch}}]{1979Natur.277..437T}
{Taylor}, J.~H., {Fowler}, L.~A., \& {McCulloch}, P.~M. 1979, \nat, 277, 437,
  \dodoi{10.1038/277437a0}

\bibitem[{Taylor(2021)}]{taylor2021nanohertz}
Taylor, S.~R. 2021, Nanohertz Gravitational Wave Astronomy (Boca Raton, FL: CRC
  Press)

\bibitem[{{Taylor} {et~al.}(2018){Taylor}, {Baker}, {Hazboun}, {Simon}, \&
  {Vigeland}}]{enterprise_ext}
{Taylor}, S.~R., {Baker}, P.~T., {Hazboun}, J.~S., {Simon}, J.~J., \&
  {Vigeland}, S.~J. 2018, enterprise extensions.
\newblock \url{https://github.com/nanograv/enterprise_extensions}

\bibitem[{{Taylor} \& {Gair}(2013)}]{2013PhRvD..88h4001T}
{Taylor}, S.~R., \& {Gair}, J.~R. 2013, \prd, 88, 084001,
  \dodoi{10.1103/PhysRevD.88.084001}

\bibitem[{{Taylor} {et~al.}(2017){Taylor}, {Lentati}, {Babak}, {Brem}, {Gair},
  {Sesana}, \& {Vecchio}}]{tlb+17}
{Taylor}, S.~R., {Lentati}, L., {Babak}, S., {et~al.} 2017, \prd, 95, 042002,
  \dodoi{10.1103/PhysRevD.95.042002}

\bibitem[{{Taylor} {et~al.}(2022){Taylor}, {Simon}, {Schult}, {Pol}, \&
  {Lamb}}]{2022PhRvD.105h4049T}
{Taylor}, S.~R., {Simon}, J., {Schult}, L., {Pol}, N., \& {Lamb}, W.~G. 2022,
  \prd, 105, 084049, \dodoi{10.1103/PhysRevD.105.084049}

\bibitem[{{Taylor} {et~al.}(2020){Taylor}, {van Haasteren}, \&
  {Sesana}}]{2020PhRvD.102h4039T}
{Taylor}, S.~R., {van Haasteren}, R., \& {Sesana}, A. 2020, \prd, 102, 084039,
  \dodoi{10.1103/PhysRevD.102.084039}

\bibitem[{{Tiburzi} {et~al.}(2016){Tiburzi}, {Hobbs}, {Kerr}, {Coles}, {Dai},
  {Manchester}, {Possenti}, {Shannon}, \& {You}}]{thk+2016}
{Tiburzi}, C., {Hobbs}, G., {Kerr}, M., {et~al.} 2016, \mnras, 455, 4339,
  \dodoi{10.1093/mnras/stv2143}

\bibitem[{{Tomczak} {et~al.}(2014){Tomczak}, {Quadri}, {Tran}, {Labb{\'e}},
  {Straatman}, {Papovich}, {Glazebrook}, {Allen}, {Brammer}, {Kacprzak},
  {Kawinwanichakij}, {Kelson}, {McCarthy}, {Mehrtens}, {Monson}, {Persson},
  {Spitler}, {Tilvi}, \& {van Dokkum}}]{Tomczak+2014}
{Tomczak}, A.~R., {Quadri}, R.~F., {Tran}, K.-V.~H., {et~al.} 2014, \apj, 783,
  85, \dodoi{10.1088/0004-637X/783/2/85}

\bibitem[{{Vallisneri}(2020)}]{libstempo}
{Vallisneri}, M. 2020, {libstempo: Python wrapper for Tempo2}.
\newblock \doeprint{2002.017}

\bibitem[{{Vallisneri} {et~al.}(2020){Vallisneri}, {Taylor}, {Simon},
  {Folkner}, {Park}, {Cutler}, {Ellis}, {Lazio}, {Vigeland}, {Aggarwal},
  {Arzoumanian}, {Baker}, {Brazier}, {Brook}, {Burke-Spolaor}, {Chatterjee},
  {Cordes}, {Cornish}, {Crawford}, {Cromartie}, {Crowter}, {DeCesar},
  {Demorest}, {Dolch}, {Ferdman}, {Ferrara}, {Fonseca}, {Garver-Daniels},
  {Gentile}, {Good}, {Hazboun}, {Holgado}, {Huerta}, {Islo}, {Jennings},
  {Jones}, {Jones}, {Kaplan}, {Kelley}, {Key}, {Lam}, {Levin}, {Lorimer},
  {Luo}, {Lynch}, {Madison}, {McLaughlin}, {McWilliams}, {Mingarelli}, {Ng},
  {Nice}, {Pennucci}, {Pol}, {Ransom}, {Ray}, {Siemens}, {Spiewak}, {Stairs},
  {Stinebring}, {Stovall}, {Swiggum}, {van Haasteren}, {Witt}, \&
  {Zhu}}]{2020ApJ...893..112V}
{Vallisneri}, M., {Taylor}, S.~R., {Simon}, J., {et~al.} 2020, \apj, 893, 112,
  \dodoi{10.3847/1538-4357/ab7b67}

\bibitem[{{van Haasteren} {et~al.}(2009){van Haasteren}, {Levin}, {McDonald},
  \& {Lu}}]{2009MNRAS.395.1005V}
{van Haasteren}, R., {Levin}, Y., {McDonald}, P., \& {Lu}, T. 2009, \mnras,
  395, 1005, \dodoi{10.1111/j.1365-2966.2009.14590.x}

\bibitem[{{van Haasteren} \& {Vallisneri}(2014)}]{2014PhRvD..90j4012V}
{van Haasteren}, R., \& {Vallisneri}, M. 2014, \prd, 90, 104012,
  \dodoi{10.1103/PhysRevD.90.104012}

\bibitem[{Vehtari {et~al.}(2021)Vehtari, Gelman, Simpson, Carpenter, \&
  B{\"u}rkner}]{vgs21}
Vehtari, A., Gelman, A., Simpson, D., Carpenter, B., \& B{\"u}rkner, P.-C.
  2021, Bayesian Analysis, 16, 667 , \dodoi{10.1214/20-BA1221}

\bibitem[{{Vigeland} {et~al.}(2018){Vigeland}, {Islo}, {Taylor}, \&
  {Ellis}}]{vite18}
{Vigeland}, S.~J., {Islo}, K., {Taylor}, S.~R., \& {Ellis}, J.~A. 2018, \prd,
  98, 044003, \dodoi{10.1103/PhysRevD.98.044003}

\bibitem[{Virtanen {et~al.}(2020)Virtanen, Gommers, Oliphant, Haberland, Reddy,
  Cournapeau, Burovski, Peterson, Weckesser, Bright, {van der Walt}, Brett,
  Wilson, Millman, Mayorov, Nelson, Jones, Kern, Larson, Carey, Polat, Feng,
  Moore, {VanderPlas}, Laxalde, Perktold, Cimrman, Henriksen, Quintero, Harris,
  Archibald, Ribeiro, Pedregosa, {van Mulbregt}, \& {SciPy 1.0
  Contributors}}]{2020SciPy-NMeth}
Virtanen, P., Gommers, R., Oliphant, T.~E., {et~al.} 2020, Nature Methods, 17,
  261, \dodoi{10.1038/s41592-019-0686-2}

\bibitem[{Wilcox(2012)}]{wilcox2012introduction}
Wilcox, R. 2012, Introduction to Robust Estimation and Hypothesis Testing,
  Statistical Modeling and Decision Science (Elsevier Science)

\bibitem[{{Wyithe} \& {Loeb}(2003)}]{wl03}
{Wyithe}, J. S.~B., \& {Loeb}, A. 2003, \apj, 590, 691, \dodoi{10.1086/375187}

\bibitem[{{Xu} {et~al.}(2023){Xu}, {Chen}, {Gio}, {et~al.}}]{cpta23}
{Xu}, H., {Chen}, S., {Gio}, Y., {et~al.} 2023, in preparation

\bibitem[{{Zic} {et~al.}(2022){Zic}, {Hobbs}, {Shannon}, {Reardon},
  {Goncharov}, {Bhat}, {Cameron}, {Dai}, {Dawson}, {Kerr}, {Manchester},
  {Mandow}, {Marshman}, {Russell}, {Thyagarajan}, \& {Zhu}}]{zhs+22}
{Zic}, A., {Hobbs}, G., {Shannon}, R.~M., {et~al.} 2022, \mnras, 516, 410,
  \dodoi{10.1093/mnras/stac2100}

\end{thebibliography}

\appendix

\section{Additional data set details}
\label{app:data_set_details}

\begin{figure}[t]
\begin{center}
    \includegraphics[width=\columnwidth]{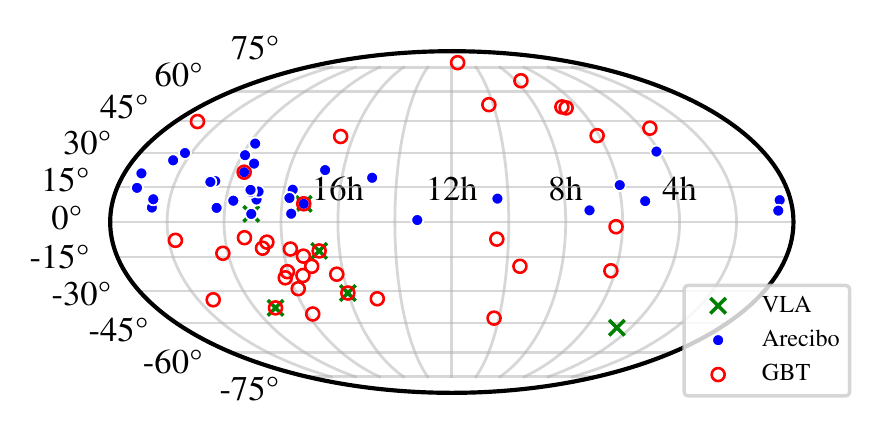}
    \includegraphics[width=\columnwidth]{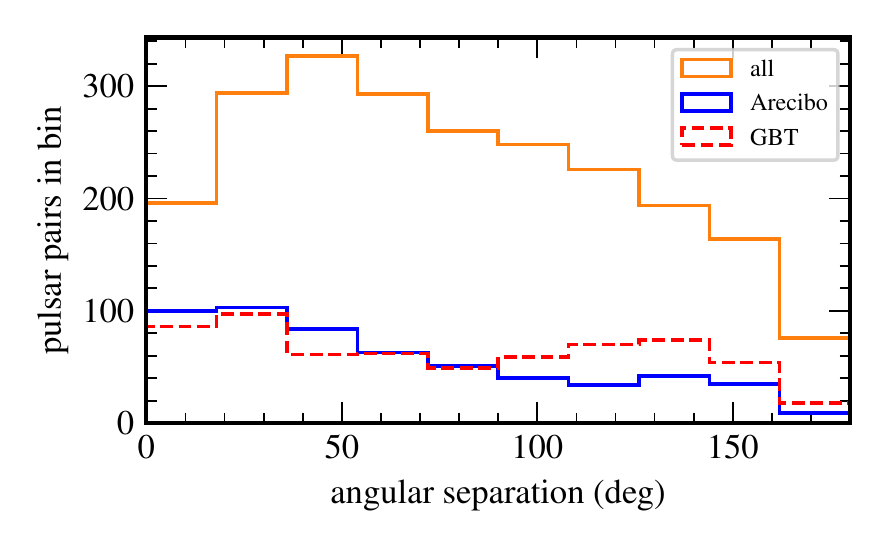}
\end{center}
	\vspace{-0.25in}
    \caption{\textit{Top:} Sky locations of the 67 pulsars used in the 15-year GWB analysis. Markers indicate which telescopes observed the pulsar. \textit{Bottom:} Distribution of angular separations probed by the pulsars in the full data set (orange), the Arecibo data set (blue), and the GBT data set (red). Because Arecibo and GBT mostly observed pulsars at different declinations, there are few inter-telescope pairs at small angular separations, resulting in a deficit of pairs for the full data set in the first bin.}
    \label{fig:ang_sep}
\end{figure}

The observations included in the NANOGrav 15-year data set were performed between July 2004 and August 2020 with the 305-m Arecibo Observatory (Arecibo), the 100-m Green Bank Telescope (GBT), and, since 2015, the 27 25-m antennae of the Very Large Array (VLA).
We used Arecibo to observe the 33 pulsars that lie within its declination range ($0^\circ < \delta < +39^\circ$); GBT to observe the pulsars that lie outside of Arecibo's range, plus J1713$+$0747 and B1937$+$21, for a total of 36 pulsars; the VLA to observe the seven pulsars J0437$-$4715, J1600$-$3053, J1643$-$1224, J1713+0747, J1903+0327, J1909$-$3744, and B1937$+$21. Six of these were also observed with Arecibo, GBT, or both; J0437$-$4715 was only visible to the VLA.
\autoref{fig:ang_sep} shows the sky locations of the 67 pulsars used for the GWB search (top) and the distribution of angular separations for the pulsar pairs (bottom).

Initial observations were performed with the ASP (Arecibo) and GASP (GBT) systems, with 64-MHz bandwidth \citep{Demorest2007}. 
Between 2010 and 2012, we transitioned to the PUPPI (Arecibo) and GUPPI (GBT) systems, with bandwidths up to 800 MHz \citep{DuPlain2008, Ford2010}.
We observe pulsars in two different radio-frequency bands in order to measure pulse dispersion from the interstellar medium: at Arecibo, we use the 1.4 GHz receiver plus either the 430 MHz or 2.1 GHz receiver (and the 327 MHz receiver for early observations of J2317+1439); at GBT, we use the 820 MHz and 1.4 GHz receivers; at the VLA, we use the 1.4 GHz and 3 GHz receivers with the YUPPI system.

In \S\ref{subsec:cross-validation} we analyze also two split-telescope data sets: 33 pulsars for Arecibo, and 35 for GBT (excluding J0614$-$3329, which was observed for less than three years).
For the two pulsars timed by both telescopes (J1713$+$0747 and B1937$+$21), we partition the timing data between the telescopes and obtain independent timing solutions for each.
We do not analyze a VLA-only data set, which would have shorter observation spans and significantly reduced sensitivity.

\section{Bayesian Methods \& Diagnostics}
\label{sec:bayesapp}


\begin{table*}[ht]
\begin{center}
\scriptsize
\caption{Prior distributions used in all analyses performed in this paper.}
\label{tab:priors}
\begin{tabular}{llll}
\hline\hline
parameter & description & prior & comments \\
\hline

\multicolumn{4}{c}{white noise} \\[1pt]
$E_{k}$ & EFAC per backend/receiver system & Uniform $[0, 10]$ & single-pulsar analysis only \\
$Q_{k}$ [s] & EQUAD per backend/receiver system & log-Uniform $[-8.5, -5]$ & single-pulsar analysis only \\
$J_{k}$ [s] & ECORR per backend/receiver system & log-Uniform $[-8.5, -5]$ & single-pulsar analysis only \\
\hline

\multicolumn{4}{c}{intrinsic red noise} \\[1pt]
$A_{\rm red}$ & red-noise power-law amplitude& log-Uniform $[-20, -11]$ & one parameter per pulsar  \\
$\gamma_{\rm red}$ & red-noise power-law spectral index & Uniform $[0, 7]$ & one parameter per pulsar \\
\hline

\multicolumn{4}{c}{all common processes, free spectrum} \\[1pt]
$\rho_{i}$ [s$^{2}$] & power-spectrum coefficients at $f=i/T$ & log-Uniform in $\rho_{i}$ $[-18,-8]$ & one parameter per frequency\\
\hline

\multicolumn{4}{c}{all common processes, power-law spectrum} \\[1pt]
$A$ & common process strain amplitude & log-Uniform $[-18, -14]$ ($\gamma=13/3$) & one parameter for PTA \\
& & log-Uniform $[-18, -11]$ ($\gamma$ varied) & one parameter for PTA \\
$\gamma$ & common process power-law spectral index & delta function ($\gamma=13/3$)& fixed \\
& & Uniform $[0,7]$ & one parameter for PTA \\
\hline

\multicolumn{4}{c}{all common processes, broken--power-law spectrum} \\[1pt]
$A$ & broken--power-law amplitude & log-Uniform $[-18, -11]$ & one parameter for PTA \\
$\gamma$ & broken--power-law low-freq.\ spectral index & Uniform $[0,7]$ & one parameter per PTA \\
$\delta$ & broken--power-law high-freq.\ spectral index & delta function ($\delta=0$) & fixed \\
$f_{\rm bend}$ [Hz] & broken--power-law bend frequency & log-Uniform  [$-8.7$,$-7$] & one parameter for PTA \\
$\ell$ & broken--power-law high-freq.\ transition sharpness & delta function ($\ell=0.1$) & fixed \\
\hline

\multicolumn{4}{c}{all common processes, $t$-process spectrum} \\[1pt]
$A$ & power-law amplitude & log-Uniform $[-18, -11]$ & one parameter for PTA \\
$\gamma$ & power-law spectral index & Uniform $[0,7]$ & one parameter per PTA \\
$x_{i}$ & modification factor & Inverse Gamma Distribution & one parameter per frequency \\
\hline

\multicolumn{4}{c}{all common processes, turnover spectrum} \\[1pt]
$A$ & turnover power-law amplitude & log-Uniform $[-18, -11]$ & one parameter for PTA \\
$\gamma$ & turnover power-law high-freq.\ spectral index & Uniform $[0,7]$ & one parameter per PTA \\
$\kappa$ & turnover power-law low-freq.\ spectral index & Uniform $[0,7]$ & one parameter per PTA \\
$f_{0}$ [Hz] & turnover power-law bend frequency & log-Uniform  [$-9$,$-7$] & one parameter for PTA \\
\hline

\multicolumn{4}{c}{all common processes, cross-correlation spline model} \\[1pt]
$y$ & normalized cross-correlation values at spline & Uniform $[-0.9,0.9]$ & seven parameters for PTA \\
& knots $(10^{-3}, 25, 49.3, 82.5, 121.8, 150, 180)^\circ$ & \\

\hline

\end{tabular}
\end{center}
\end{table*}

The prior probability distributions assumed for all analyses in this paper are listed in \autoref{tab:priors}. We use Markov chain Monte Carlo (MCMC) techniques to sample randomly from the joint posterior distribution of our model parameters.
Marginal distributions are obtained simply by considering only the parameter of interest in each sample.
To assess convergence of our MCMC runs beyond visual inspection we use the Gelman--Rubin statistic, requiring $\hat{R} < 1.01$ for all parameters \citep{gelmanrubin1992,vgs21}.
We performed most runs discussed in this paper with the \texttt{PTMCMC} sampler \citep{ptmcmc} and postprocessed samples with \texttt{chainconsumer} \citep{Hinton2016}.

In \citetalias{abb+20} we use an analytic approximation for the uncertainty of marginalized-posterior statistics \citep{wilcox2012introduction}. Here we instead adopt a boostrap approach: we resample the original MCMC samples (with replacement) to generate new sets that act as independent sampling realizations. We then calculate the distributions of the desired summary statistics (e.g., quantiles, marginalized posterior values) over these sets. From these distributions, we determine central values and uncertainties (either medians and $68\%$ confidence intervals, or means and standard deviations).

We rely on a variety of techniques to perform Bayesian model comparison.
The first is thermodynamic integration \citep[e.g.,][]{ogata1989monte,gelman1998simulating}, which computes Bayesian evidence integrals directly through parallel tempering:
we run $N_\beta$ MCMC chains that explore variants of the likelihood function raised to different exponents $\beta$, then approximate the evidence for model $\mathcal{H}$ as
\begin{equation}
\ln p(d|\mathcal{H}) = \int_0^1 \langle \ln p(d|\theta) \rangle_\beta \,\,\mathrm{d}\beta \approx \frac{1}{N_\beta} \sum_\beta \langle \ln p(d|\theta) \rangle_\beta,
\end{equation}
where all likelihoods and posteriors are computed within model $\mathcal{H}$, $\theta$ denotes all of the model's parameters, and the expectation $\langle \ln p(d|\theta) \rangle_\beta$ is approximated by MCMC with respect to the posterior $p_\beta(\theta|d) \propto p(d|\theta,\mathcal{H})^\beta p(\theta,\mathcal{H})$.
The inverse temperatures $\beta$ are spaced geometrically, as is the default in \texttt{PTMCMC}.

To compare nested models, which differ by ``freezing'' a subset of parameters, we also use the Savage--Dickey density ratio~\citep{dickey1971}:
if models $\mathcal{H}$ and $\mathcal{H}_0$ differ by the fact that (say) $\theta_0$ is frozen to 0 in the latter, then $p(d|\mathcal{H}_0) / p(d|\mathcal{H}) = p(\theta_0 = 0|d, \mathcal{H}) / p(\theta_0 = 0|\mathcal{H})$.

When comparing \textit{disjoint} models with different likelihoods (e.g., \modelhd\ versus \modelcurn), we use product-space sampling \citep{cc95,g01}. This method treats model comparison as a parameter estimation problem, where we sample the union of the unique parameters of all models, plus a model-indexing parameter that activates the relevant likelihood function and parameter space of one of the sub-models. Bayes factors are then obtained by counting how often the model index falls in each activation region and taking ratios of those counts.

In some situations, it can be difficult to sample a computationally expensive model directly. In these cases, we sample a computationally cheaper approximate distribution and reweight those posterior samples to estimate the posterior for the computationally expensive model~\citep{HourihaneMeyers2022}.
The reweighted posterior can be used in the thermodynamic-integration or Savage--Dickey methods.
In addition, the mean of the weights yields the Bayes factor between the expensive and approximate models, which may be of direct interest (e.g., \modelhd\ can be approximated by \modelcurn).
We estimate Bayes-factor uncertainties using bootstrapping and, for product-space sampling, with the Markov-model techniques of \cite{cornish2015bayeswave} and \cite{heck2019quantifying}.

\section{Broken power-law model}
\label{sec:appbroken}

As shown in \citetalias{abb+20}, the simultaneous Bayesian estimation of white measurement noise and of red-noise processes described by power laws biases the recovery of the spectral index of the latter \citep{lcc+2017,Hazboun:2019vhv}.
Just as in \citetalias{abb+20} and \citet{2022MNRAS.510.4873A}, we impose a high-frequency cutoff on the red-noise processes. To choose the cutoff frequency, we perform inference on our data with a \modelcurngamma\ model modified so that the common process has power spectral density
\begin{equation}
	S(f) = \frac{A^2}{12\pi^2} \left( \frac{f}{f_\mathrm{ref}}\right)^{-\gamma} \left[ 1 + \left(\frac{f}{f_{\rm break}} \right)^{1/\ell} \right]^{\ell \gamma}\,\, f_\mathrm{ref}^{-3};
\label{eq:brokenpl}
\end{equation}
then set the cutoff to the MAP $f_{\rm break}$.
\autoref{eq:brokenpl} is fairly generic, allowing for separate spectral indices at low ($\gamma$) and high ($\delta$) frequencies. The break frequency $f_{\rm break}$ dictates where the broken power law changes spectral index, while $\ell$ (which we set to 0.1) controls the smoothness of the transition.

The marginal posterior for $f_\mathrm{break}$, obtained in the factorized-likelihood approximation using the techniques of \citet{2023arXiv230315442L}, has median and $90\%$ credible region of $3.2_{-1.2}^{+5.4}\times10^{-8}$~Hz, and a MAP value of $2.75 \times 10^{-8}$ Hz.
The latter is close to $f_{14} = 14/T$ in our frequency basis (with $T$ the total span of the data set), so we use 14 frequencies to model common-spectrum noise processes (see \S\ref{sec:data} and \citetalias{abb+20}).

\section{${t}$-process spectrum model}
\label{sec:apptprocess}

\begin{figure*}[t]
\begin{center}
    \includegraphics{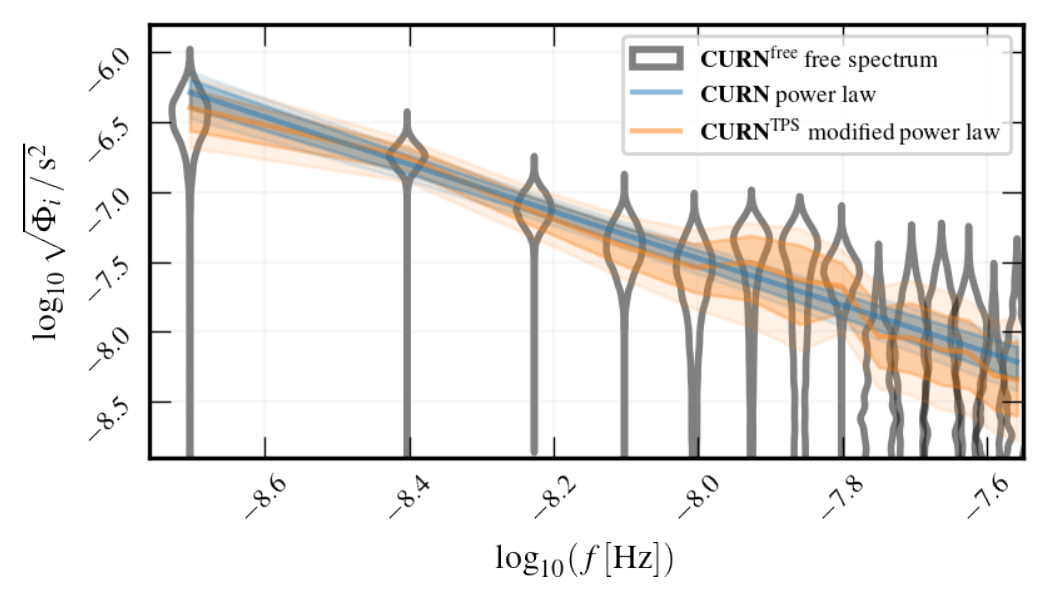}
    \includegraphics{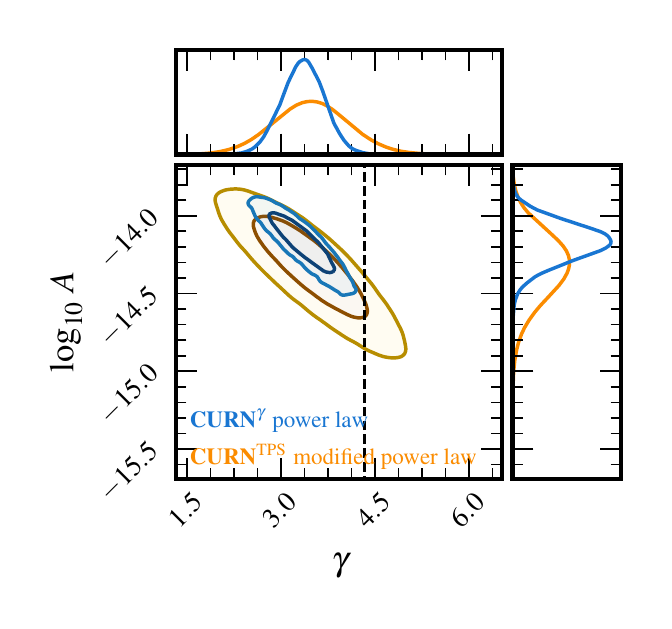}
\end{center}
\vspace{-12pt}
    \caption{Power-law (\modelcurngamma, blue) and $t$-process power-law (\modelcurntps, orange) spectral posteriors. \textbf{Left}: reconstructed spectra, compared to free-spectral bin-variance posteriors (\modelcurntps, violin plots). \textbf{Right}: joint $(\log_{10} A, \gamma)$ posteriors.
The ``fuzzy'' $t$-process allows local deviations from a perfect power law, producing wider constraints that are more consistent with $\gamma=13/3$ (dashed line).}
    \label{fig:tps}
\end{figure*}

The free-spectrum analysis of our data (\S\ref{subsec:spectral} and \autoref{fig:periodogram}) shows that the frequency bins at $f_1$, $f_6$, $f_7$, and $f_8$ appear to be in tension with a pure power law, skewing the estimation of $\gamma$ and reducing the \modelhdgw\ vs.\ \modelcurngw\ Bayes factor.
Assuming that those frequency components reflect unmodeled systematics or stronger-than-expected statistical fluctuations, we can make our inference more robust to such outliers with a ``fuzzy'' power-law model that allows the individual $\Phi_i$ to vary more freely around their expected values.
To wit, we introduce the $t$-process spectrum (TPS)
\begin{equation}
	\Phi_{\mathrm{TPS},i} = x_i \Phi_{\mathrm{powerlaw},i} \quad \text{with} \quad
	x \sim \mathrm{invgamma}(x_i; 1, 1),
\end{equation}
where $\Phi_{\mathrm{powerlaw}, i}$ follows \autoref{eq:powerlaw} and $x$ follows the inverse gamma distribution with parameters $\alpha = \beta = 1$; the resulting Gaussian mixture yields a Student's-$t$ distribution for the $\Phi_{\mathrm{TPS}, i}$.
\autoref{fig:tps} shows \modelcurngamma\ power-law posteriors and \modelcurntps\ modified power-law posteriors, obtained in the factorized-likelihood approximation \citep{2022PhRvD.105h4049T,2023arXiv230315442L} and compared to \modelcurnfree\ bin variances.
The TPS model is spread more widely and deviates from the perfect power law at bins $f_1$, $f_6$, $f_7$, and $f_8$, as expected.
The right panel of \autoref{fig:tps} shows the joint $\log_{10} A, \gamma$ posteriors for \modelcurngamma\ and \modelcurntps.
The latter is more consistent with steeper power laws, and it includes $\gamma=13/3$ at $1\sigma$ credibility.

\section{Turnover model}
\label{sec:appturnover}

The final parameterized spectral model that we investigate is motivated by the idea that the dynamics of SMBHBs are influenced by their environments at sub-parsec separations \citep{2002ApJ...567L...9A,shm+04,mm05}.
These interactions affect binary evolution and the resulting spectrum of the GWB.
The process of bringing two SMBHs together after galaxy mergers involves a complex chain of interactions: despite significant theoretical work, the lack of observational constraints makes it difficult to draw any conclusions. PTAs, however, provide a unique opportunity to probe the timescale over which two SMBHs evolve from the merger of their galaxies to a bound binary that produces GW signals in the PTA sensitivity band.

When dynamical interactions dominate orbital evolution, binaries will traverse the GW spectrum more quickly, reducing GW emission compared to a GW-driven inspiral.
This kind of behavior is captured by the \emph{turnover model} \citep{scm2015}:
\begin{equation}
	S(f) = \frac{A^2}{12\pi^2} \left( \frac{f}{f_\mathrm{ref}}\right)^{-\gamma} \left[ 1 + \left(\frac{f_0}{f} \right)^{\kappa} \right]^{-1}\,\, f_\mathrm{ref}^{-3}.
\end{equation}
This is qualitatively similar to the broken power law discussed earlier, except that here $f_0$ represents the GW frequency at which typical binary evolution transitions from environmentally dominated (at lower frequencies and wider separations) to GW-dominated (at higher frequencies and smaller separations). The parameter $\kappa$ controls the shape of the spectrum below $f_0$, and depends on the orbital-evolution mechanism. Note that the actual turning point of the spectrum is not at $f_0$ but at $f_\mathrm{bend} = f_0 \times (3\kappa/4 - 1)^{1/\kappa}$ \citepalias{abb+16}.

Applying this model to our data, we find hints of departures from a pure power law: the transition frequency $f_0$ lies below $10$~nHz with $65\%$ credibility, while the bend frequency lies below $10$~nHz with $75\%$ credibility. Nevertheless, Bayesian comparison of this \modelcurnturnover\ model with \modelcurngamma\ reports an inconclusive Bayes factor of $1.46 \pm 0.02$ in favor of \modelcurnturnover. Furthermore, the estimation of \modelcurnturnover\ parameters is 
sensitive to DM modeling (see \S\ref{subsec:dm_models}). 
While the spectra are broadly consistent whether we use DMX or DMGP to model DM fluctuations, 
there are differences in the power at certain frequencies that lead to differences in the turnover parameters. 
This is discussed in greater detail in \citet{aaa+23_smbhb}.

\section{Sky scrambles}
\label{sec:appskyscrambles}

\begin{figure*}
\begin{center}
    \includegraphics[width=0.9\textwidth]{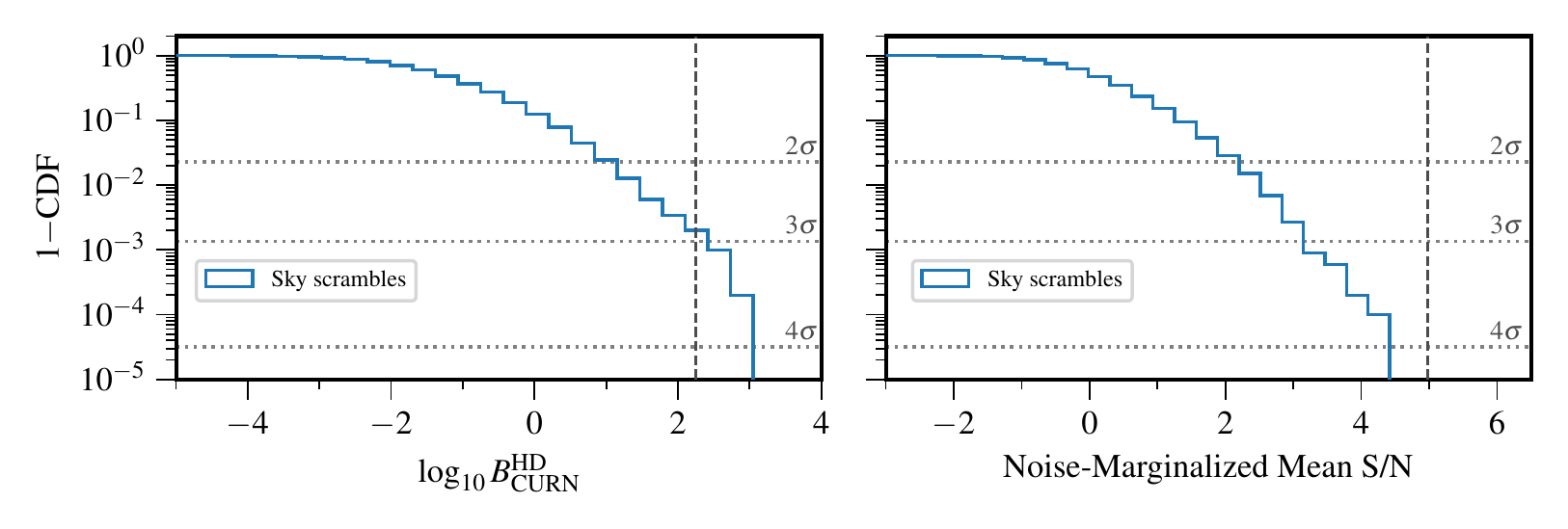} 
\end{center} \vspace{-18pt}
    \caption{Empirical background distribution of \modelhdgamma-to-\modelcurngamma\ Bayes factor (left, see \S\ref{sec:bayes}) and noise-marginalized optimal statistic (right, see \S\ref{sec:optimal}), as computed in 5,000 sky scrambles, 
    which erases the dependence of inter-pulsar correlations on the angular separation between the pulsars.
Dotted lines indicate Gaussian-equivalent 2$\sigma$, 3$\sigma$, and 4$\sigma$ thresholds. The dashed vertical lines indicate the values of the detection statistics for the unscrambled data set. We find $p=1.6 \times 10^{-3}$ (approx.\ $3\sigma$) for the Bayesian analysis, and $p < 10^{-4}$ ($>3\sigma$) for the optimal-statistic analysis. 
    \label{fig:skyscrambles}}
\end{figure*}

In the sky-scramble method \citep{cs16}, inter-pulsar correlations are analyzed as if the pulsars occupied random sky positions, with the purpose of creating a background distribution of PTA detection statistics for null-hypothesis testing, in alternative to phase shifts (\citealt{tlb+17}; see \S\ref{sec:bayes} and \S\ref{sec:optimal}). 
If a correlated signal is present in the data, phase shifts and sky scrambles actually test \textit{different} 
null hypotheses: phase shifts test the hypothesis that no inter-pulsar correlations are present, 
while sky scrambles assume that inter-pulsar correlations are present at the level measured in the data, 
but test the hypothesis that these correlations have no dependence on angular separation.

As is the convention in the literature, we require that scrambled overlap reduction functions (ORFs)
be independent of each other and of the unscrambled ORF using a match statistic,
\begin{equation}
	\bar{M} = \frac{\sum_{a,b \neq a} \Gamma_{ab} \Gamma'_{ab}}{\sqrt{ \left( \sum_{a, b \neq a} \Gamma_{ab} \Gamma_{ab} \right) \left( \sum_{a, b \neq a} \Gamma'_{ab} \Gamma'_{ab} \right)}} \,,
\end{equation}
where $\Gamma_{ab}$ and $\Gamma'_{ab}$ are two different ORFs. 
For the sky scrambles used in our analysis, the scrambled ORFs have $\bar{M} < 0.1$ 
with respect to the unscrambled ORF, and $\bar{M} < 0.17$ with each other. 
We generate 10,000 sky scrambles, owing to the difficulty in obtaining large numbers of scrambled ORFs that satisfy the match threshold; because of limitations of computational resources, we obtain our detection statistics for 5,000 of those ORFs. 
\autoref{fig:skyscrambles} shows the resulting background distributions for the \modelhdgamma-to-\modelcurngamma\ Bayes factor (left panel) and the optimal-statistic S/N (right panel). 
The Bayes factors exceed the observed value in eight of the 5,000 sky scrambles ($p = 1.6 \times 10^{-3}$), 
while none of the sky scrambles have noise-marginalized mean S/N greater than observed ($p < 10^{-4}$).

\begin{figure}
\begin{center}
    \includegraphics[width=\columnwidth]{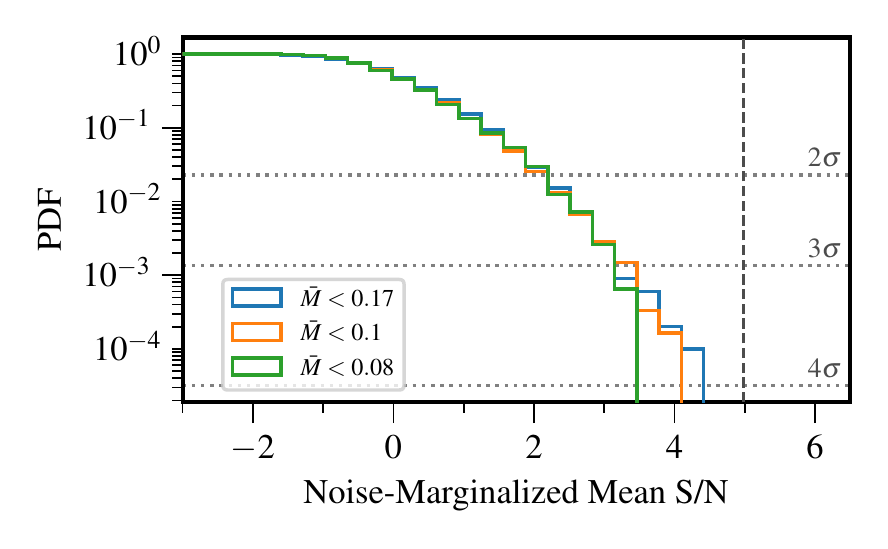} 
\end{center} \vspace{-18pt}
    \caption{Comparison between empirical background distributions for the noise-marginalized optimal statistic, 
as computed by the sky-scramble technique. We show distributions computed using a match threshold of 
$\bar{M} < 0.17$ (blue), $\bar{M} < 0.1$ (orange), and $\bar{M} < 0.08$ (green). 
Dotted lines indicate Gaussian-equivalent 2$\sigma$, 3$\sigma$, and 4$\sigma$ thresholds. The dashed vertical lines indicate the values of the detection statistics for the unscrambled data set.
We find little difference between the background distributions computed using different match thresholds, 
modulo the fact that imposing a smaller threshold yields fewer sky scrambles, 
which limits the precision to which the $p$-value can be measured.
    \label{fig:skyscrambles_match}}
\end{figure}
We note that the null distribution recovered by the sky scrambles is not very sensitive to the choice of match threshold 
for $|\bar{M}| \lesssim 0.2$. \autoref{fig:skyscrambles_match} compares the null distributions when the match threshold for all ORFs with each other and with the unscrambled ORF is set to 
$|\bar{M}| < 0.17$ (blue), $|\bar{M}| < 0.1$ (orange), and $|\bar{M}| < 0.08$ (green). 
There is very little difference among the distributions; however, imposing a smaller threshold means that fewer sky scrambles can be used (6,043 with $|\bar{M}| < 0.1$ and 1,534 with $|\bar{M}| < 0.08$, compared to 10,000 with $|\bar{M}| < 0.17$), which limits the precision with which the $p$-value can be measured. 
We find no evidence that the recovered null distribution is biased 
when including sky scrambles with matches up to 0.17.

\section{Multiple-correlation optimal statistic}
\label{sec:appmcos}


\begin{table*}
  \begin{center}
  \caption{Multiple-correlation optimal statistic best-fit coefficients $\hat{A}_k^2$, S/Ns, and AIC probabilities}
  \vspace{-6pt}
  \begin{tabular}{l|cccccc|c}
  \hline \hline
   & \multicolumn{2}{c}{HD Correlations} & \multicolumn{2}{c}{Monopole Correlations} & \multicolumn{2}{c|}{Dipole Correlations} &  \\
  Model & $\hat{A}_\mathrm{HD}^2$ & S/N & $\hat{A}_\mathrm{mo}^2$ & S/N & Mean $\hat{A}_\mathrm{di}^2$ & S/N & $p(\mathrm{AIC})$ \\
  \hline
  HD only  & $6.8(9) \times 10^{-30}$ & 4(1) & \nodata & \nodata & \nodata & \nodata & $3 \times 10^{-2}$ \\ 
  Monopole only  & \nodata & \nodata & $1.1(1) \times 10^{-30}$ & $4(1)$ & \nodata & \nodata & $6 \times 10^{-3}$ \\ 
  Dipole only  & \nodata & \nodata & \nodata & \nodata & $1.5(3) \times 10^{-30}$ & 4(1) & $8 \times 10^{-4}$ \\ 
HD + monopole & $5.5(8) \times 10^{-30}$ & 3.4(8) & $8(1) \times 10^{-31}$ & 2.9(8) & \nodata & \nodata & 1 \\			
HD + dipole & $5.5(8) \times 10^{-30}$ & 3.2(7) & \nodata & \nodata & $8(2) \times 10^{-31}$ & 1.7(7) & $6 \times 10^{-2}$ \\ 
monopole + dipole & \nodata & \nodata & $8(1) \times 10^{-31}$ & 2.7(7) & $9(2) \times 10^{-31}$ & 1.9(6) & $1 \times 10^{-2}$ \\ 
HD + monopole + dipole & $5.1(8) \times 10^{-30}$ & 2.9(6) & $7(1) \times 10^{-31}$ & 2.4(6) & $3(2) \times 10^{-31}$ & 0.6(4) & 0.48 \\ 
  \hline
  \end{tabular}
  \end{center}
  \vspace{-6pt}
  \label{tab:mc_optstat}
  \tablecomments{All values were computed for the 15-year data set, assuming a power-law power spectral density using the 14 lowest frequency components.
  Here $\hat{A}^2$, S/N, and AIC are marginalized over pulsar noise parameters with fixed $\gamma=13/3$.
  The numbers in parentheses represent the mean least-squares errors for the $\hat{A}_k^2$ coefficients and standard deviations over noise-parameter posteriors for S/Ns. 
  We compute $p(\mathrm{AIC})$ with respect to the model with the lowest mean AIC (i.e., HD + monopole).}
\end{table*}

The multiple-correlation optimal statistic (MCOS; \citealt{sv2023}) fits the inter-pulsar correlation coefficients $\rho_{ab}$ with a linear model that includes multiple components with different correlation patterns, but with the same spectral shape.
The linear-model coefficients are the squared amplitudes of the components.
Within such a model, the significance of each component can be quoted as a S/N given by its best-fit coefficient divided by the fit error.
Just as for the noise-marginalized optimal statistic \citep{vite18}, the posterior distribution of pulsar noise parameters induces a distribution of MCOS statistics. 

We fit the 15-year data with models that include \mbox{HD,} \mbox{monopole,} and dipole-correlated components in various combinations.
\autoref{tab:mc_optstat} lists the noise-marginalized amplitude estimates and S/N for all models.
The goodness-of-fit of the models can be compared using the Akaike Information Criterion (AIC; \citealt{Akaike1998}): 
\begin{equation} \label{eq:AIC}
   \mathrm{AIC} = 2k + \chi^2  \,,
\end{equation}
where $k$ is the number of model parameters and $\chi^2$ is the fit's chi-squared, computed without accounting for GW-induced $\rho_{ab}$ correlations.
(This can be thought of as a pseudo-Bayes factor, with the factor of $2k$ imposing an Occam penalty.)
The relative probability of a model compared to the most-favored model is then given by
\begin{equation} \label{eq:AIC comparison}
    p(\mathrm{AIC}) = \exp \, \left[(\mathrm{AIC_{min}} - \mathrm{AIC}) / 2 \right] \,,
\end{equation}
where $\mathrm{AIC_{min}}$ is the minimum AIC across all models.

\autoref{tab:mc_optstat} lists the AIC probabilities, computed by averaging the AIC of each model over pulsar noise parameters.
The HD-correlated model is preferred among the models with a single correlated process. 
The models with both HD and monopole correlations are preferred among all models: 
for a model with HD and monopole correlations, we find S/N of $3.4 \pm 0.8$  for HD correlations and $2.9 \pm 0.8$ for monopolar correlations, while for a model with HD, monopole, and dipole correlations, we find S/N of $2.9 \pm 0.6$ for HD correlations, $2.4 \pm 0.6$ for monopole correlations, and $0.6 \pm 0.4$ for dipole correlations (means $\pm$ standard deviations across noise-parameter posteriors).
The statistical significance of these S/Ns can be quantified empirically using simulations of 15-year--like data sets (see App.\ \ref{sec:appastrosims}), which report $p$-values $< 10^{-2}$ and $\simeq 4 \times 10^{-2}$ for the observed mean HD and monopole statistics across data replications with no spatially correlated injections.
\begin{figure}[t]
\begin{center}
    \includegraphics[width=0.85\columnwidth]{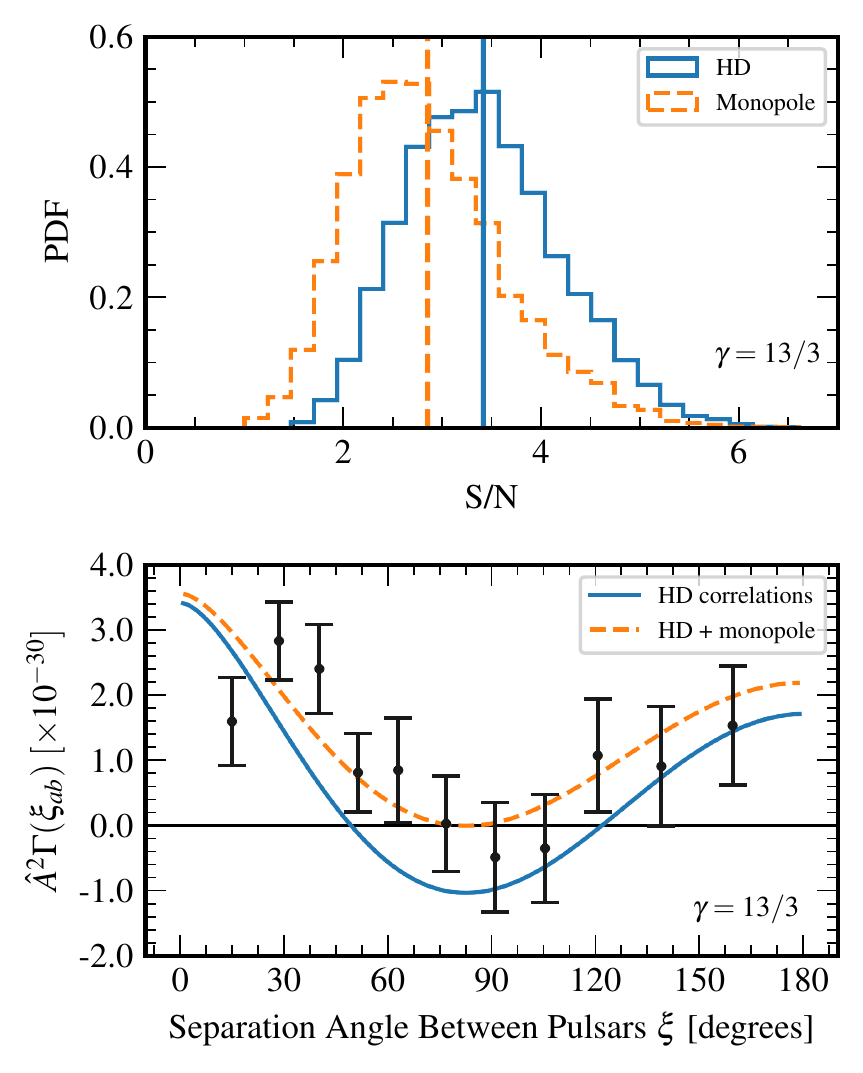}
\end{center}
	\vspace{-12pt}
    \caption{Results of the MCOS analysis, which prefers a model including both HD and monopole correlations. 
    \textbf{Top}: MCOS S/N for HD correlations (solid blue) and monopole correlations (dashed orange), marginalized over \modelcurngw\ noise-parameter posteriors. The vertical lines indicate the mean S/Ns. We find a S/N of $3.4\pm0.8$ for HD correlations and $2.9\pm0.8$ for monopole correlations. 
    \textbf{Bottom}: Binned cross-correlations $\rho_{ab}$ (black error bars), computed with MAP noise parameters from a \modelcurngw\ run. The solid blue and dashed orange curves show best-fit HD and HD+monopole correlation patterns, corresponding to $\hat{A}^2 = 6.8 \times 10^{-30}$ and to $\hat{A}_\mathrm{HD}^2 = 5.5 \times 10^{-30}$, $\hat{A}^2_\mathrm{monopole} = 8 \times 10^{-31}$, respectively.
The monopolar component accounts for the vertical shift of the cross-correlations with respect to the HD curve. 
We use the standard version of the optimal statistic that does not include inter-pulsar correlations 
to compute $\rho_{ab}$, so the points and errors do not match those shown in panel (c) of \autoref{fig:spectrum_correlations_plot}.}    \label{fig:mcos_corr}
\end{figure}

As discussed in \citet{sv2023}, the optimal statistic and the MCOS are metrics of the \emph{apparent} spatial correlation pattern of the data, but they have a limited ability to identify its actual source.
That is because a real HD signal may also excite the monopole optimal statistic and the monopole component of the MCOS; conversely, a real monopolar signal may also excite the HD optimal statistic and the HD component of the MCOS; and so on.
The S/Ns quoted in \autoref{tab:mc_optstat} quantify how often we would expect to measure the observed value of the optimal statistic if only uncorrelated noise is present, but they do not describe how often one type of correlated noise would produce a given value of the optimal statistic for a different type of correlation. 
This effect can be characterized using simulations (see App.\ \ref{sec:appastrosims}), which report a $p$-value of $0.11$ for the observed mean monopole statistic when a HD-correlated signal with the MAP 15-year amplitude is included in the simulated data sets. 
We conclude that there are some indications of a possible monopole-correlated signal in the data 
with S/N comparable to but smaller than the S/N for HD correlations; 
however, from simulations we conclude that it is possible for such a signal to appear in an MCOS analysis 
if only a HD-correlated stochastic process is present.

\section{Multiple-correlation optimal statistic simulations}

In this appendix we obtain the distribution of the MCOS over an ensemble of simulated data sets, with the goal of characterizing the probability that the observed S/Ns could have been produced by pulsar noise alone, or by a GWB with HD correlations.
Unlike our Bayesian analysis, the MCOS prefers a model that includes both HD and monopolar components. So we are especially interested in asking how frequently we may expect the observed MCOS monopole if the data contain only the GWB.
In Apps.\ \ref{sec:appastrosims} and \ref{sec:appmodelsims} we present two different types of simulations: ``astrophysical,'' where we generate synthetic data with MAP noise parameters inferred from the 15-year data set, both with and without the GWB; and ``model checking,'' where we create data replications following the \modelhdgw\ posteriors for the real data set.
Note that neither simulation attempts to account for the monochromatic character of the putative monopolar signal (see \S\ref{subsec:spectral}).

\subsection{Astrophysical simulations}
\label{sec:appastrosims}
\begin{figure}[tb]
	\begin{center}
		\includegraphics[width=\columnwidth]{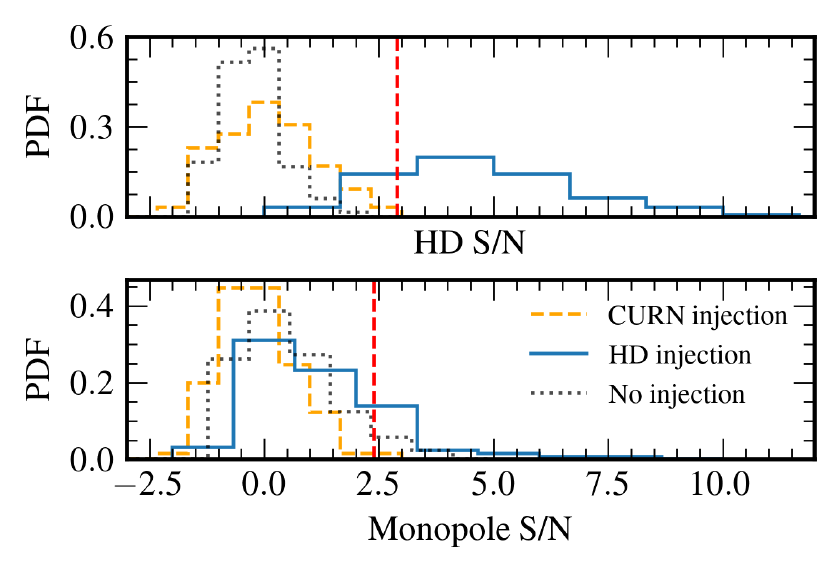}
	\end{center}
	\vspace{-12pt}
	\caption{\textbf{Top:} MCOS HD S/N values recovered in the three simulations described in App.\ \ref{sec:appastrosims}, compared to the MCOS HD S/N measured in the real data set (vertical dashed red line), which has $p$-values of $< 10^{-2}$ for simulations i and iii, and $0.64$ for simulations ii. 
	\textbf{Bottom:} MCOS monopole S/N values recovered in the three simulations, compared to the real-data MCOS monopole S/N (vertical dashed red line), which has $p$-values of $4 \times 10^{-2}$, $1.1 \times 10^{-1}$, and $< 10^{-2}$ for simulation i, ii, and iii respectively.}
	\label{fig:a4c_sims_cc}
\end{figure}

Following \citet{astro4cast}, we generate simulated data sets adopting MAP pulsar-noise parameters obtained from the real data independently for each pulsar; these ``noise runs'' include an additional power-law process to reduce contamination between the putative GWB and the pulsars' intrinsic red noise \citep{2022PhRvD.105h4049T}.
We produce 100 realizations each of three different simulations: (i) injecting no spatially correlated power-law GWB or excess uncorrelated common-spectrum noise; (ii) injecting a spatially correlated power-law GWB with amplitude $2.7 \times 10^{-15}$ and spectral index $13/3$; and (iii) injecting no GWB or common-spectrum noise, but omitting the additional power-law process in the estimation of intrinsic pulsar noise, with the goal of testing how often excess common-spectrum noise is recognized as a spatially correlated GWB.

We compute HD + monopole + dipole MCOS S/Ns for all synthetic data sets (see \autoref{fig:a4c_sims_cc}). The mean HD S/Ns observed in the real data (see App.\ \ref{sec:appmcos}) correspond to $p$-values of $< 10^{-2}$ for simulations (i) and (iii), and $0.64$ for simulation (ii). The mean monopole S/Ns observed in the real data set correspond to $p$-values of $4 \times 10^{-2}$, $1.1 \times 10^{-1}$, and $< 10^{-2}$ for simulations (i), (ii), and (iii) respectively. 
We conclude that it is unlikely that we would measure HD correlations at the level observed in real data when no correlated signal is present (simulation (i)) or when only uncorrelated common-spectrum red noise is present (simulation (iii)).
In addition, the HD S/Ns obtained from a HD-correlated GWB injection (simulation (ii)) are fully consistent with the S/N observed in real data.
By contrast, the observed monopole S/N could have been produced by intrinsic pulsar noise alone, or by a real HD signal.

\subsection{Model-checking simulations}
\label{sec:appmodelsims}
\begin{figure}[t]
	\includegraphics[width=0.45\textwidth]{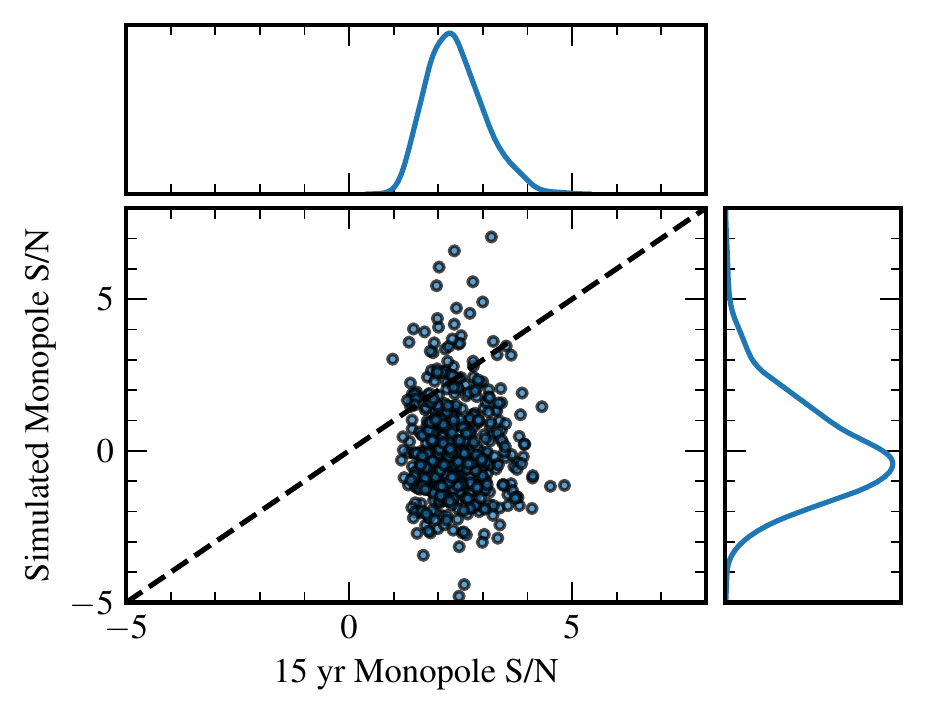}
	\caption{Distribution of real-data and replicated MCOS monopole S/Ns. Each point represents a draw ${\bm\eta}^{(k)}$ from \modelhdgw\ posterior, which is used to simulate $\tresid^{\textrm{sim},(k)}$ and to compute both S/Ns.
	The replicated monopole S/N is greater for 11\% of the simulations.}
	\label{fig:monopole_pvalue_enterprise_sims}
\end{figure}

In App.\ \ref{sec:appastrosims} we have tackled the question of monopole S/N significance using simulations based on real-data MAP estimates ${\bm \eta}^\mathrm{MAP}$ of pulsar-noise and GW parameters.
In this appendix we adopt a procedure with a stronger Bayesian flavor, evaluating the MCOS on a population of \emph{data replications} created using \modelhdgw\ as a generative model with noise hyperparameters ${\bm\eta}$ drawn from the \modelhdgw\ real-data posterior.
This can be seen also as a Bayesian model-checking exercise \citep{gelman_bayes_pvals,gelman2013bayesian}: if we find that the summary statistic of interest (the monopole MCOS) has a much more extreme value in real data than in data replications, we should suspect that the data model (here \modelhdgw) is missing something.

We perform the test by drawing 500 parameter vectors $\{{\bm\eta}^{(k)}\}$ from the \modelhdgw\ real-data posterior; for each ${\bm\eta}^{(k)}$ we simulate a data set $\tresid^{\textrm{sim},(k)}\sim p(\tresid | \etavec^{(k)})$ and compare $\mathrm{MCOS}(\tresid^{\textrm{sim},(k)};\etavec^{(k)})$ with $\mathrm{MCOS}(\tresid;\etavec^{(k)})$.
Our notation emphasizes the dependence of the MCOS on the pulsar noise parameters through the $\mathbf{P}$ matrices in \autoref{eq:rhoab}. 
\autoref{fig:monopole_pvalue_enterprise_sims} shows the resulting distribution of monopole S/Ns.
The replicated monopole S/N is greater than its observed counterpart for 11\% of the draws. Thus, it is plausible that the MCOS could measure the observed monopole S/N in data that contain only a HD-correlated GWB. Conversely, the observed monopole S/N does not by itself suggest that \modelhdgw\ is misspecified.

\end{document}